\documentclass[eqnobysec,12pt]{iopart}
\usepackage{bm}
\usepackage{graphicx}

\newcommand{\beq}{\begin{equation}}
\newcommand{\eeq}{\end{equation}}
\newcommand{\beqa}{\begin{eqnarray}}
\newcommand{\eeqa}{\end{eqnarray}}
\newcommand{\eq}[1]{Eq.~(\ref{#1})}
\newcommand{\be}{\begin{eqnarray}}
\newcommand{\ee}{\end{eqnarray}}
\newcommand{\nn}{\nonumber \\ }

\begin{document}

\title[Chiral EFT based nuclear forces]{Chiral EFT based nuclear forces: Achievements and challenges}

\author{R Machleidt and F Sammarruca}

\address{Department of Physics, University of Idaho, Moscow, Idaho 83844, U.S.A.}
\ead{machleid@uidaho.edu, fsammarr@uidaho.edu}

\begin{abstract}
During the past two decades, chiral effective field theory has become a popular tool to derive nuclear forces from first principles. Two-nucleon interactions have been worked out up to  sixth order of chiral perturbation theory and three-nucleon forces up to fifth order.
Applications of some of these forces have been conducted 
in nuclear few- and many-body
systems---with a certain degree of success. 
But in spite of these achievements, we are still faced with great challenges. 
Among them is the issue of a proper 
uncertainty quantification of predictions obtained when applying these forces
in {\it ab initio} calculations of nuclear structure and reactions. A related problem is 
the order by order convergence of the chiral expansion.
We start this review with a pedagogical introduction and then present the current status of the field of chiral nuclear forces.
This is followed by a discussion of representative examples for the application of chiral two- and three-body forces in the nuclear many-body system including convergence issues.
\end{abstract}

%\keywords{Low-energy QCD, effective field theory, chiral perturbation theory, nucleon-nucleon scattering.}

\submitto{\PS}
\maketitle

\section{Historical Perspective}

In 1975, when A. Bohr, B. Mottelson, and L. Rainwater were honored with the Nobel Prize, the Reid potential~\cite{Rei68} was the most popular
nucleon-nucleon ($NN$) potential within the international nuclear physics community. 
It was applied in most miscroscopic nuclear structure
calculations produced in the 1970's. The Reid potential is a phenomenological potential that
was considered very quantitative by the standards of the time and easy to use, which explains its popularity. However, attempts to derive the $NN$ interaction on fundamental grounds had been around for quite a while. Since the mid 1960's, one-boson exchange potentials were being developed~\cite{BS64}, which by the mid 1970's 
assumed a quantitative character comparable to the Reid potential~\cite{Erk74,HM75,Mac89}. Moreover, research that went beyond the simple one-boson-exchange assumption (which always includes a `fictitious' $\sigma$-boson) was also under way.
The most notable work of this kind became known as the Paris~\cite{Lac80} and the Bonn potentials~\cite{MHE87}.

Since the more sophisticated meson models seemed to have a sound theoretical foundation and, in addition, were quantitatively very successful, it appeared that they were the solution of the nuclear force problem. However, with the discovery (in the 1970's) that the fundamental theory of strong interactions is quantum chromodynamics (QCD) and not meson theory, all ``meson theories'' had to be viewed as models, and the attempts to derive the nuclear force from first principals had to start all over again.

The problem with a derivation of nuclear forces from QCD is two-fold. First, each nucleon consists of three valence quarks, quark-antiquark pairs, and gluons such that the system of two nucleons is a complicated many-body problem. Second, the force between quarks, which is created by the exchange of gluons, has the feature of being very strong at the low energy-scale that is characteristic of nuclear physics. This extraordinary strength makes it difficult to find converging expansions. Therefore, during the first round of new attempts, QCD-inspired quark models became popular. The positive aspect of these models is that they try to explain nucleon structure (which consists of three quarks) and nucleon-nucleon interactions (six-quark systems) on an equal footing. Some of the gross features of the two-nucleon force, like the ``hard core'', are explained successfully in such models. However, from a critical point of view, it must be noted that these quark-based approaches are yet another set of models and not a theory. Alternatively, one may try to solve the six-quark problem with brute computing power, by putting the six-quark system on a four dimensional lattice of discrete points which represents the three dimensions of space and one dimension of time. This method has become known as lattice QCD and is making progress. However, such calculations are computationally very expensive and cannot be used as a standard nuclear physics tool.

Around 1980, a major breakthrough occurred when the nobel laureate Steven Weinberg applied the concept of an effective field theory (EFT) to low-energy QCD~\cite{Wei79,Wei91}. He simply wrote down the most general Lagrangian that is consistent with all the properties of low-energy QCD, since that would make this theory equivalent to low-energy QCD. A particularly important property is 
$SU(2)_R\times SU(2)_L$ symmetry,
the so-called chiral symmetry, which is ``spontaneously'' broken. 
Massless spin-$\frac12$ fermions posses chirality, which means that their spin and momentum are either parallel
to each other (``right-handed'') or anti-parallel (``left-handed'') and remain so forever. Since the quarks, which nucleons are made of (``up'' and ``down'' quarks), are almost massless, approximate chiral symmetry is a given. Naively, this symmetry should have the consequence that one finds in nature mesons of the same mass, but with positive and negative parity. However, this is not the case and such failure is termed a ``spontaneous'' breaking of the symmetry. According to a theorem first proven by Goldstone, the spontaneous breaking of a symmetry
implies the existence of a particle, here, the pseudoscalar pion. Thus, the pion becomes the main player in generating the nuclear force. The interaction of pions with nucleons is weak
at low energies as compared to the interaction of gluons with quarks. Therefore, pion-nucleon processes can be calculated without problem. Moreover, this effective field theory can be expanded in powers of momentum over ``scale'', where scale denotes the ``chiral symmetry breaking scale'' which is about 1 GeV. This scheme is also known as chiral perturbation theory (ChPT) and allows to calculate the various terms that make up the nuclear potential systematically power by power, or order by order. Another advantage of the chiral EFT approach is its ability to generate not only the force between two nucleons, but also many-nucleon forces, on the same footing~\cite{Wei92}.
In modern theoretical nuclear physics, the chiral EFT approach is becoming increasingly popular and is applied with great success~\cite{ME11,EHM09,Mei16,Mac14,Mac16}.

This article is organized as follows.
In Sec.~\ref{sec_EFT}, we will present a pedagogical introduction into the EFT approach to low-energy QCD, including the development of effective Lagrangians. 
Section~\ref{sec_overview} provides a broad overview of the hierarchy of nuclear forces as they emerge from EFT. Sections~\ref{sec_pions} to \ref{sec_pot2} then spell out in detail the development of the two-nucleon forces from long-range to short-range and the construction of quantitative $NN$ potentials. 
Chiral many-body forces are presented in Sec.~\ref{sec_manyNF}.
Applications of chiral forces
in the many-body problem and convergence issues are discussed in Sec.~\ref{sec_manybody}, and Sec.~\ref{sec_concl}
concludes the article.

\begin{table}
\caption{Abbreviations and acronyms used in this article.}
%\lineup
\begin{indented}
\item[] \begin{tabular}{@{}lll}
\br
   Abbreviation/Acronym   &  Explanation  \\
\mr
CCWZ & Callan, Coleman, Wess, and Zumino~\cite{CCWZ}\\
ChPT & chiral perturbation theory  \\
CMS & center-of-mass system  \\
ct & contact (term) & \\
EFT & effective field theory \\
EoS & equation of state \\
FFG & free Fermi gas \\
GW & George Washington (University) \\
HI & heavy ion \\
IANM & isospin-asymmetric nuclear matter \\
KH & Karlsruhe (University) \\
LEC & low-energy constant \\
LO & leading order \\
LS & Lippmann-Schwinger \\
NLO & next-to-leading order \\
NM & neutron matter \\
$NN$ & nucleon-nucleon \\
NNLO, N$^2$LO & next-to-next-to-leading order \\
N$^3$LO, ... & next-to-next-to-next-to-leading order, ... \\
PREX & lead radius experiment \\
PWA & partial-wave analysis \\
QCD & quantum chromodynamics \\
SFR & spectral function renormalization~\cite{EGM04} \\
SNM & symmetric nuclear matter \\
$SU(n)$ & special unitary group in $n$ dimensions \\
VPI & Virginia Polytechnic Institute \\
1PE & one-pion exchange \\
2PE & two-pion exchange \\
3PE & three-pion exchange \\
2NF & two-nucleon force \\
3NF & three-nucleon force \\
4NF & four-nucleon force \\
\br
\end{tabular}
\end{indented}
\label{tab_acro}
\end{table}

\section{Effective field theory for low-energy QCD
\label{sec_EFT}}

Quantum chromodynamics provides the theoretical framework
to describe strong interactions, namely interactions involving  
quarks and gluons. According to QCD, objects which carry color
interact weakly at short distances and strongly at large
distances, where the separation between the two regimes is
about 1 fm. Naturally, short distances and long distances can be
associated with high and low energies, respectively, causing the     
quarks to be confined into hadrons, which carry no color. 
At the same time, the weak nature of the force at high energies 
results into what is known as ``asymptotic freedom". 
(We note that these behaviors originate from the fact that 
QCD is a non-Abelian gauge field theory
with color $SU(3)$ the underlying gauge group.) 
Therefore, QCD is perturbative at high energy, 
but strongly coupled at low-energy.             
The energies typical for nuclear physics are low and, thus,
nucleons are appropriate degrees of freedom. The nuclear force 
can then be regarded as a      
residual color interaction acting between nucleons in a way similar to how
the van der Waals forces bind neutral molecules.
If described in terms of quark and gluon degrees of freedom, the interaction  
between nucleons is an extremely complex problem, which can be confronted 
with the computational methods known as lattice QCD.
In a recent paper~\cite{Org15}, the 
nucleon-nucleon system is investigated at a pion mass of about 450 MeV.
Over the range of energies that are studied, the scattering phase shifts in the $^1S_0$ and 
$^3S-D_1$ channels are found to be similar to those in nature and indicate a repulsive
short-range component of the interaction.
 This result is then extrapolated
to the physical pion mass with the help of chiral perturbation theory. The pion mass
of 450 MeV is still too large to allow for reliable extrapolations, but the feasibility has been
demonstrated and more progress can be expected for the near future.
In a lattice calculation of a very different kind, the nucleon-nucleon ($NN$) potential
was studied in Ref.~\cite{Hat12}. The central component of this potential exhibits 
repulsion at the core as well as intermediate-range attraction. This is encouraging, but 
one must keep in mind that the pion masses employed in this study are still quite large.
In summary, although calculations within lattice QCD are being performed and improved, 
 they are computationally very costly, and thus they are useful, in practice, only to explore a few cases.  
 Clearly, a different approach is necessary to address a full variety of nuclear structure problems. 

The concept of an effective field theory shows an alternative and realistic way to proceed.
The first step towards the development of an EFT is the identification of appropriate scales. 
The large difference between the masses of
the pions and the masses of the vector mesons, like $\rho(770)$ and $\omega(782)$, provides a clue.
From that observation, one is prompted to take the pion mass as the identifier of the soft scale, 
$Q \sim m_\pi$,
while the rho mass sets the hard scale, $\Lambda_\chi \sim m_\rho$, often referred to 
as the chiral-symmetry breaking scale.
It is then natural to 
consider an expansion in terms of                                            
$Q/\Lambda_\chi$. With regard to the choice of  
degrees of freedom, we observed earlier that, as far as 
conventional nuclear physics is concerned, 
quarks and gluons are ineffective and thus             
nucleons and pions should be taken as the appropriate degrees of freedom. 
Of course, we do not wish to construct yet one more phenomenological model and,
therefore, our EFT must be firmly linked with QCD. 
This strong link is present if we require the EFT to be consistent with 
the symmetries of QCD.                     
The meaning and relevance of such statement is expressed in the so-called `folk theorem' by
Weinberg~\cite{Wei79}:
\begin{quote}
If one writes down the most general possible Lagrangian, including {\it all}
terms consistent with assumed symmetry principles,
and then calculates matrix elements with this Lagrangian to any given order of
perturbation theory, the result will simply be the most general possible 
S-matrix consistent with analyticity, perturbative unitarity,
cluster decomposition, and the assumed symmetry principles.
\end{quote}
        
To summarize, the development of a proper EFT must proceed as follows:                      
\begin{enumerate}
\item
Identify the low- and high-energy scales, and the degrees of freedom suitable       
for (low-energy) nuclear physics.
\item
Recognize the symmetries of low-energy QCD and
explore the mechanisms responsible of their breakings.                         
\item
Build the most general Lagrangian which respects those
(broken) symmetries.                        
\item
Formulate a scheme to organize contributions in order 
of their importance. 
Clearly, this amounts to performing an expansion in 
terms of (low) momenta. 
\item
Using the expansion mentioned above, evaluate Feynman diagrams
to any desired accuracy.
\end{enumerate}
           
In what follows, we will discuss each of the steps above. 
Note that the first one has already been addressed, so 
we will move directly to the second one.

\subsection{Symmetries of low-energy QCD}

Our purpose here is to provide a compact introduction into 
(low-energy) QCD, with particular attention to the symmetries
and their breakings. 
For more details the reader is referred to                 
Refs.~\cite{ME11,Sch03}.

\subsubsection{Chiral symmetry}

We begin with the QCD Lagrangian,         
\begin{equation}
{\cal L}_{\rm QCD} = 
\bar{q} (i \gamma^\mu {\cal D}_\mu - {\cal M})q
 - \frac14 
{\cal G}_{\mu\nu,a}
{\cal G}^{\mu\nu}_{a} 
\label{eq_LQCD}
\end{equation}
with the gauge-covariant derivative
\begin{equation}
{\cal D}_\mu = \partial_\mu -ig\frac{\lambda_a}{2}
{\cal A}_{\mu,a}
\label{eq_Dm}
\end{equation}
and the gluon field strength tensor\footnote{For $SU(N)$ group indices, we use 
Latin letters, $\ldots,a,b,c,\ldots,i,j,k,\dots$,
and, in general, do not distinguish between subscripts and superscripts.}
\begin{equation}
{\cal G}_{\mu\nu,a} =
\partial_\mu {\cal A}_{\nu,a}
-\partial_\nu {\cal A}_{\mu,a}
 + g f_{abc}
{\cal A}_{\mu,b}
{\cal A}_{\nu,c} \,.
\label{eq_Gmn}
\end{equation}
In the above, $q$ denotes the quark fields and ${\cal M}$
the quark mass matrix. Further, $g$ is the
strong coupling constant and ${\cal A}_{\mu,a}$
are the gluon fields. 
Moreover,
$\lambda_a$ are the
Gell-Mann matrices and $f_{abc}$
the structure constants of the $SU(3)_{\rm color}$
Lie algebra $(a,b,c=1,\dots ,8)$;
summation over repeated indices is always implied.
The gluon-gluon term in the last equation arises
from the non-Abelian nature of the gauge theory
and is the reason for the peculiar features
of the color force.

The current masses of the up $(u)$, down $(d)$, and
strange (s) quarks are in a $\overline{MS}$ scheme at a scale of $\mu \approx 2$ GeV~\cite{PDG}:
\begin{eqnarray}
m_u &=& 2.3\pm 0.7 \mbox{ MeV} ,
\label{eq_umass} \\
m_d &=& 4.8\pm 0.5 \mbox{ MeV} ,
\label{eq_dmass} \\
m_s &=& 95\pm 5 \mbox{ MeV} .
\label{eq_smass}
\end{eqnarray}
These masses are small as compared to
a typical hadronic scale
such as the mass of a light hadron other than a Goldstone bosons, e.g., 
$m_\rho=0.78 \mbox{ GeV} \approx 1 \mbox{ GeV}$.

Thus it is relevant to discuss the
QCD Lagrangian in the case when the quark masses
vanish:  
\begin{equation}
{\cal L}_{\rm QCD}^0 = \bar{q} i \gamma^\mu {\cal D}_\mu
q - \frac14 
{\cal G}_{\mu\nu,a}
{\cal G}^{\mu\nu}_{a} \,.
\end{equation}
Right- and left-handed quark fields are defined as
\begin{equation}
q_R=P_Rq \,, \;\;\;
q_L=P_Lq \,,
\end{equation}
with 
\begin{equation}
P_R=\frac12(1+\gamma_5) \,, \;\;\;
P_L=\frac12(1-\gamma_5) \,.
\end{equation}
Then the Lagrangian can be rewritten as 
\begin{equation}
{\cal L}_{\rm QCD}^0 = 
\bar{q}_R i \gamma^\mu {\cal D}_\mu q_R 
+\bar{q}_L i \gamma^\mu {\cal D}_\mu q_L 
- \frac14 
{\cal G}_{\mu\nu,a}
{\cal G}^{\mu\nu}_{a} \, .
\end{equation}
This equation revels that
{\it the right- and left-handed components of
massless quarks do not mix} in the QCD Lagrangian. 
For the two-flavor case, this is
$SU(2)_R\times SU(2)_L$ 
symmetry, also known as {\it chiral symmetry}.
However, this symmetry is broken in two ways: explicitly and spontaneously.

\subsubsection{Explicit symmetry breaking}

The mass term  
 $- \bar{q}{\cal M}q$
in the QCD Lagrangian Eq.~(\ref{eq_LQCD}) 
breaks chiral symmetry explicitly. To better see this,
let's rewrite ${\cal M}$ for the two-flavor case,
\begin{eqnarray}
{\cal M} & = & 
\left( \begin{array}{cc}
            m_u & 0 \\
              0  & m_d 
           \end{array} \right)  \nonumber \\
  & = & \frac12 (m_u+m_d) 
\left( \begin{array}{cc}
            1 & 0 \\
              0  & 1 
           \end{array} \right) 
+ \frac12 (m_u-m_d) 
\left( \begin{array}{cc}
            1 & 0 \\
              0  & -1 
           \end{array} \right)  \nonumber \\
 & = & \frac12 (m_u+m_d) \; I + \frac12 (m_u-m_d) \; \tau_3 \,.
\label{eq_mmatr}
\end{eqnarray}
The first term in the last equation in invariant under $SU(2)_V$
(isospin symmetry) and the second term vanishes for
$m_u=m_d$.
Therefore, isospin is an exact symmetry if 
$m_u=m_d$.
However, both terms in Eq.~(\ref{eq_mmatr}) break chiral symmetry.
Since the up and down quark masses
[Eqs.~(\ref{eq_umass}) and (\ref{eq_dmass})]
are small as compared to
the typical hadronic mass scale of $\sim 1$ GeV,
the explicit chiral symmetry breaking due to non-vanishing
quark masses is very small.

\subsubsection{Spontaneous symmetry breaking}

A (continuous) symmetry is said to be {\it spontaneously
broken} if a symmetry of the Lagrangian 
is not realized in the ground state of the system.
There is evidence that the (approximate) chiral
symmetry of the QCD Lagrangian is spontaneously 
broken---for dynamical reasons of nonperturbative origin
which are not fully understood at this time.
The most plausible evidence comes from the hadron spectrum.

From chiral symmetry,
one naively expects the existence of degenerate hadron
multiplets of opposite parity, i.e., for any hadron of positive
parity one would expect a degenerate hadron state of negative 
parity and vice versa. However, these ``parity doublets'' are
not observed in nature. For example, take the $\rho$-meson which is
a vector meson of negative parity ($J^P=1^-$) and mass 
776 MeV. There does exist a $1^+$ meson, the $a_1$, but it
has a mass of 1230 MeV and, therefore, cannot be perceived
as degenerate with the $\rho$. On the other hand, the $\rho$
meson comes in three charge states (equivalent to
three isospin states), the $\rho^\pm$ and the $\rho^0$,
with masses that differ by at most a few MeV. Thus,
in the hadron spectrum,
$SU(2)_V$ (isospin) symmetry is well observed,
while axial symmetry is broken:
$SU(2)_R\times SU(2)_L$ is broken down to $SU(2)_V$.

A spontaneously broken global symmetry implies the existence
of (massless) Goldstone bosons.
The Goldstone bosons are identified with the isospin
triplet of the (pseudoscalar) pions, 
which explains why pions are so light.
The pion masses are not exactly zero because the up
and down quark masses
are not exactly zero either (explicit symmetry breaking).
Thus, pions are a truly remarkable species:
they reflect spontaneous as well as explicit symmetry
breaking.
Goldstone bosons interact weakly at low energy.
They are degenerate with the vacuum 
and, therefore, interactions between them must
vanish at zero momentum and in the chiral limit
($m_\pi \rightarrow 0$).

\subsection{Chiral effective Lagrangians 
\label{sec_Lpi} }

The next step in our EFT program is to build the most general
Lagrangian consistent with the (broken) symmetries discussed
above.
An elegant formalism for the construction of such Lagrangians
was developed by 
Callan, Coleman, Wess, and Zumino (CCWZ)~\cite{CCWZ}
who developed the foundations 
of non-linear realizations of chiral symmetry from the point of view
of group theory.\footnote{An accessible introduction into the rather
involved CCWZ formalism can be found in Ref.~\cite{Sch03}.}
The Lagrangians we give below are built upon the CCWZ
formalism. 

We already addressed the fact that the appropriate degrees of freedom are
pions (Goldstone bosons) and nucleons.
Because pion interactions must      
vanish at zero momentum transfer and in the limit of    
$m_\pi \rightarrow 0$, namely the chiral limit, the                         
Lagrangian is expanded in powers of derivatives
and pion masses. More precisely, the Lagrangian is expanded in powers of 
$Q/\Lambda_\chi$ where $Q$ stands for a (small) momentum
or pion mass and 
$\Lambda_\chi \approx 1$ GeV is identified with the              
hard scale.                                  
These are the basic steps behind the chiral perturbative expansion.    

Schematically, we can write the effective Lagrangian as                   
\begin{equation}
{\cal L}
=
{\cal L}_{\pi\pi} 
+
{\cal L}_{\pi N} 
+
{\cal L}_{NN} 
 + \, \ldots \,,
\end{equation}
where ${\cal L}_{\pi\pi}$
deals with the dynamics among pions, 
${\cal L}_{\pi N}$ 
describes the interaction
between pions and a nucleon,
and ${\cal L}_{NN}$  contains two-nucleon contact interactions
which consist of four nucleon-fields (four nucleon legs) and no
meson fields.
The ellipsis stands for terms that involve two nucleons plus
pions and three or more
nucleons with or without pions, relevant for nuclear
many-body forces 
(an example for this in lowest order are the last two terms of Eq.~(\ref{eq_LD1}), below).
The individual Lagrangians are organized order by order:
\begin{equation}
{\cal L}_{\pi\pi} 
 = 
{\cal L}_{\pi\pi}^{(2)} 
 + {\cal L}_{\pi\pi}^{(4)} 
 + \ldots \,,
\end{equation}
\begin{equation}
{\cal L}_{\pi N} 
= 
{\cal L}_{\pi N}^{(1)} 
+
{\cal L}_{\pi N}^{(2)} 
+
{\cal L}_{\pi N}^{(3)} 
+
{\cal L}_{\pi N}^{(4)} 
+
{\cal L}_{\pi N}^{(5)} 
+ \ldots ,
\end{equation}
and
\begin{equation}
\label{eq_LNN}
{\cal L}_{NN} =
{\cal L}^{(0)}_{NN} +
{\cal L}^{(2)}_{NN} +
{\cal L}^{(4)}_{NN} + 
{\cal L}^{(6)}_{NN} + 
\ldots \,,
\end{equation}
where the superscript refers to the number of derivatives or 
pion mass insertions (chiral dimension)
and the ellipsis stands for terms of higher dimensions.

Above, we have organized the Lagrangians by the number
of derivatives or pion-masses. This is 
the standard way, appropriate particularly for
considerations of $\pi$-$\pi$ and $\pi$-$N$ scattering.
As it turns out (cf.\ Section~\ref{sec_chpt}), 
for interactions among nucleons,
sometimes one makes use of the so-called
index of the interaction,
\begin{equation}
\Delta  \equiv   d + \frac{n}{2} - 2  \, ,
\label{eq_Delta}
\end{equation}
where $d$ is the number of derivatives or pion-mass insertions 
and $n$ the number of nucleon field operators (nucleon legs).
We will now write down the Lagrangian in terms
of increasing values of the parameter $\Delta$ and
we will do so using the so-called heavy-baryon formalism which we indicate
by a ``hat''~\cite{Fet00}.

The leading-order Lagrangian reads,
\begin{eqnarray}
\widehat{\cal L}^{\Delta=0} &=&
	\frac{1}{2} 
	\partial_\mu \mbox{\boldmath{$\pi$}} \cdot 
	\partial^\mu \mbox{\boldmath{$\pi$}}
     -  \frac{1}{2} m_\pi^2 \mbox{\boldmath{$\pi$}}^2
\nonumber \\ &&
     +  \frac{1-4\alpha}{2f_\pi^2} 
        (\mbox{\boldmath{$\pi$}} \cdot \partial_\mu \mbox{\boldmath{$\pi$}})
        (\mbox{\boldmath{$\pi$}} \cdot \partial^\mu \mbox{\boldmath{$\pi$}})
     -  \frac{\alpha}{f_\pi^2} 
        \mbox{\boldmath{$\pi$}}^2
	\partial_\mu \mbox{\boldmath{$\pi$}} \cdot 
	\partial^\mu \mbox{\boldmath{$\pi$}}
     +\;  \frac{8\alpha-1}{8f_\pi^2} 
        m_\pi^2 \mbox{\boldmath{$\pi$}}^4
\nonumber \\ &&
+ \bar{N} \left[ 
i \partial_0 
- \frac{g_A}{2f_\pi} \; \mbox{\boldmath $\tau$} \cdot 
 ( \vec \sigma \cdot \vec \nabla ) \mbox{\boldmath $\pi$} 
- \frac{1}{4f_\pi^2} \; \mbox{\boldmath $\tau$} \cdot 
 ( \mbox{\boldmath $\pi$}
\times \partial_0 \mbox{\boldmath $\pi$})
\right] N
\nonumber \\ &&
+ \bar{N} \left\{
 \frac{g_A(4\alpha-1)}{4f_\pi^3} \;
(\mbox{\boldmath $\tau$} \cdot 
\mbox{\boldmath $\pi$}) 
\left[ \mbox{\boldmath $\pi$} \cdot 
 ( \vec \sigma \cdot \vec \nabla )
\mbox{\boldmath $\pi$} \right]
 \right. \nonumber \\ &&   \left.
+ \; \frac{g_A\alpha}{2f_\pi^3} \;
\mbox{\boldmath $\pi$}^2 
\left[ \mbox{\boldmath $\tau$} \cdot 
 ( \vec \sigma \cdot \vec \nabla )
\mbox{\boldmath $\pi$} 
\right]
\right\} N 
\nonumber \\ &&
-\frac{1}{2} C_S \bar{N} N \bar{N} N 
-\frac{1}{2} C_T (\bar{N} \vec \sigma N) \cdot (\bar{N} \vec \sigma N) 
\; + \; \ldots \,,
\label{eq_LD0}
\end{eqnarray}
and subleading Lagrangians are,
\begin{eqnarray}
\widehat{\cal L}^{\Delta=1} &=&
 \bar{N} \left\{
 \frac{{\vec \nabla}^2}{2M_N} 
-\frac{ig_A}{4M_Nf_\pi} 
\mbox{\boldmath $\tau$} \cdot 
\left[
\vec \sigma \cdot
\left( \stackrel{\leftarrow}{\nabla} 
\partial_0 \mbox{\boldmath $\pi$}
 -
\partial_0 \mbox{\boldmath $\pi$}
\stackrel{\rightarrow}{\nabla} \right)
\right]
\right.
\nonumber \\ &&
\left.
- \frac{i}{8M_N f_\pi^2}
\mbox{\boldmath $\tau$} \cdot 
\left[
\stackrel{\leftarrow}{\nabla} 
\cdot
( \mbox{\boldmath $\pi$} \times \vec\nabla \mbox{\boldmath $\pi$} )
   -   
( \mbox{\boldmath $\pi$} \times \vec\nabla \mbox{\boldmath $\pi$} )
\cdot
\stackrel{\rightarrow}{\nabla} 
\right]
\right\} N 
\nonumber \\ &&
+ \bar{N} \left[
 4c_1m_\pi^2
-\frac{2 c_1}{f_\pi^2} \, m_\pi^2\, \mbox{\boldmath $\pi$}^2 
\, + \, 
\left( c_2 - \frac{g_A^2}{8M_N}\right) 
\frac{1}{f_\pi^2}
(\partial_0 \mbox{\boldmath{$\pi$}} \cdot 
 \partial_0 \mbox{\boldmath{$\pi$}})
\right.  \nonumber \\ &&  \left.
 + \, \frac{c_3}{f_\pi^2}\,
(\partial_\mu \mbox{\boldmath{$\pi$}} \cdot 
\partial^\mu \mbox{\boldmath{$\pi$}})
%\right.  \nonumber \\ &&  \left.
 - \, \left( c_4 + \frac{1}{4M_N} \right) 
\frac{1}{2f_\pi^2}
\epsilon^{ijk} \epsilon^{abc} \sigma^i \tau^a
(\partial^j \pi^b) (\partial^k \pi^c)
 \right] N 
\nonumber \\ &&
- \frac{D}{4f_\pi} (\bar{N}N) \bar{N} \left[ 
\mbox{\boldmath $\tau$} 
\cdot ( \vec \sigma \cdot \vec \nabla )
\mbox{\boldmath $\pi$} 
\right] N
\nn &&
-\frac12 E
(\bar{N}N)
(\bar{N}
\mbox{\boldmath $\tau$} 
N)
\cdot
(\bar{N}
\mbox{\boldmath $\tau$} 
N)
\; + \; \ldots \,,
\label{eq_LD1}
\\
\widehat{\cal L}^{\Delta=2} &=&
\; {\cal L}^{(4)}_{\pi\pi} \; +
\; \widehat{\cal L}^{(3)}_{\pi N} \; + \; \widehat{\cal L}^{(2)}_{NN}
\; + \; \ldots \,,
\label{eq_LD2}
\\
\widehat{\cal L}^{\Delta=3} &=&
\; \widehat{\cal L}^{(4)}_{\pi N} 
\; + \; \ldots \,,
\label{eq_LD3}
\\
\widehat{\cal L}^{\Delta=4} &=&
\; \widehat{\cal L}^{(5)}_{\pi N} \; +
\; \widehat{\cal L}^{(4)}_{NN}
\; + \; \ldots \,.
\label{eq_LD4}
\\
\widehat{\cal L}^{\Delta=6} &=&
\; \widehat{\cal L}^{(6)}_{NN}\; +
\;  \ldots \,,
\label{eq_LD6}
\end{eqnarray}
where we included terms relevant for a calculation of the two-nucleon force up to sixth order.
The Lagrangians $\widehat{\cal L}^{(3)}_{\pi N}$ and
$\widehat{\cal L}^{(4)}_{\pi N}$ 
can be found in Ref.~\cite{KGE12}
and $NN$ contact Lagrangians are given below.
The pion fields are denoted by $\boldsymbol{\pi}$ and the heavy baryon nucleon field by $N$  ($\bar{N}=N^\dagger$).
Furthermore, $g_A$, $f_\pi$, $m_\pi$, and $M_N$ are the axial-vector coupling constant,
pion decay constant, pion mass, and nucleon mass, respectively. Numerical values for these quantities will be given later.
The $c_i$ are low-energy constants (LECs) from the dimension two $\pi N$ Lagrangian 
and $\alpha$ is a parameter that appears in the expansion 
of a $SU(2)$ matrix $U$ in powers
of the pion fields, see Ref.~\cite{ME11} for more details. Results are independent of $\alpha$.

The lowest order (or leading order) $NN$ Lagrangian has no derivatives 
and reads~\cite{Wei91}
\begin{equation}
\label{eq_LNN0}
\widehat{\cal L}^{(0)}_{NN} =
-\frac{1}{2} C_S \bar{N} N \bar{N} N 
-\frac{1}{2} C_T (\bar{N} \vec \sigma N) \cdot (\bar{N} \vec \sigma N) \, ,
\end{equation}
where 
$C_S$ and $C_T$ are free paramters to be determined by 
fitting to the $NN$ data.

The second order $NN$ Lagrangian can be stated as
follows~\cite{ORK96}
\begin{eqnarray}
\label{eq_LNN2}
\widehat{\cal L}^{(2)}_{NN} &=&
-C'_1 \left[(\bar{N} \vec \nabla N)^2+ (\overline{\vec \nabla N} N)^2 \right]
-C'_2 (\bar{N} \vec \nabla N)\cdot (\overline{\vec \nabla N} N)
\nonumber \\ &&
-C'_3 \bar{N} N \left[\bar N \vec \nabla^2 N+\overline{\vec \nabla^2 N} N \right]
\nonumber \\ &&
-i C'_4 \left[
\bar N \vec \nabla N \cdot 
(\overline{\vec \nabla N} \times \vec \sigma N) +
\overline{(\vec \nabla N)} N \cdot 
(\bar N \vec \sigma \times \vec \nabla N) \right]
\nonumber \\ &&
-i C'_5 \bar N N(\overline{\vec \nabla N} \cdot \vec \sigma \times \vec \nabla N)
-i C'_6 (\bar N \vec \sigma N)\cdot (\overline{\vec \nabla N} \times \vec \nabla N)
\nonumber \\ &&
-\left(C'_7 \delta_{ik} \delta_{jl}+C'_8 \delta_{il} \delta_{kj}
+C'_9 \delta_{ij} \delta_{kl}\right)
\nonumber \\ && \times
\left[\bar N \sigma_k \partial_i N \bar N \sigma_l \partial_j N +
\overline{\partial_i N} \sigma_k N \overline{\partial_j N} \sigma_l N \right]
\nonumber \\ &&
-\left(C'_{10} \delta_{ik} \delta_{jl}+C'_{11} \delta_{il} \delta_{kj}+C'_{12} \delta_{ij} \delta_{kl}\right)
\bar N \sigma_k \partial_i N \overline{\partial_j N} \sigma_l N
\nonumber \\ &&
-\left(\frac{1}{2} C'_{13} (\delta_{ik} \delta_{jl}+
\delta_{il} \delta_{kj}) 
+C'_{14} \delta_{ij} \delta_{kl} \right)
\nn && \times
\left[\overline{\partial_i N} \sigma_k \partial_j N + \overline{\partial_j N} \sigma_k \partial_i N\right]
\bar N \sigma_l N \, .
\end{eqnarray}
Similar to $C_S$ and $C_T$ of Eq.~(\ref{eq_LNN0}), the $C'_i$ of Eq.~(\ref{eq_LNN2}) are free parameters   
which are determined in a fit to the $NN$ data.
Clearly, the contact Lagrangians grow considerably in size and complexity 
as the order increases. Therefore we do not provide here explicit expressions for 
$\widehat{\cal L}^{(4)}_{NN}$ and
$\widehat{\cal L}^{(6)}_{NN}$.
The $NN$ contact potentials derived from some of the $NN$ Lagrangians
are given in Sec.~\ref{sec_ct}.

\section{Nuclear forces from EFT: Overview
\label{sec_overview}}

We proceed here with discussing the various steps towards a derivation
of nuclear forces from EFT.
In this section, we will discuss the expansion
we are using in more details as well as the various 
Feynman diagrams as they emerge at each order. 

\subsection{Chiral perturbation theory and power counting
\label{sec_chpt}}

An infinite number of Feynman diagrams can be evaluated from 
the effective Langrangians and so one needs to be able to organize these
diagrams in order of their importance. 
Chiral perturbation theory provides such organizational scheme. 

In ChPT, 
graphs are analyzed
in terms of powers of small external momenta over the large scale:
$(Q/\Lambda_\chi)^\nu$,
where $Q$ is generic for a momentum (nucleon three-momentum or
pion four-momentum) or a pion mass and $\Lambda_\chi \sim 1$ GeV
is the chiral symmetry breaking scale (hadronic scale, hard scale).
Determining the power $\nu$ 
has become known as power counting.

For the moment, we will consider only so-called irreducible
graphs.
By definition, an irreducible graph is a diagram that
cannot be separated into two
by cutting only nucleon lines.
Following the Feynman rules of covariant perturbation theory,
a nucleon propagator carries the dimension $Q^{-1}$,
a pion propagator $Q^{-2}$,
each derivative in any interaction is $Q$,
and each four-momentum integration $Q^4$.
This is also known as naive dimensional analysis.
Applying then some topological identities, one obtains
for the power of an irreducible diagram
involving $A$ nucleons~\cite{ME11}
\begin{equation} \nu = -2 +2A - 2C + 2L 
+ \sum_i \Delta_i \, ,
\label{eq_nu} 
\end{equation}
with
\begin{equation}
\Delta_i  \equiv   d_i + \frac{n_i}{2} - 2  \, .
\label{eq_Deltai}
\end{equation}
In the two equations above: for each
vertex $i$, $C$ represents the number of individually connected parts of the diagram while
$L$ is the number of loops;                  
$d_i$ indicates how many derivatives or pion masses are present 
and $n_i$ the number of nucleon fields.                  
The summation extends over all vertices present in that particular diagram.
Notice also that chiral symmetry implies $\Delta_i \geq 0$. 
Interactions among pions have at least two derivatives
($d_i\geq 2, n_i=0$), while 
interactions between pions and a nucleon have one or more 
derivatives  
($d_i\geq 1, n_i=2$). Finally, pure contact interactions
among nucleons ($n_i=4$)
have $d_i\geq0$.
In this way, a low-momentum expansion based on chiral symmetry 
can be constructed.                   

Naturally,                                            
the powers must be bounded from below for the expansion
to converge. This is in fact the case, 
with $\nu \geq 0$. 

Furthermore, the power formula 
Eq.~(\ref{eq_nu}) 
allows to predict
the leading orders of connected multi-nucleon forces.
Consider a $m$-nucleon irreducibly connected diagram
($m$-nucleon force) in an $A$-nucleon system ($m\leq A$).
The number of separately connected pieces is
$C=A-m+1$. Inserting this into
Eq.~(\ref{eq_nu}) together with $L=0$ and 
$\sum_i \Delta_i=0$ yields
$\nu=2m-4$. Thus, two-nucleon forces ($m=2$) appear
at $\nu=0$, three-nucleon forces ($m=3$) at
$\nu=2$ (but they happen to cancel at that order),
and four-nucleon forces at $\nu=4$ (they don't cancel).
More about this in the next sub-section.

For later purposes, we note that for an irreducible 
$NN$ diagram ($A=2$, $C=1$), the
power formula collapses to the very simple expression
\begin{equation}
\nu =  2L + \sum_i \Delta_i \,.
\label{eq_nunn}
\end{equation}

To summarize, at each order                            
$\nu$ we only have a well defined number of diagrams, 
which renders the theory feasible from a practical standpoint.
The magnitude of what has been left out at order $\nu$ can be estimated (in a 
very simple way) from 
$(Q/\Lambda_\chi)^{\nu+1}$. The ability to calculate observables (in 
principle) to any degree of accuracy gives the theory 
its predictive power.

\subsection{The ranking of nuclear forces}

\begin{figure}[t]\centering
%\vspace*{-0.5cm}
\scalebox{0.90}{\includegraphics{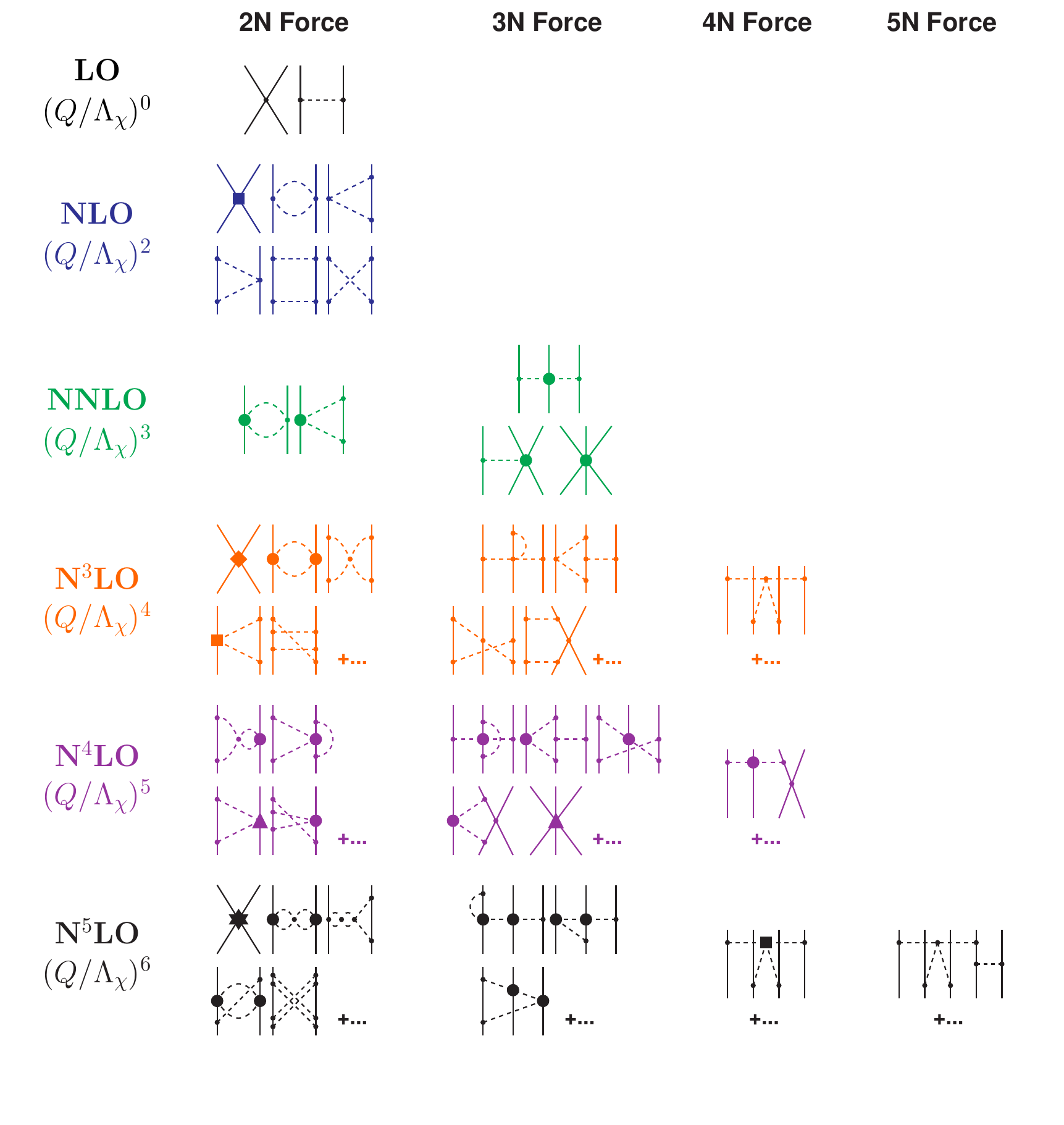}}
\vspace*{-1.5cm}
\caption{Hierarchy of nuclear forces in ChPT. Solid lines
represent nucleons and dashed lines pions. 
Small dots, large solid dots, solid squares, triangles, diamonds, and stars
denote vertices of index $\Delta= \, $ 0, 1, 2, 3, 4, and 6, respectively. 
Further explanations are
given in the text.}
\label{fig_hi}
\end{figure}

As shown in Fig.~\ref{fig_hi}, nuclear forces appear in ranked
orders in accordance with 
the power counting scheme. 

The lowest power is $\nu = 0$, also known as the leading order (LO).
At LO we have only two contact contributions with no momentum dependence 
($\sim Q^0$). They are signified by the 
four-nucleon-leg diagram 
with a small-dot vertex shown in the first row of 
Fig.~\ref{fig_hi}.
Besides this, we have the
static one-pion exchange (1PE), also shown 
in the first row of                 
Fig.~\ref{fig_hi}.              

In spite of its simplicity, this rough description                      
contains some of the main attributes of the $NN$ force. 
First, through the 1PE it generates the tensor component of the force
known to be crucial for the two-nucleon bound state. Second, 
it predict correctly 
$NN$ phase parameters for high partial waves.               
At LO, the two terms which result from a partial-wave expansion of the contact term
impact states of zero orbital angular momentum and produce attraction at 
short- and intermediate-range.                              

Notice that that there are no terms with power 
$\nu=1$, as they would violate parity conservation 
and time-reversal invariance.

The next order is then
$\nu=2$, next-to-leading order, or NLO.

Note that the two-pion exchange (2PE) makes its first appearance at this order,
and thus it is referred to as the 
``leading 2PE''. As is well known from decades of nuclear physics, 
this contribution is essential for a realistic account of the intermediate-range attraction.    
However, the leading 2PE has insufficient strength, for the following reason: 
the loops present in the diagrams which involve pions 
carry the power $\nu=2$ [cf.\ Eq.~(\ref{eq_nunn})],
and so only                                  
$\pi NN$ and $\pi \pi NN$ vertices with $\Delta_i = 0$ are allowed at this order. 
These vertices are known to be weak.
Moreover, seven new contacts appear at this order which 
impact $L = 0$ and $L = 1$ states. (As always, two-nucleon contact terms are indicated 
by four-nucleon-leg diagrams and a vertex of appropriate shape, in this case a solid square.) 
At this power, the appropriate operators                                  
include spin-orbit, central,
spin-spin, and tensor terms, namely all the spin and isospin
operator structures needed for a realistic description of the 
2NF, although the medium-range attraction still lacks 
sufficient strength.                             

At the next order, 
$\nu=3$ or next-to-next-to-leading order (NNLO), 
the 2PE contains the so-called          
$\pi\pi NN$ seagull vertices with two derivatives.                    
These vertices (proportional to 
the $c_i$ LECs and denoted by a large solid dot
in Fig.~\ref{fig_hi}), 
bring in correlated 2PE
and intermediate $\Delta(1232)$-isobar contributions.
Consistent with what meson theory of                               
the nuclear force~\cite{Lac80,MHE87} has shown since a long time 
concerning the importance of these effects,            
at this order the 2PE finally provides medium-range     
attraction of realistic strength, bringing the description of the $NN$ force
to an almost quantitative level. 
No new contacts become available at NNLO. 

The discussion above reveals how                      
two- and many-nucleon forces are generated                      
and increase in number as we move to higher orders.
Three-nucleon forces appear at NLO,                          
but their net contribution vanishes at this order~\cite{Wei92}.
The first non-zero 3NF contribution is found 
at NNLO~\cite{Kol94,Epe02b}. It is therefore easy to understand why  
3NF are very weak as compared to the 2NF which contributes already at 
$(Q/\Lambda_\chi)^0$.

For $\nu =4$, or next-to-next-to-next-to-leading
order (N$^3$LO), we display some representative diagrams in 
Fig.~\ref{fig_hi}. There is a large attractive one-loop 2PE contribution (the bubble diagram with two large solid dots $\sim c_i^2$), which slightly over-estimates the 2NF attraction
at medium range. 
Two-pion-exchange graphs with two loops are seen at this order, together with 
three-pion exchange (3PE), which was determined to be very weak 
at N$^3$LO~\cite{Kai00a,Kai00b}.
The most important feature at this order is the presence 
of 15 additional contacts $\sim Q^4$, signified 
by the four-nucleon-leg diagram in the figure with the diamond-shaped vertex. 
These contacts impact states with orbital angular momentum up to $L = 2$, 
and are the reason for the                            
 quantitative description of the
two-nucleon force (up to approximately 300 MeV
in terms of laboratory energy) 
at this order~\cite{ME11,EM03}.
More 3NF diagrams show up 
at N$^3$LO, as well as the first contributions to 
four-nucleon forces (4NF).         
We then see that forces involving more and more nucleons appear for the
first time at higher and higher orders, which 
gives theoretical support to the fact that          
2NF $\gg$ 3NF $\gg$ 4NF
\ldots.

Further 2PE and 3PE occur at N$^4$LO (fifth order). The contribution to the 2NF 
at this order has been first calculated  by Entem {\it et al.}~\cite{Ent15a}. It turns out to be moderately repulsive, thus
compensating for the attractive surplus generated at N$^3$LO by the bubble diagram with two solid dots. The long- and intermediate-range 3NF contributions at this order have been evaluated~\cite{KGE12,KGE13}, but not yet applied in nuclear structure calculations. They are
expected to be sizeable. Moreover, a new set of 3NF contact terms appears~\cite{GKV11}.
The N$^4$LO 4NF has not been derived yet. Due to the subleading $\pi\pi N N$ seagull vertex (large solid dot $\sim c_i$), this 4NF could be sizeable.

Finally turning to N$^5$LO (sixth order): The dominant 2PE and 3PE contributions to the 2NF have been derived by Entem {\it et al.} in Ref.~\cite{Ent15b}, which represents
the most sophisticated investigation ever conducted in chiral EFT for the $NN$ system. The effects are small indicating the desired trend towards convergence of the chiral expansion for the 2NF. 
Moreover, a new set of 26 $NN$ contact terms $\sim Q^6$ occurs that contributes up to $F$-waves (represented by the $NN$ diagram with a star in Fig.~\ref{fig_hi})
bringing the total number of $NN$ contacts to 50~\cite{EM03a}.
The three-, four-, and five-nucleon forces of this order have not yet been derived.

This section has provided an overview.
In the following sections, we will present more details. 

\section{Pion-exchange contributions to the $NN$ interaction}
\label{sec_pions}

The various pion-exchange contributions to the $NN$ potential may be analyzed
according to the number of pions being exchanged between the two
nucleons:
\begin{equation}
V = V_{1\pi} + V_{2\pi} + V_{3\pi} + V_{4\pi} + \ldots \,,
\end{equation}
where the meaning of the subscripts is obvious
and the ellipsis represents $5\pi$ and higher pion exchanges. For each of the above terms, 
we assume a low-momentum expansion:
\begin{eqnarray}
V_{1\pi} & = & V_{1\pi}^{(0)} + V_{1\pi}^{(2)} 
+ V_{1\pi}^{(3)} + V_{1\pi}^{(4)} + V_{1\pi}^{(5)} + V_{1\pi}^{(6)} + \ldots 
\label{eq_1pe_orders}
\\
V_{2\pi} & = & V_{2\pi}^{(2)} + V_{2\pi}^{(3)} + V_{2\pi}^{(4)} + V_{2\pi}^{(5)} + V_{2\pi}^{(6)}  
+ \ldots \\
V_{3\pi} & = & V_{3\pi}^{(4)} + V_{3\pi}^{(5)} + V_{3\pi}^{(6)} + \ldots \\
V_{4\pi} & = & V_{4\pi}^{(6)} + \ldots \,,
\end{eqnarray}
where the superscript denotes the order $\nu$ of the expansion
and the ellipses stand for contributions of seventh
and higher orders.
Due to parity and time-reversal,
there are no first order contributions.
Moreover, since $n$ pions create $L=n-1$ loops, 
the leading
order for $n$-pion exchange ocurrs at $\nu=2n-2$
[cf.\ Eq.~(\ref{eq_nunn})].

Order by order, the pion-exchange part of the $NN$ potential builds up as follows:
\beqa
V_{\rm LO} & \equiv & V^{(0)} =
V_{1\pi}^{(0)} 
\label{eq_VLO}
\\
V_{\rm NLO} & \equiv & V^{(2)} = V_{\rm LO} +
V_{1\pi}^{(2)} +
V_{2\pi}^{(2)} 
\label{eq_VNLO}
\\
V_{\rm NNLO} & \equiv & V^{(3)} = V_{\rm NLO} +
V_{1\pi}^{(3)} + 
V_{2\pi}^{(3)} 
\label{eq_VNNLO}
\\
V_{\rm N3LO} & \equiv & V^{(4)} = V_{\rm NNLO} +
V_{1\pi}^{(4)} + 
V_{2\pi}^{(4)} +
V_{3\pi}^{(4)} 
\label{eq_VN3LO}
\\
V_{\rm N4LO} & \equiv & V^{(5)} = V_{\rm N3LO} +
V_{1\pi}^{(5)} + 
V_{2\pi}^{(5)} +
V_{3\pi}^{(5)} 
\label{eq_VN4LO}
\\
V_{\rm N5LO} & \equiv & V^{(6)} = V_{\rm N4LO} +
V_{1\pi}^{(6)} + 
V_{2\pi}^{(6)} +
V_{3\pi}^{(6)} +
V_{4\pi}^{(6)} 
\label{eq_VN5LO}
\eeqa
where 
LO stands for leading order, NLO for next-to-leading order, etc..

The explicit expressions for the potentials will be stated in terms of contributions to the 
momentum-space $NN$ amplitudes in the center-of-mass system (CMS), 
which arise from the following general decomposition:
\begin{eqnarray} 
V({\vec p}~', \vec p) &  = &
 \:\, V_C \:\, + \bm{\tau}_1 \cdot \bm{\tau}_2 \, W_C 
\nonumber \\ & + &
\left[ \, V_S \:\, + \bm{\tau}_1 \cdot \bm{\tau}_2 \, W_S 
\,\:\, \right] \,
\vec\sigma_1 \cdot \vec \sigma_2
\nonumber \\ &+& 
\left[ \, V_{LS} + \bm{\tau}_1 \cdot \bm{\tau}_2 \, W_{LS}    
\right] \,
\left(-i \vec S \cdot (\vec q \times \vec k) \,\right)
\nonumber \\ &+& 
\left[ \, V_T \:\,     + \bm{\tau}_1 \cdot \bm{\tau}_2 \, W_T 
\,\:\, \right] \,
\vec \sigma_1 \cdot \vec q \,\, \vec \sigma_2 \cdot \vec q  
\nonumber \\ &+& 
\left[ \, V_{\sigma L} + \bm{\tau}_1 \cdot \bm{\tau}_2 \, 
      W_{\sigma L} \, \right] \,
\vec\sigma_1\cdot(\vec q\times \vec k\,) \,\,
\vec \sigma_2 \cdot(\vec q\times \vec k\,)
%\bigg\}
\, ,
%\nonumber \\ && 
\label{eq_nnamp}
\end{eqnarray}
where ${\vec p}\,'$ and $\vec p$ denote the final and initial nucleon momenta in the CMS, 
respectively. Moreover, $\vec q = {\vec p}\,' - \vec p$ is the momentum transfer, 
$\vec k =({\vec p}\,' + \vec p)/2$ the average momentum, and $\vec S =(\vec\sigma_1+
\vec\sigma_2)/2 $ the total spin, with $\vec \sigma_{1,2}$ and $\bm{\tau}_{1,2}$ the spin 
and isospin operators of nucleon 1 and 2, respectively.
For on-shell scattering, $V_\alpha$ and $W_\alpha$ ($\alpha=C,S,LS,T,\sigma L$) can be 
expressed as functions of $q= |\vec q\,|$ and $p=|{\vec p}\,'| = |\vec p\,|$, only. 
  
We will now discuss the contributions order by order.

\subsection{Leading order (LO)}
\label{sec_lo}
At leading order, there is only the $1\pi$-exchange contribution, cf.\ Fig.~\ref{fig_hi}.
The charge-independent $1\pi$-exchange is given by
\begin{equation}
V_{1\pi}^{\rm(CI)} ({\vec p}~', \vec p) = - 
%\frac{1}{(2\pi)^3} \,
\frac{g_A^2}{4f_\pi^2}
\: 
\bm{\tau}_1 \cdot \bm{\tau}_2 
\:
\frac{
\vec \sigma_1 \cdot \vec q \,\, \vec \sigma_2 \cdot \vec q}
{q^2 + m_\pi^2} 
\,.
\label{eq_1PEci}
\end{equation}
Higher order corrections to the $1\pi$-exchange  are taken care of by  mass
and coupling constant renormalizations which, in turn, are accounted for 
by working with the physical values.
 Note also that, on 
shell, there are no relativistic corrections. Thus, we apply  $1\pi$-exchange in the form
\eq{eq_1PEci} through all orders.

We use $g_A=1.290$ (instead of $g_A=1.276$~\cite{Liu10})
to account for the so-called Goldberger-Treiman discrepancy. 
Via the Goldberger-Treiman relation,
$g_{\pi NN} =
 g_A  M_N/f_\pi$, 
our value for $g_A$ 
together with $f_\pi=92.4$ MeV and $M_N=938.918$ MeV
implies
$g_{\pi NN}^2/4\pi = 13.67$
which is consistent with the empirical values
obtained from $\pi N$ and $NN$ 
data analysis~\cite{Pav00,Arn00}.

For results presented below, we will be specifically calculating neutron-proton ($np$) scattering 
and take the charge-dependence of the $1\pi$-exchange into account.
Thus, the $1\pi$-exchange  potential that we actually apply reads
\begin{equation}
V_{1\pi}^{(np)} ({\vec p}~', \vec p) 
= -V_{1\pi} (m_{\pi^0}) + (-1)^{I+1}\, 2\, V_{1\pi} (m_{\pi^\pm})
\,,
\label{eq_1penp}
\end{equation}
where $I=0,1$ denotes the total isospin of the two-nucleon system and
\begin{equation}
V_{1\pi} (m_\pi) \equiv - \,
%\frac{1}{(2\pi)^3} \,
\frac{g_A^2}{4f_\pi^2} \,
\frac{
\vec \sigma_1 \cdot \vec q \,\, \vec \sigma_2 \cdot \vec q}
{q^2 + m_\pi^2} 
\,.
\end{equation}
We use $m_{\pi^0}=134.9766$ MeV and
 $m_{\pi^\pm}=139.5702$ MeV.
Formally speaking, the charge-dependence of the 1PE exchange is of order 
NLO~\cite{ME11}, but we include it already at leading order to make the comparison with
the $np$ phase shifts more meaningful.

\subsection{Next-to-leading order (NLO)}
\label{sec_nlo}

The $NN$ diagrams that occur at NLO (cf.\ Fig.~\ref{fig_hi})
contribute in the following way~\cite{KBW97}:
\begin{eqnarray} 
W_C &=&{L(\tilde{\Lambda};q)\over384\pi^2 f_\pi^4} \left[4m_\pi^2(1+4g_A^2-5g_A^4)
+q^2(1+10g_A^2-23g_A^4) - {48g_A^4 m_\pi^4 \over w^2} \right] ,  
\nn
 &&
\label{eq_2C}
\\   
V_T &=& -{1\over q^2} V_{S} \; = \; -{3g_A^4 \over 64\pi^2 f_\pi^4} L(\tilde{\Lambda};q)\,,
\label{eq_2T}
\end{eqnarray}  
where the (regularized) logarithmic loop function is given by:
\begin{equation} 
L(\tilde{\Lambda};q) = {w\over 2q} 
\ln {\frac{\tilde{\Lambda}^2(2m_\pi^2+q^2)-2m_\pi^2 q^2+\tilde{\Lambda}\sqrt{
\tilde{\Lambda}^2-4m_\pi^2}\, q\,w}{2m_\pi^2(\tilde{\Lambda}^2+q^2)}}
\label{eq_L}
\end{equation}
with $ w = \sqrt{4m_\pi^2+q^2}$.
$\tilde{\Lambda}$ 
denotes the cutoff of the spectral-function renormalization (SFR)~\cite{EGM04}.
Note that
\begin{equation}
\lim_{\tilde{\Lambda} \rightarrow \infty} L(\tilde{\Lambda};q) =  {w\over q} 
\ln {\frac{w+q}{2m_\pi}} \,,
\end {equation}
is the logarithmic loop function of dimensional regularization.

\subsection{Next-to-next-to-leading order (NNLO)}
\label{sec_nnlo}

The NNLO contribution (cf.\ the 2NF diagrams of the NNLO row in Fig.~\ref{fig_hi}) is given by~\cite{KBW97}:
\begin{eqnarray} 
V_C &=&  {3g_A^2 \over 16\pi f_\pi^4} \left[2m_\pi^2(c_3- 2c_1)+c_3 q^2 \right](2m_\pi^2+q^2) 
A(\tilde{\Lambda};q) \,, \label{eq_3C}
\\
W_T &=&-{1\over q^2}W_{S} =-{g_A^2 \over 32\pi f_\pi^4} c_4 w^2  A(\tilde{\Lambda};q)\,.
\label{eq_3T}
\end{eqnarray}   
The loop function that appears in the above expressions,
regularized by spectral-function cut-off $\tilde{\Lambda}$, is
\begin{equation} 
A(\tilde{\Lambda};q) = {1\over 2q} \arctan{q (\tilde{\Lambda}-2m_\pi) \over q^2
+2\tilde{\Lambda} m_\pi} \,, \label{eq_A}
\end{equation}
and 
\begin{equation}
\lim_{\tilde{\Lambda} \rightarrow \infty} A(\tilde{\Lambda};q) =  
{1\over 2q} \arctan{q \over 2m_\pi} 
\end {equation}
yields the loop function used in dimensional regularization.

\subsection{Next-to-next-to-next-to-leading order (N$^3$LO)}
\label{sec_n3lo}

The number of diagrams involved is now dramatically increasing.
Therefore, we will provide additional figures showing the full complexity
of the diagrams representing the nuclear forces at higher orders.

\begin{figure}\centering
\vspace*{-0.5cm}
%\hspace*{-1.5cm}
\scalebox{0.65}{\includegraphics{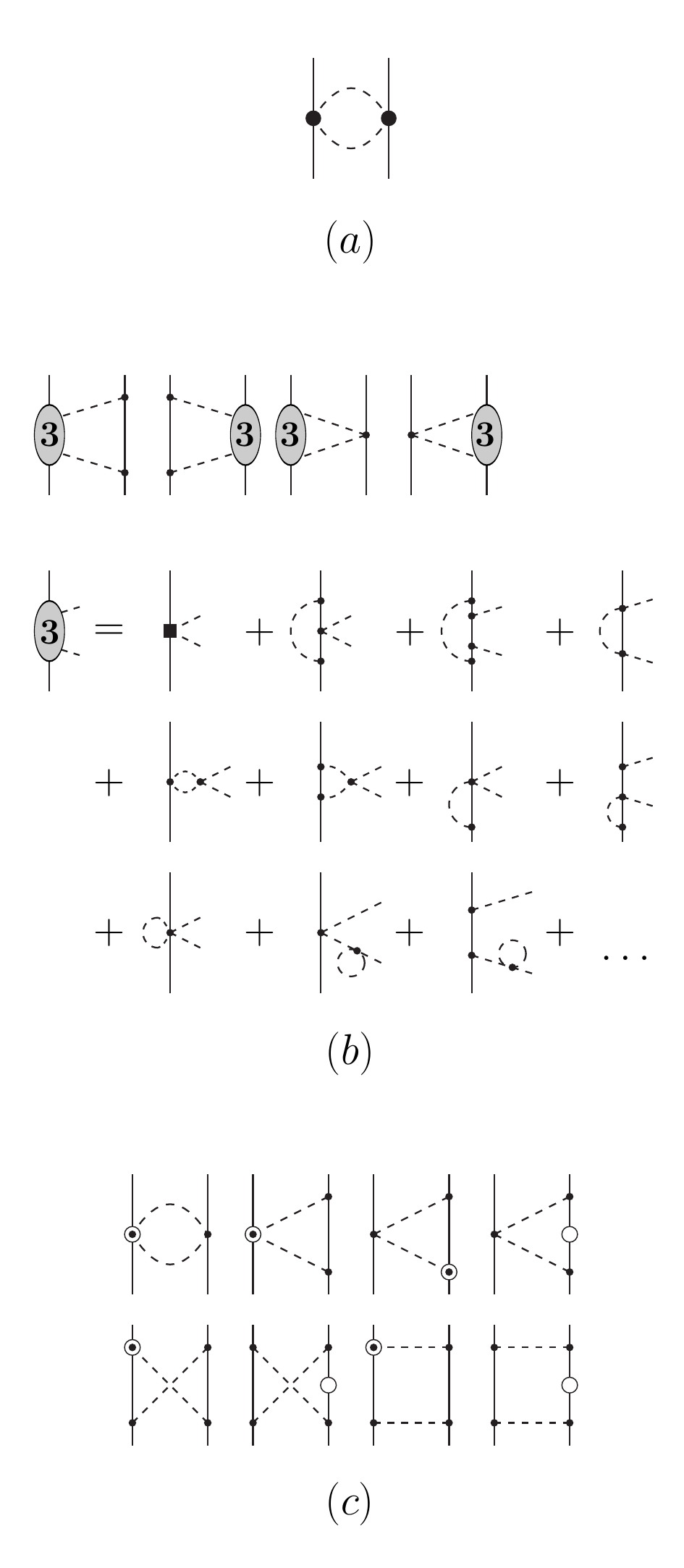}}
\vspace*{-0.5cm}
\caption{N$^3$LO two-pion exchange contributions
with (a) the N$^3$LO football diagram, (b) the leading 2PE two-loop contributions,
and (c) the relativistic corrections of NLO diagrams.
Notation as in Fig.~\ref{fig_hi}.
Shaded ovals 
represent complete $\pi N$-scattering amplitudes with their order 
specified by the number in the oval.
Open circles denote relativistic $1/M_N$ corrections.}
\label{fig_diagn3lo2pe}
\end{figure}

The 2PE contributions at N$^3$LO are shown in Fig.~\ref{fig_diagn3lo2pe}.
They consist of three parts, which we will discuss one by one.

\subsubsection{Football diagram at N$^3$LO}

The football diagram at N$^3$LO, Fig.~\ref{fig_diagn3lo2pe}(a), generates~\cite{Kai01a}:
\begin{eqnarray}  
V_C & = & {3\over 16 \pi^2 f_\pi^4 } 
\left[\left( {c_2 \over 6} w^2 +c_3(2m_\pi^2+q^2) -4c_1 m_\pi^2 \right)^2 
+{c_2^2 \over 45 } w^4 \right]  L(\tilde{\Lambda};q) \,, 
\label{eq_4c2C}
\\
W_T  &=&  -{1\over q^2} W_S 
     = {c_4^2 \over 96 \pi^2 f_\pi^4 }  w^2 L(\tilde{\Lambda};q)
\,.
\label{eq_4c2T}
\end{eqnarray}

\subsubsection{Leading two-loop contributions}

The leading order $2\pi$-exchange two-loop diagrams are shown in Fig.~\ref{fig_diagn3lo2pe}(b).
In terms of spectral functions, the results are~\cite{Kai01a}:
\begin{eqnarray} 
{\rm Im}\, V_C(i\mu) &=& {3g_A^4 (2m_\pi^2-\mu^2) \over \pi \mu (4f_\pi)^6} \Bigg[ (m_\pi^2-2\mu^2) 
\left( 2m_\pi +{2m_\pi^2 -\mu^2 \over2\mu} \ln{\mu+2m_\pi \over \mu-2m_\pi} \right) 
   \Bigg. \nn && \Bigg.
+4g_A^2 m_\pi(2m_\pi^2-\mu^2) \Bigg] \,,
\\ 
{\rm Im}\, W_C(i\mu) &=& {2\kappa \over 3\mu (8\pi f_\pi^2)^3} \int_0^1 dx\, 
\Big[ g_A^2(\mu^2-2m_\pi^2) +2(1-g_A^2)\kappa^2x^2 \Big]
\nonumber \\ && \times \Bigg\{ \,96 \pi^2 f_\pi^2 \left[ (2m_\pi^2-\mu^2)(\bar{d}_1 
+\bar{d}_2) -2\kappa^2x^2 \bar{d}_3+4m_\pi^2 \bar{d}_5 \right] 
\nonumber \\ && +\left[ 4m_\pi^2 (1+2g_A^2) -\mu^2(1+5g_A^2)\right] 
{\kappa\over \mu} \ln {\mu +2\kappa\over 2m_\pi} \,
\Bigg.  \nonumber \\ && \Bigg.
+\,{\mu^2 \over 12} (5+13g_A^2) -2m_\pi^2 (1+2g_A^2) 
-\,3\kappa^2x^2 
\nonumber \\ && 
+6 \kappa x \sqrt{m_\pi^2 +\kappa^2 x^2} \ln{ \kappa x +\sqrt{m_\pi^2 
+\kappa^2 x^2}\over  m_\pi}
\nonumber \\ && +g_A^4\left(\mu^2 -2\kappa^2 x^2 -2m_\pi^2\right) 
\Bigg.  \nonumber \\ && \Bigg. \times
\left[ {5\over 6} +{m_\pi^2\over \kappa^2 x^2} 
-\left( 1 +{m_\pi^2\over \kappa^2 x^2} \right)^{3/2} 
\ln{ \kappa x +\sqrt{m_\pi^2 +\kappa^2 x^2}\over  m_\pi} \right] \Bigg\} \,,   
\\
{\rm Im}\, V_S(i\mu) &=&\mu^2\,{\rm Im}\, V_T(i\mu) = 
{g_A^2\mu \kappa^3 \over 8\pi f_\pi^4} \left(\bar{d}_{15}-\bar{d}_{14}\right) 
\nonumber \\ &&+{2g_A^6\mu \kappa^3 \over (8\pi f_\pi^2)^3} 
\int_0^1 dx(1-x^2)\left[ {1\over 6}-{m_\pi^2 \over \kappa^2x^2} 
+\left( 1+{m_\pi^2 \over \kappa^2x^2} \right)^{3/2} 
 \right. \nn  && \left. \times
\ln{ \kappa x +\sqrt{m_\pi^2 +\kappa^2 x^2}\over  m_\pi}
\right] \,,
\\
{\rm Im}\, W_S(i\mu) &=& \mu^2 \,{\rm Im}\, W_T(i\mu) = 
{g_A^4(4m_\pi^2-\mu^2) \over \pi (4f_\pi)^6} \Bigg[ \left( m_\pi^2 -{\mu^2 \over 4} \right)
\ln{\mu+2m_\pi \over \mu-2m_\pi} 
 \Bigg. \nn && \Bigg. 
+(1+2g_A^2)\mu  m_\pi\Bigg] \,, 
\end{eqnarray}
where $\kappa = \sqrt{\mu^2/4-m_\pi^2}$.

The momentum space amplitudes $V_\alpha(q)$ and $W_\alpha(q)$
are obtained from
the above expressions by means of 
subtracted dispersion integrals:
\begin{eqnarray} 
V_{C,S}(q) &=& 
-{2 q^{m+3} \over \pi} \int_{nm_\pi}^{\tilde{\Lambda}} d\mu \,
{{\rm Im\,}V_{C,S}(i \mu) \over \mu^{m+2} (\mu^2+q^2) }\,, 
\label{eq_disp1}
\\
V_T(q) &=& 
{2 q^{m+1} \over \pi} \int_{nm_\pi}^{\tilde{\Lambda}} d\mu \,
{{\rm Im\,}V_T(i \mu) \over \mu^m (\mu^2+q^2) }\,, 
\label{eq_disp2}
\end{eqnarray}
and similarly for $W_{C,S,T}$.
We use $m=3$ for the dispersion integrals that contribute at N$^3$LO and N$^4$LO, and $m=5$ at N$^5$LO.
Moreover, $n=2$ is applied for two-pion exchange and $n=3$ for
three-pion exchange.
For $\tilde{\Lambda} \rightarrow \infty$ the above dispersion integrals yield the
results of dimensional regularization, while for finite $\tilde{\Lambda} \geq nm_\pi$
we have what has become known  as spectral-function regularization (SFR) \cite{EGM04}. The 
purpose of the finite scale $\tilde{\Lambda}$ is to constrain the imaginary parts to the  
low-momentum region where chiral effective field theory is applicable.

\subsubsection{Leading relativistic corrections}

Counting $Q/M_N \sim Q^2/\Lambda^2_\chi$, the relativistic corrections of the NLO diagrams, which are shown in Fig.~\ref{fig_diagn3lo2pe}(c),
are of order N$^3$LO and are given by~\cite{ME11}:
\begin{eqnarray}
V_C &=& \frac{3 g_A^4}{128 \pi f_\pi^4 M_N} 
\bigg[\frac{m_\pi^5}{2w^2}+(2m_\pi^2+q^2)(q^2-m_\pi^2) A(\tilde{\Lambda};q) \bigg]
\,,
\label{eq_3EM1}
\\
W_C &=& \frac{g_A^2}{64 \pi f_\pi^4 M_N} 
\Bigg\{ \frac{3g_A^2m_\pi^5}{2\omega^2} +\big[g_A^2 (3m_\pi^2+2q^2) - 2m_\pi^2-q^2\big]
 \Bigg. \nn && \Bigg.  \times 
(2m_\pi^2+q^2) A(\tilde{\Lambda};q) \Bigg\}
\,,
\\
V_T &=& -\frac{1}{q^2} V_S = \frac{3 g_A^4}{256 \pi f_\pi^4 M_N} 
(5m_\pi^2+2q^2) A(\tilde{\Lambda};q)\,,
\\
W_T &=& -\frac{1}{q^2} W_S = \frac{g_A^2}{128 \pi f_\pi^4 M_N} 
\big[g_A^2 (3m_\pi^2+q^2)-w^2 \big] A(\tilde{\Lambda};q) \,,
\label{eq_3EM4}
\\
V_{LS} &=&  {3g_A^4  \over 32\pi f_\pi^4 M_N} \, (2m_\pi^2+q^2) A(\tilde{\Lambda};q)
 \,,\\  
W_{LS} &=& {g_A^2(1-g_A^2)\over 32\pi f_\pi^4 M_N} \, w^2 A(\tilde{\Lambda};q) \,.
\end{eqnarray}

\subsubsection{Leading three-pion exchange contributions}

\begin{figure}\centering
%\vspace*{-1cm}
%\hspace*{-1.5cm}
\scalebox{0.75}{\includegraphics{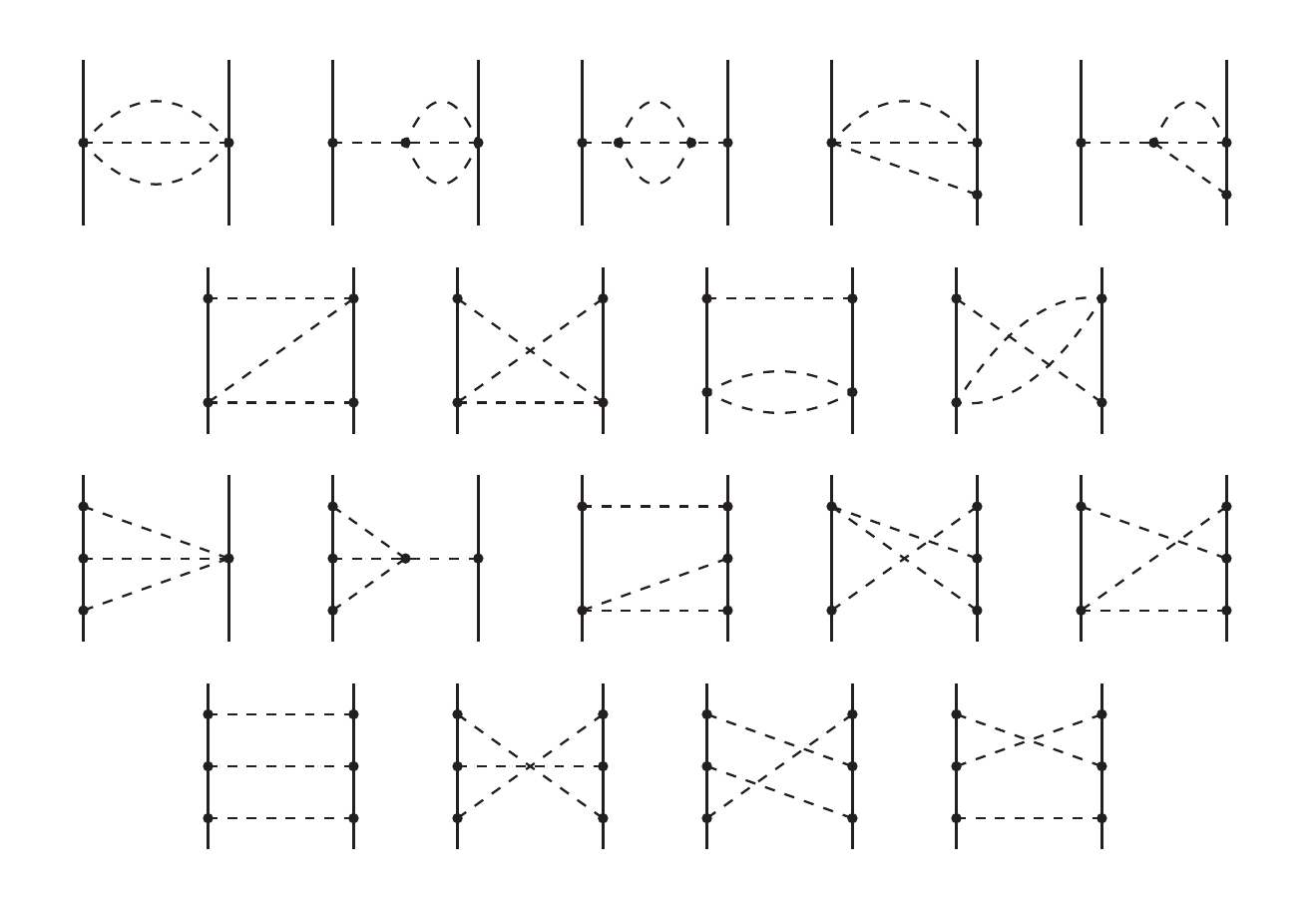}}
\vspace*{-0.3cm}
\caption{N$^3$LO three-pion exchange contributions.
Notation as in Fig.~\ref{fig_hi}.
(Figure reproduced from Ref.~\cite{ME11}.)}
\label{fig_diagn3lo3pe}
\end{figure}

 The leading $3\pi$-exchange contributions that occur at N$^3$LO
are shown in Fig.~\ref{fig_diagn3lo3pe}. They
have been calculated in Refs.~\cite{Kai00a,Kai00b} and are found to be negligible.
Therefore, we omit them.

\subsection{Next-to-next-to-next-to-next-to-leading order (N$^4$LO)}
\label{sec_n4lo}

At this order, we have two- and three-pion exchange contributions, which we will now discuss one by one.

\subsubsection{Two-pion exchange contributions at N$^4$LO}

\begin{figure}\centering
%\vspace*{-0.5cm}
%\hspace*{-1.5cm}
\scalebox{0.65}{\includegraphics{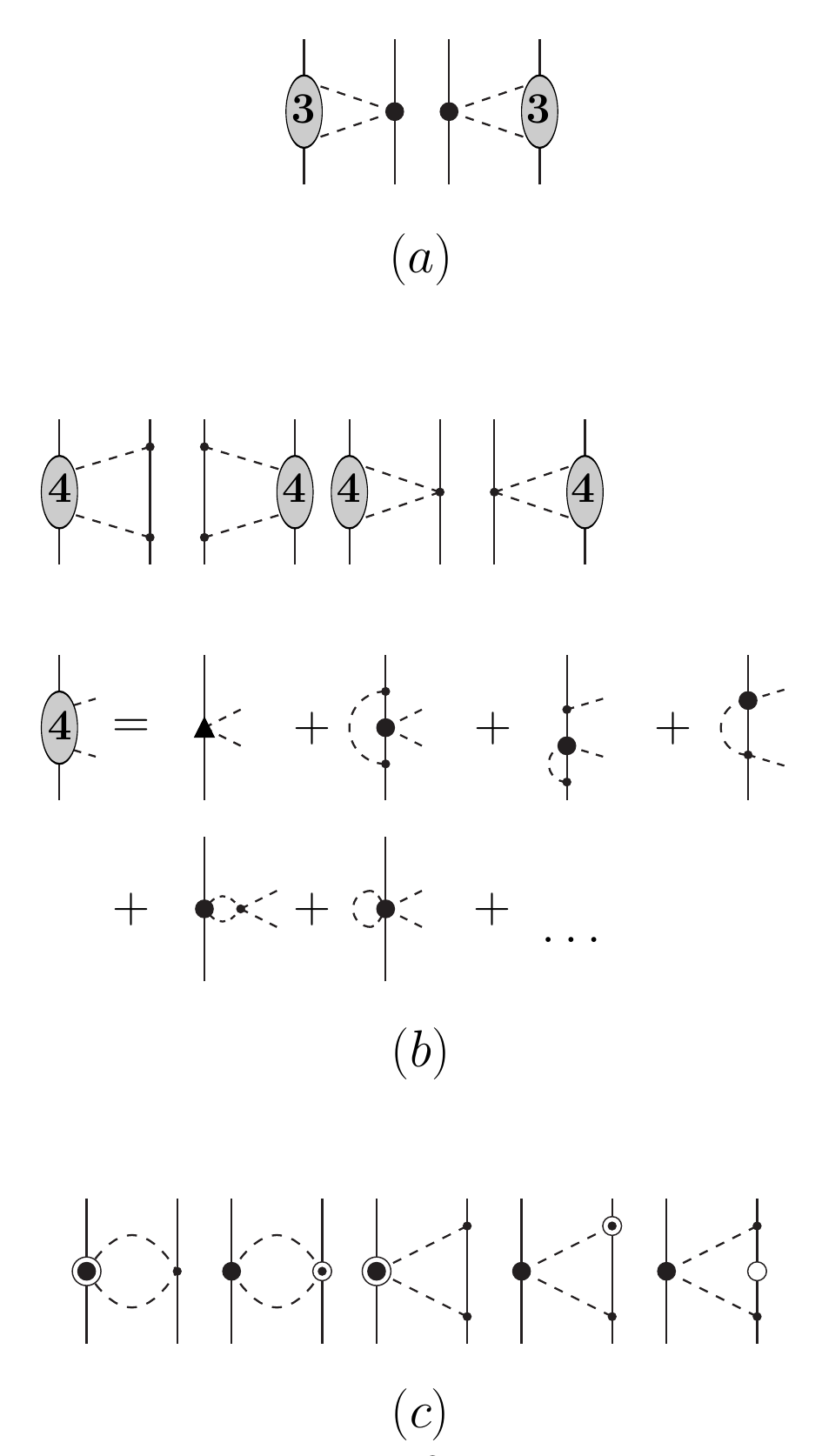}}
%\vspace*{-0.5cm}
\caption{N$^4$LO two-pion-exchange contributions.
(a) The leading one-loop $\pi N$ amplitude is folded with the chiral 
$\pi\pi NN$ vertices proportional to $c_i$. 
(b) The
one-loop $\pi N$ amplitude
proportional to $c_i$ 
is folded with the leading order chiral $\pi N$
amplitude.
(c) Relativistic corrections of NNLO diagrams.
Notation as in Figs.~\ref{fig_hi} and \ref{fig_diagn3lo2pe}.
}
\label{fig_diagn4lo2pe}
\end{figure}

The $2\pi$-exchange contributions that occur at N$^4$LO are displayed graphically in 
Fig.~\ref{fig_diagn4lo2pe}. 
We can distinguish between three groups of diagrams.

First, there are the N$^4$LO $2\pi$-exchange two-loop contributions of class (a), Fig.~\ref{fig_diagn4lo2pe}(a). For this class the spectral functions are obtained by 
integrating the product of the leading one-loop $\pi N$ amplitude and the chiral 
$\pi\pi NN$ vertex proportional to $c_i$ over the Lorentz-invariant $2\pi$-phase space. 

Second, we have the N$^4$LO $2\pi$-exchange two-loop contributions of class (b), Fig.~\ref{fig_diagn4lo2pe}(b).
Here, the product of the one-loop $\pi N$ amplitude
proportional to $c_i$ (see Ref.~\cite{KGE12} for details)
and the leading order chiral $\pi N$ amplitude is integrated over the $2\pi$-phase space.

The analytic expressions for the spectral functions of class (a) and (b) are very involved, which
is why we do not reprint them here. The interested reader is referred to Ref.~\cite{Ent15a}.

Finally, there also some relativistic corrections.
This group consists of diagrams with one vertex proportional to $c_i$
and one $1/M_N$ correction.
A few representative graphs are shown in Fig.~\ref{fig_diagn4lo2pe}(c).
Since in this investigation we count $Q/M_N \sim (Q/\Lambda_\chi)^2$, these relativistic 
corrections  are formally of order N$^4$LO. 
The result for this group of diagrams is~\cite{Kai01a}:
\begin{eqnarray} 
V_C & = & {g_A^2\, L(\tilde{\Lambda};q) \over 32 \pi^2 M_N f_\pi^4 } \left[ 
(6c_3-c_2) q^4 +4(3c_3-c_2-6c_1)q^2 m_\pi^2
\right. \nn && \left.
+6(2c_3-c_2)m_\pi^4- 24(2c_1+c_3)m_\pi^6 w^{-2} 
\right] \,,
\label{eq_4cMC}
\\
W_C &=& -{c_4 \over 192 \pi^2 M_N f_\pi^4 } 
\left[ g_A^2 (8m_\pi^2+5q^2) + w^2 \right] q^2 \,  L(\tilde{\Lambda};q)
\,, \\
W_T  &=&  -{1\over q^2} W_S = {c_4 \over 192 \pi^2 M_N f_\pi^4 } 
\left[ w^2-g_A^2 (16m_\pi^2+7q^2) \right]  L(\tilde{\Lambda};q)
\label{eq_4cMS}
\,,  \\
V_{LS}& = & {c_2 \, g_A^2 \over 8 \pi^2 M_N f_\pi^4 } 
\, w^2 L(\tilde{\Lambda};q) 
\,, \\
W_{LS}  &=& 
-{c_4  \over 48 \pi^2 M_N f_\pi^4 } 
\left[ g_A^2 (8m_\pi^2+5q^2) + w^2 \right]  L(\tilde{\Lambda};q)
\,.
\label{eq_4cMLS}
\end{eqnarray}

\subsubsection{Three-pion exchange contributions at N$^4$LO}

\begin{figure}\centering
%\vspace*{-1cm}
%\hspace*{-1.5cm}
\scalebox{0.75}{\includegraphics{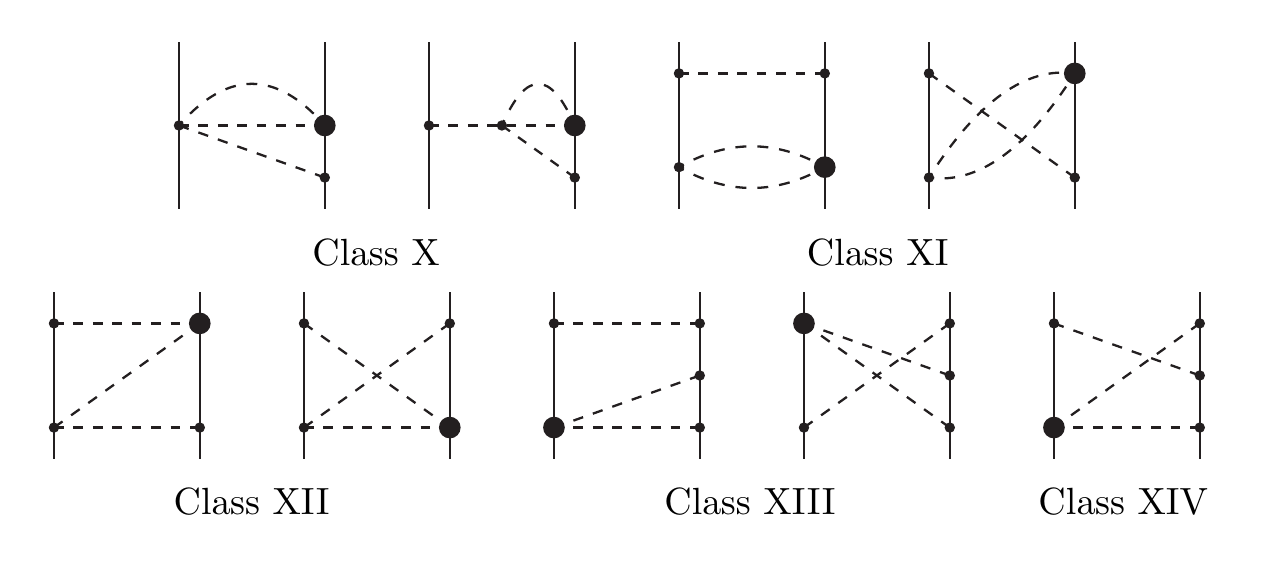}}
\vspace*{-0.3cm}
\caption{N$^4$LO three-pion exchange contributions.
Roman numerals refer to sub-classes following the scheme introduced in  
Refs.~\cite{Kai01,Ent15a}.
Notation as in Fig.~\ref{fig_hi}.
(Figure reproduced from Ref.~\cite{Ent15a}.)}
\label{fig_diagn4lo3pe}
\end{figure}

The $3\pi$-exchange of order N$^4$LO is shown in Fig.~\ref{fig_diagn4lo3pe}.
The spectral functions for these diagrams have been calculated in Ref.~\cite{Kai01}.
We use here the classification scheme introduced in that reference and note that class XI
vanishes. Moreover, we find that the class X and part of class XIV make only 
negligible contributions. Thus, we include in our calculations only class XII and XIII, and 
the $V_S$ contribution of class XIV.
For the very involved expressions, we refer the interested reader to Ref.~\cite{Ent15a}.

\subsection{Next-to-next-to-next-to-next-to-next-to-leading order (N$^5$LO)}
\label{sec_n5lo}

At N$^5$LO, we are faced with two-, three-, and four-pion exchange contributions.

\subsubsection{Two-pion exchange contributions at N$^5$LO}

\begin{figure}\centering
\vspace*{-0.3cm}
%\hspace*{-1.5cm}
\scalebox{0.75}{\includegraphics{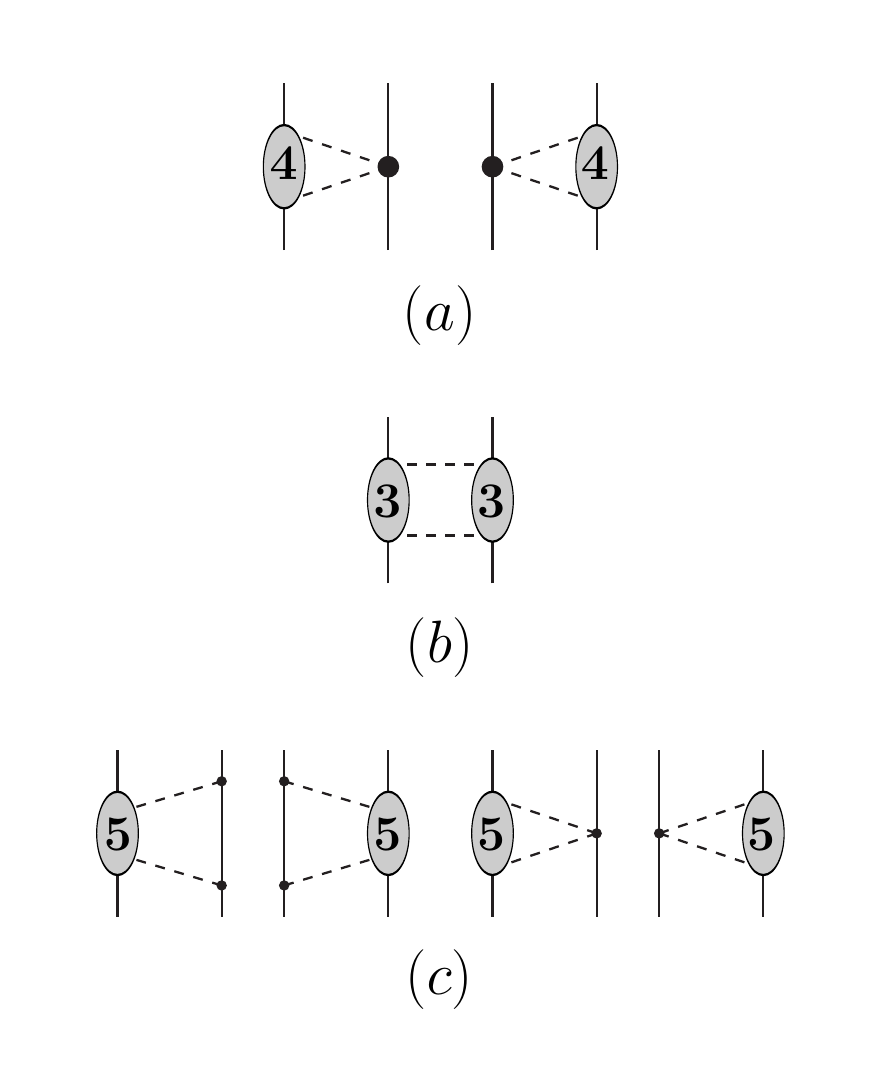}}
\vspace*{-0.5cm}
\caption{N$^5$LO two-pion-exchange contributions.
(a) The subleading one-loop $\pi N$-amplitude is folded with the chiral 
$\pi\pi NN$-vertices proportional to $c_i$. (b) The leading one-loop 
$\pi N$-amplitude is folded with itself. (c) The leading two-loop 
$\pi N$-amplitude is folded with the tree-level $\pi N$-amplitude. 
Notation as in Figs.~\ref{fig_hi} and \ref{fig_diagn3lo2pe}.
(Figure reproduced from Ref.~\cite{Ent15b}.)
}
\label{fig_diagn5lo2pe}
\end{figure}

The $2\pi$-exchange contributions that occur at N$^5$LO are displayed 
graphically in Fig.~\ref{fig_diagn5lo2pe}. We will now discuss each class separately.

The N$^5$LO $2\pi$-exchange two-loop contributions, denoted by class (a),  
are shown in Fig.~\ref{fig_diagn5lo2pe}(a). For this class the spectral functions 
are obtained by integrating the product of the subleading one-loop 
$\pi N$-amplitude (see Ref.~\cite{KGE12} for details) and the chiral 
$\pi\pi NN$-vertex proportional to $c_i$ over the Lorentz-invariant 
$2\pi$-phase space~\cite{Ent15b}.

A first set of $2\pi$-exchange contributions at three-loop order, denoted by 
class (b), is displayed in Fig.~\ref{fig_diagn5lo2pe}(b). Here, the leading one-loop $\pi N$-scattering amplitude is multiplied 
with itself and integrated over the $2\pi$-phase space~\cite{Ent15b}.

Further $2\pi$-exchange three-loop contributions at N$^5$LO, denoted by 
class (c), are shown in Fig.~\ref{fig_diagn5lo2pe}(c). For these, the two-loop 
$\pi N$-scattering amplitude (which is of order five) would have to be folded 
with the tree-level $\pi N$-amplitude. To our knowledge, the two-loop elastic 
$\pi N$-scattering amplitude has never been evaluated in some decent 
analytical form. Note that the loops involved in the class (c) contributions 
include only leading order chiral $\pi N$-vertices. According to our 
experience such contributions are typically small. For these reasons,
class (c) is neglected.

Besides the above, there are also some relativistic $1/M_N^2$-corrections.
This group consists of the $1/M_N^2$-corrections to the leading chiral 
$2\pi$-exchange diagrams. Since we count $Q/M_N \sim (Q/\Lambda_\chi)^2$, these 
relativistic corrections  are formally of sixth order (N$^5$LO). 
The expressions for the corresponding $N\!N$-amplitudes can be found in
Ref.~\cite{Kai02}.

\subsubsection{Three-pion exchange contributions at N$^5$LO}

\begin{figure}\centering
%\vspace*{-1cm}
%\hspace*{-1.5cm}
\scalebox{0.75}{\includegraphics{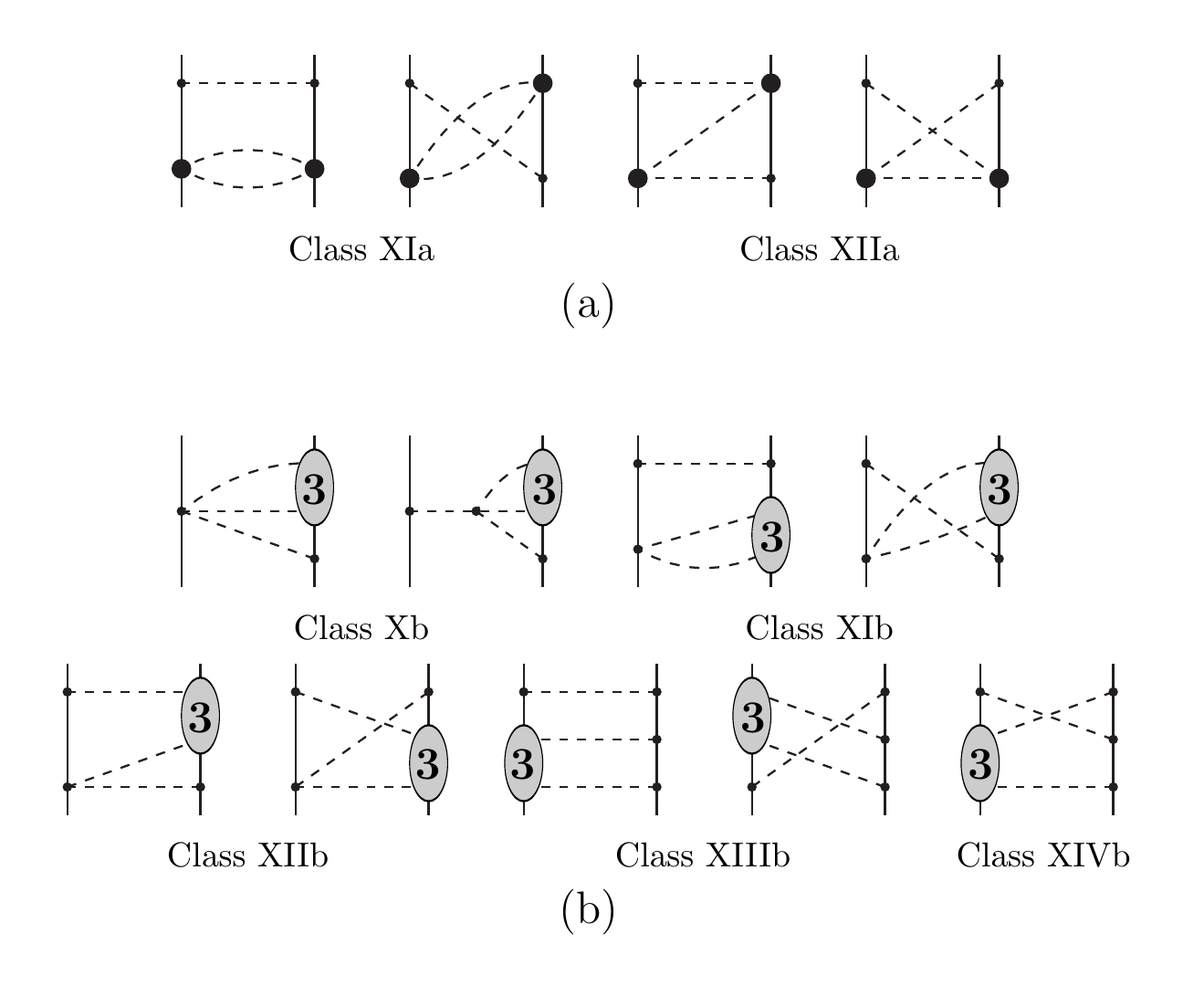}}
\vspace*{-0.5cm}
\caption{N$^5$LO three-pion exchange contributions.
(a) Diagrams proportional to $c_i^2$.
(b) Diagrams involving the one-loop $\pi N$-amplitude.
 Notation as in Figs.~\ref{fig_hi}, \ref{fig_diagn3lo2pe}, and \ref{fig_diagn4lo3pe}.
 (Figure reproduced from Ref.~\cite{Ent15b}.)}
\label{fig_diagn5lo3pe}
\end{figure}

The $3\pi$-exchange contributions of order N$^5$LO are shown in 
Fig.~\ref{fig_diagn5lo3pe}. We can distinguish between two classes.

Class (a) consists of the diagrams displayed in Fig.~\ref{fig_diagn5lo3pe}(a). They 
are characterized by the presence of one subleading $\pi\pi NN$-vertex in each 
nucleon line. Using a notation introduced in Refs.~\cite{Kai01,Ent15a},
we distinguish between the various sub-classes of diagrams by roman numerals.

Class (b) is shown in Fig.~\ref{fig_diagn5lo3pe}(b). Each $3\pi$-exchange 
diagram of this class includes the one-loop $\pi N$-amplitude (completed by 
the low-energy constants $\bar d_j$). Only those parts of the 
$\pi N$-scattering amplitude, which are either independent of the pion 
CMS-energy or depend on it linearly could be treated with the 
techniques available. The contributions are, in  general, small. Results presented below
include only the larger portions within this class. The omitted pieces are 
about one order of magnitude smaller. To facilitate a better understanding, 
we have subdivided this class into sub-classes labeled by roman numerals, 
following  Refs.~\cite{Kai01,Ent15a}.

The very involved analytic expressions for the spectral function can be found in Ref.~\cite{Ent15b}.

\subsubsection{Four-pion exchange at N$^5$LO}
The exchange of four pions between two nucleons occurs for the first time at 
N$^5$LO. The pertinent diagrams involve three loops and only leading order 
vertices, which explains the sixth power in small momenta. Three-pion 
exchange with just leading order vertices turned out to be negligibly 
small~\cite{Kai00a,Kai00b}, and so we expect four-pion exchange with 
leading order vertices to be even smaller. Therefore, we can safely neglect 
this contribution.

\section{Perturbative $NN$ scattering in peripheral partial waves}
\label{sec_pertNN}

We will now discuss $NN$ scattering involving states of high orbital angular 
momentum. 
We recall that peripheral scattering is the best tool to explore the
$NN$ force beyond short distances. Due to the high angular momentum 
``barrier", the contribution from short-range terms is marginal.    
In fact, since the contact terms at N$^4$LO  
do not contribute for $L\geq 3$, there exists the unique opportunity
to study the nucleon-nucleon force when it is controlled entirely by
pion exchanges, which carry the signature of chiral symmetry. 
In short, states with 
$L\geq 3$ are a suitable ground to test the predictive power of chiral EFT.
The LECs can be taken from $\pi N$ analysis, leaving no free parameters.
Furthermore, the scattering phases in high angular momentum states are small,
suggesting that a perturbative treatment would be appropriate.
On the other hand, the latter is not suitable for the central partial waves,
which require a 
non-perturbative approach to the solution of the 
Lippmann-Schwinger equation, with all its model (cutoff) dependence.

The perturbative $K$-matrix for $np$ scattering
is calculated as follows:
\begin{eqnarray}
K({\vec p}~',\vec p) &=& V_{1\pi}^{(np)}({\vec p}~',\vec p\,)
+V_{2\pi, \rm it}^{(np)}({\vec p}~',{\vec p}\,) 
+V_{3\pi, \rm it}^{(np)}({\vec p}~',{\vec p}\,) 
+ V({\vec p}~',{\vec p}\,) 
\label{eq_kmat}
\end{eqnarray}
with $V_{1\pi}^{(np)} ({\vec p}~', \vec p)$ as in Eq.~(\ref{eq_1penp}), and
$V_{2\pi, \rm it}^{(np)} ({\vec p}~',{\vec p})$ representing the once iterated
1PE given by
\begin{equation}
V_{2\pi, \rm it}^{(np)} ({\vec p}~',{\vec p}\,)  = {\cal P}\!\!\int {d^3p'' \over 
(2\pi)^3} \:\frac{M_N^2}{E_{p''}} \: \frac{V_{1\pi}^{(np)}({\vec p}~',{\vec p}~'')
\,V_{1\pi}^{(np)}({\vec p}~'',\vec p\,)} {{ p}^{2}-{p''}^{2}}\,, \label{eq_2piit}
\end{equation}
where ${\cal P}$ denotes the principal value integral and $E_{p''}=\sqrt{M_N^2+{p''}^2}$.
A calculation at LO includes only the first term on the right hand side of 
Eq.~(\ref{eq_kmat}), $ V_{1\pi}^{(np)} ({\vec p}~', \vec p) $, while calculations at NLO 
or higher order also include the second term on the right hand side,
$V_{2\pi, \rm it}^{(np)} ({\vec p}~',{\vec p})$.
At NNLO, the twice iterated 1PE should be included as well; and at higher orders further iterations
should be accounted for.
However, we found that the difference between the once iterated 1PE and the 
infinitely iterated 1PE is so small that it could not be identified on the scale
of our phase shift figures. For that reason, we omit iterations of 1PE beyond what is
contained in $V_{2\pi, \rm it}^{(np)} ({\vec p}~',{\vec p})$.
Furthermore, $V_{3\pi, \rm it}^{(np)}({\vec p}~',\vec p\,)$ 
stands for terms where irreducible 2PE is iterated with 1PE.

\begin{table}
\caption{Low-energy constants as determined in Ref.~\cite{KGE12}.
The sets `GW' and `KH' are based upon the $\pi N$ partial wave analyses of 
Refs.~\cite{Arn06} and \cite{Koc86}, respectively. 
The $c_i$  appear in Eq.~(\ref{eq_LD1}) and
are in units of GeV$^{-1}$.
The $\bar{d}_i$ and $\bar{e}_i$ 
belong to  $\widehat{\cal L}^{(3)}_{\pi N}$ and   $\widehat{\cal L}^{(4)}_{\pi N}$
[cf.\ Eqs.~(\ref{eq_LD2}) and (\ref{eq_LD3})] and are in units of
  GeV$^{-2}$ and GeV$^{-3}$, respectively.}
%\lineup
\begin{indented}
\item[] \begin{tabular}{@{}crr}
\hline
\hline
           &  GW &  KH \\
\hline
$c_1$ & --1.13 & --0.75 \\
$c_2$ & 3.69 & 3.49 \\
$c_3$ & --5.51 & --4.77 \\
$c_4$ & 3.71 & 3.34 \\
$\bar{d}_1 + \bar{d}_2$ & 5.57 & 6.21 \\
$\bar{d}_3$ & --5.35 & --6.83 \\
$\bar{d}_5$ & 0.02 & 0.78 \\
$\bar{d}_{14} - \bar{d}_{15}$ & --10.26 & --12.02 \\
$\bar{e}_{14}$ & 1.75 & 1.52 \\
$\bar{e}_{15}$ & --5.80 & --10.41 \\
$\bar{e}_{16}$ & 1.76 & 6.08 \\
$\bar{e}_{17}$ & --0.58 & --0.37 \\
$\bar{e}_{18}$ & 0.96 & 3.26 \\
\hline
\hline
\end{tabular}
\end{indented}
\label{tab_lecs}
\end{table}

Finally, the fourth term on the r.h.s.\ of Eq.~(\ref{eq_kmat}), $ V({\vec p}~',{\vec p}) $,
stands for the irreducible multi-pion exchange contributions that occur at the order at
which the calculation is conducted.
In multi-pion exchanges,
we use the average pion mass $m_\pi = 138.039$ MeV and, thus,
neglect the charge-dependence due to pion-mass splitting
in irreducible multi-pion diagrams.

Throughout this paper, we use
\beq
M_N  =  \frac{2M_pM_n}{M_p+M_n} = 938.9183 \mbox{ MeV.}
\eeq
 Based upon relativistic kinematics,
the CMS on-shell momentum $p$ is related to
the kinetic energy of the incident neutron 
in the laboratory system (``Lab.\ Energy''), $T_{\rm lab}$, by
\beq
p^2  =  \frac{M_p^2 T_{\rm lab} (T_{\rm lab} + 2M_n)}
               {(M_p + M_n)^2 + 2T_{\rm lab} M_p}  
\,,
\eeq
with $M_p=938.2720$ MeV and $M_n=939.5654$ MeV
 the proton and neutron masses, respectively.

The $K$-matrix, Eq.~(\ref{eq_kmat}), is decomposed into partial waves following 
Ref.~\cite{EAH71}
and phase shifts are then calculated via
\begin{equation}
\tan \delta_L (T_{\rm lab}) = -\frac{M_N^2p }{16\pi^2E_p} \, p \, K_L(p,p)
\,.
\end{equation}
For more details concerning the evaluation of phase shifts, including the case of coupled 
partial waves, see Ref.~\cite{Mac93} or the appendix of \cite{Mac01}.

\begin{figure*}\centering
\vspace*{-2cm}
%\hspace*{-0.9cm}
\scalebox{0.65}{\includegraphics{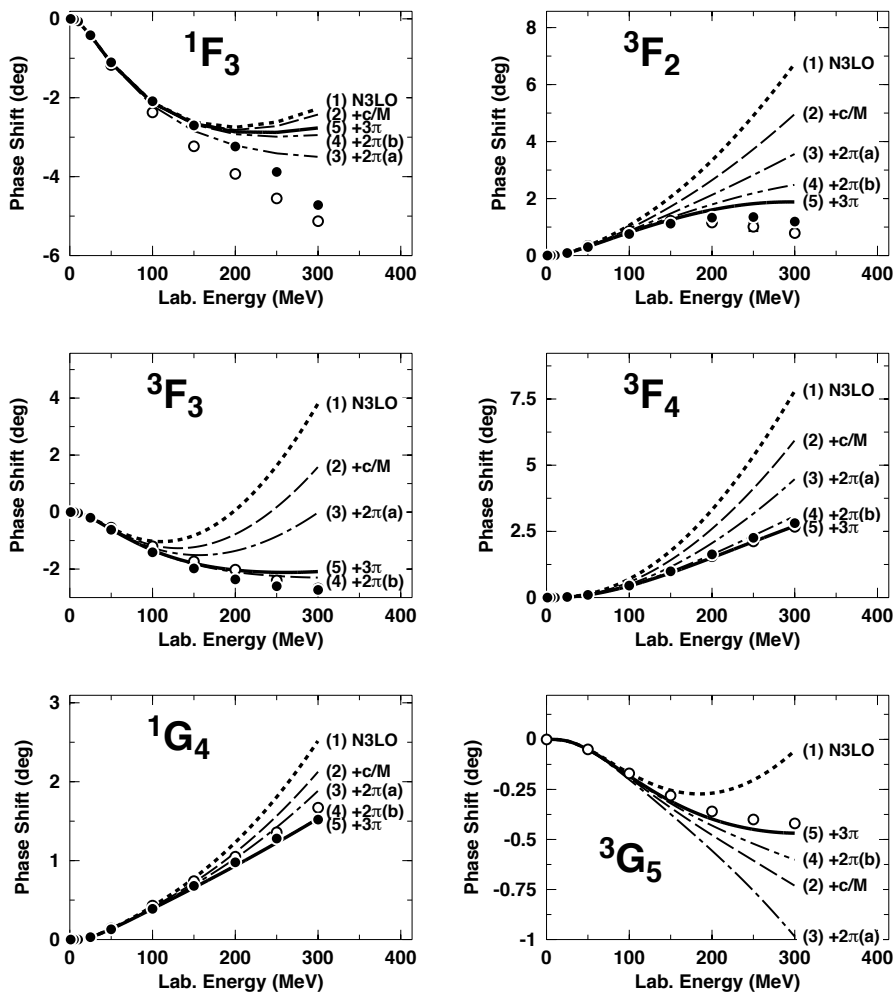}}
\vspace*{-3.0cm}
\caption{Effect of individual N$^4$LO (fifth-order) contributions on the neutron-proton phase shifts 
of some selected peripheral partial waves. 
The individual contributions are added up successively in the order given in parenthesis
next to each curve. Curve (1) is N$^3$LO and curve (5) is the complete N$^4$LO.
The KH LECs are used and $\tilde{\Lambda}$=1.5 GeV.
The filled and open circles represent the results from the Nijmegan multi-energy $np$ phase-shift analysis~\cite{Sto93} and the VPI/GW single-energy $np$ analysis SM99~\cite{SM99}, respectively.
(Figure reproduced from Ref.~\cite{Ent15a}.)
\label{fig_ph1}}
\end{figure*}

Chiral symmetry establishes a link between the dynamics in the $\pi N$-system 
and the $N\!N$-system (through common low-energy constants). In order to check 
the consistency, we use the LECs for subleading $\pi N$-couplings as 
determined in analyses of low-energy elastic $\pi N$-scattering.
Appropriate analyses for our purposes are contained in 
Refs.~\cite{KGE12}, where $\pi N$-scattering has been calculated at fourth 
order using the same power-counting of relativistic $1/M_N$-corrections as in 
the present work. Reference~\cite{KGE12} performed two fits, one to the 
GW~\cite{Arn06} and one to the KH~\cite{Koc86} partial wave analysis 
resulting in the two sets of LECs listed in Table~\ref{tab_lecs}.

The contributions up to N$^3$LO and their impact on peripheral $NN$ scattering
have been discussed and demonstrated in detail in Ref.~\cite{ME11} and, therefore, we will not repeat that demonstration here. But we will discuss the recent progress that has been made in the
calculation of orders beyond N$^3$LO.

We start with the individual N$^4$LO (fifth-order) contributions.
For this purpose, we display in 
Fig.~\ref{fig_ph1} 
phase shifts
for six important peripheral partial waves, namely,
$^1F_3$, $^3F_2$, $^3F_3$, $^3F_4$, $^1G_4$, and $^3G_5$.
In each frame, the following curves
are shown:
\begin{description}
\item[(1)]
N$^3$LO.
\item[(2)] The previous curve plus
the $c_i/M_N$ corrections (denoted by `c/M'),
Fig.~\ref{fig_diagn4lo2pe}(c).
\item[(3)] The previous curve plus
the N$^4$LO $2\pi$-exchange two-loop contributions of class (a),
Fig.~\ref{fig_diagn4lo2pe}(a).
\item[(4)] The previous curve plus
the N$^4$LO $2\pi$ two-loop contributions of class (b),
Fig.~\ref{fig_diagn4lo2pe}(b).
\item[(5)] The previous curve plus
the N$^4$LO $3\pi$-exchange contributions,
Fig.~\ref{fig_diagn4lo3pe}.
\end{description}
In summary, the various curves add up
successively
the individual N$^4$LO contributions 
in the order indicated in the curve labels.
The last curve in this series, curve (5),
is the full N$^4$LO result.
In these calculations, a SFR cutoff $\tilde{\Lambda}=1.5$ GeV is applied 
[cf.\ Eqs.~(\ref{eq_disp1}) and (\ref{eq_disp2})] 
and the KH LECs (cf.\ Table~\ref{tab_lecs}) are used.

\begin{figure*}\centering
%\vspace*{-1cm}
%\hspace*{-0.3cm}
\scalebox{0.59}{\includegraphics{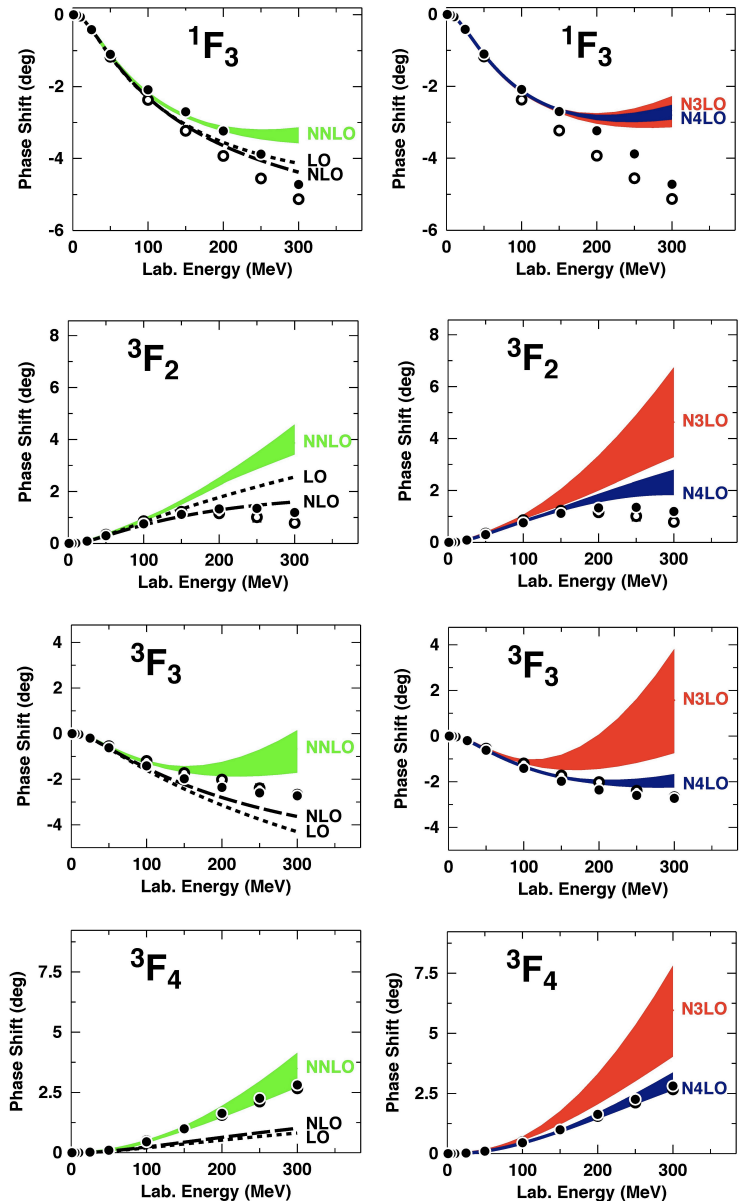}}
%\vspace*{-2.5cm}
\caption{
Phase-shifts of neutron-proton scattering at various orders up to N$^4$LO. 
The colored bands show the variation of the predictions when the
SFR cutoff $\tilde{\Lambda}$ is changed over the range 0.7 to 1.5 GeV.
The KH LECs are applied.
Empirical phase shifts as in Fig.~\ref{fig_ph1}.
(Figure reproduced from Ref.~\cite{Ent15a}.)}
\label{fig_ph2}
\end{figure*}

\begin{figure*}\centering
%\vspace*{-1cm}
%\hspace*{-0.55cm}
\scalebox{0.59}{\includegraphics{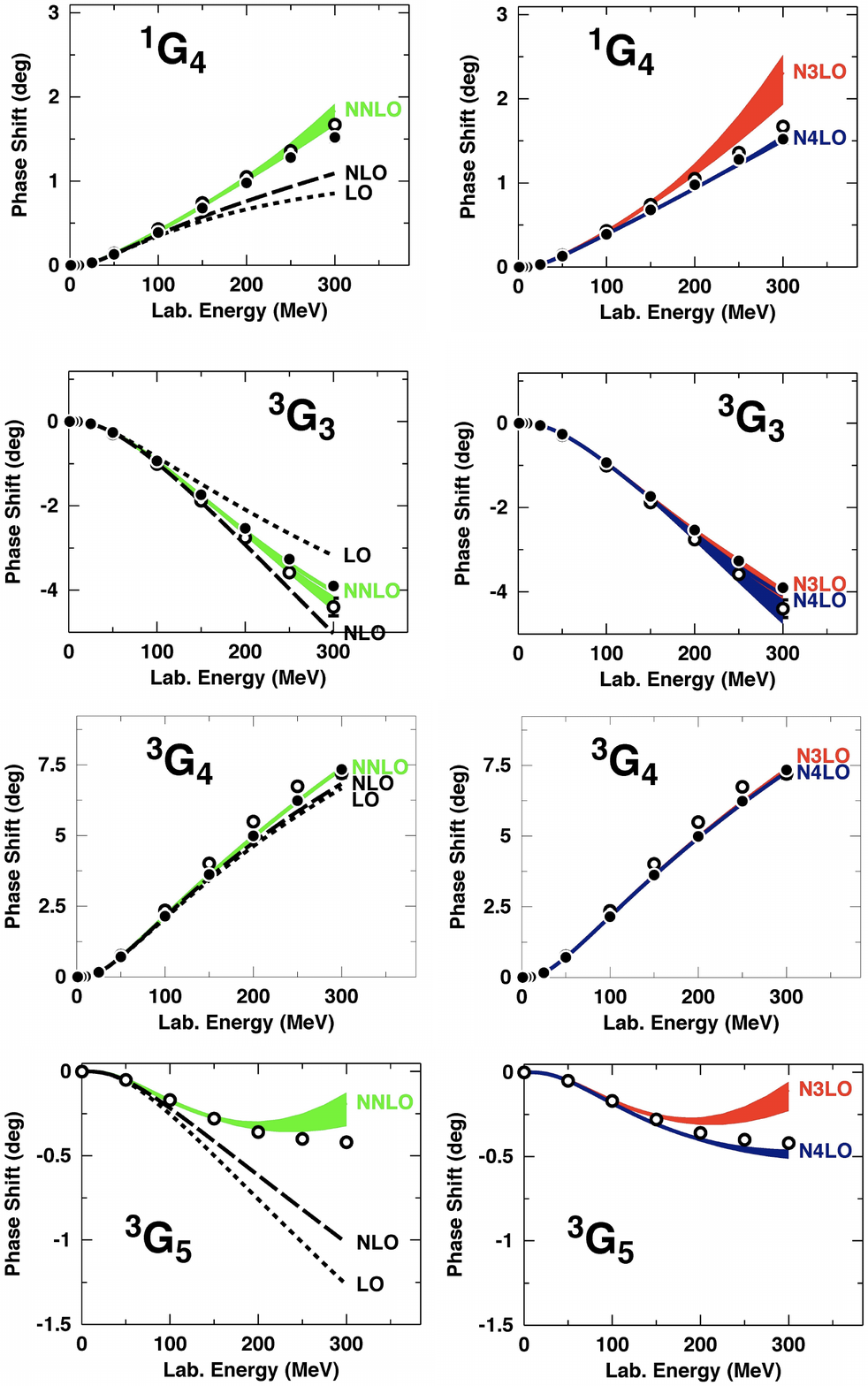}}
%\vspace*{-2.5cm}
\caption{
Same as Fig.~\ref{fig_ph2}, but for $G$-waves.
(Figure reproduced from Ref.~\cite{Ent15a}.)
\label{fig_ph3}}
\end{figure*}

From Fig.~\ref{fig_ph1}, we make the following observations. In triplet $F$-waves,
the $c_i/M_N$ corrections as well as the 2PE two-loops, class (a) and (b),
are all repulsive and of about the same strength. As a consequence, the problem of
the excessive attraction, that N$^3$LO is beset with, is overcome.
A similar trend is seen in $^1G_4$.
An exception is $^1F_3$, where the class (b) contribution is attractive leading to
phase shifts above the data for energies higher than 150 MeV. 

Now turning to the N$^4$LO 3PE contributions [curve (5) in Fig.~\ref{fig_ph1}]: 
they are substantially smaller than the 2PE two-loop ones, in all peripheral partial waves. 
This can be interpreted as an indication of convergence
with regard to the number of pions being exchanged between two nucleons---a trend
that is very welcome. Further, note that the total 3PE contribution is a very comprehensive one,
cf.\ Fig.~\ref{fig_diagn4lo3pe}. It is the sum of ten terms
which, individually, can be fairly large.
However, destructive interference between them leads to the small net result.

For all $F$ and $G$ waves (except $^1F_3$), the final N$^4$LO result
is close to the empirical phase shifts. Notice that this includes
also $^3G_5$, which posed persistent problems at N$^3$LO~\cite{EM02}.

It is also of interest to know how predictions change with variations
of $\tilde{\Lambda}$ within a reasonable range. We have, therefore, varied $\tilde{\Lambda}$ between 0.7 and 1.5 GeV
and show the predictions for all $F$ and $G$ waves in Figs.~\ref{fig_ph2} and \ref{fig_ph3},
respectively, in terms of colored bands. It is seen that, at N$^3$LO, the variations of the predictions
are very large and always too attractive while, at N$^4$LO, the variations are small
and the predictions are close to the data or right on the data.
 Figs.~\ref{fig_ph2} and \ref{fig_ph3} also include the lower orders (LO, NLO, and NNLO)
such that a comparison of the relative size of the order-by-order  contributions
is possible. We observe that there is not much of a convergence, since obviously the magnitudes
of the NNLO, N$^3$LO, and N$^4$LO contributions are about the same.

\begin{figure*}\centering
\vspace*{-2cm}
%\hspace*{-0.8cm}
\scalebox{0.70}{\includegraphics{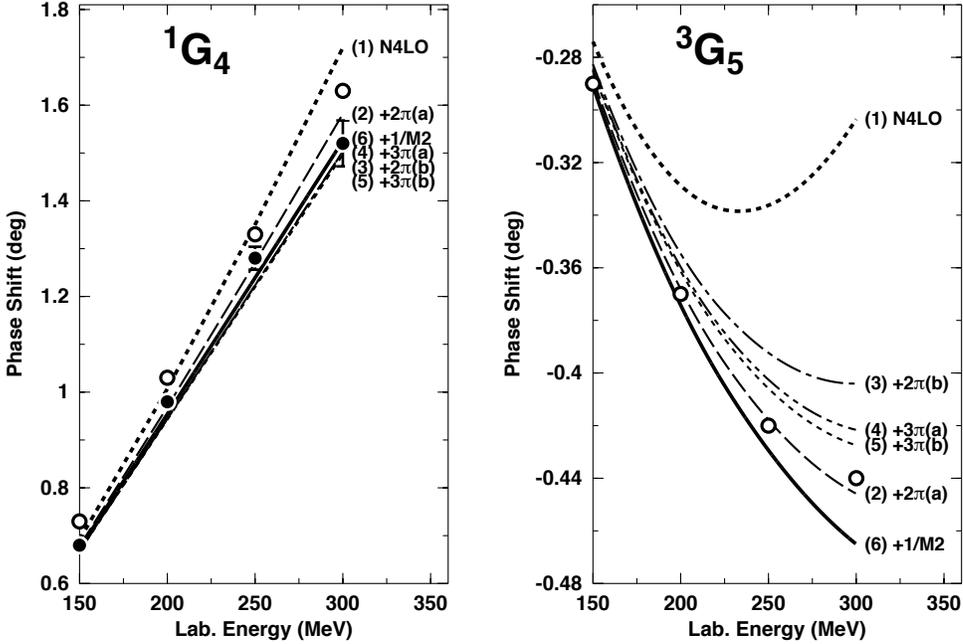}}
\vspace*{-8.75cm}
\caption{Effect of individual N$^5$LO (sixth-order) contributions on the neutron-proton 
phase shifts of two $G$-waves. The individual contributions are 
added up successively in the order given in parentheses next to each curve. 
Curve (1) is N$^4$LO and curve (6) contains all N$^5$LO contributions
calculated in Ref.~\cite{Ent15b}. A SFR cutoff $\tilde{\Lambda}=800$ MeV is applied
and the GW LECs are used.
The filled and open circles represent the results from the Nijmegen 
multi-energy $np$ phase-shift analysis~\cite{Sto93} and the GW $np$-analysis 
SP07~\cite{SP07}, respectively.
(Figure reproduced from Ref.~\cite{Ent15b}.)
\label{fig_ph4}}
\end{figure*}

To obtain more insight into the convergence issue, we need to proceed
 to the next order, which is N$^5$LO.
As shown in Figs.~\ref{fig_diagn5lo2pe} and \ref{fig_diagn5lo3pe}, 
the sixth-order corrections consist of several 
contributions. As in the case of N$^4$LO, we will first show how the individual N$^5$LO
contributions impact $N\!N$-phase-shifts in peripheral waves.
In Fig.~\ref{fig_ph4}, we display phase-shifts
for two peripheral partial waves, namely, $^1G_4$, and $^3G_5$.
The following curves are shown:
\begin{description}
\item[(1)]
N$^4$LO.
\item[(2)] The previous curve plus
the N$^5$LO $2\pi$-exchange contributions of class (a),
Fig.~\ref{fig_diagn5lo2pe}(a).
\item[(3)] The previous curve plus
the N$^5$LO $2\pi$-exchange contributions of class (b),
Fig.~\ref{fig_diagn5lo2pe}(b).
\item[(4)] The previous curve plus
the N$^5$LO $3\pi$-exchange contributions of class (a),
Fig.~\ref{fig_diagn5lo3pe}(a).
\item[(5)] The previous curve plus
the N$^5$LO $3\pi$-exchange contributions of class (b),
Fig.~\ref{fig_diagn5lo3pe}(b).
\item[(6)] The previous curve plus
the $1/M_N^2$-corrections (denoted by `1/M2')~\cite{Kai02}.
\end{description}
The last curve in 
this series, curve (6), includes all N$^5$LO contributions calculated in Ref.~\cite{Ent15b}.
For all curves of this figure, a SFR cutoff $\tilde{\Lambda}=800$ MeV 
[cf.\ Eqs.~(\ref{eq_disp1}) and (\ref{eq_disp2})] 
is employed and the GW (cf.\ Table~\ref{tab_lecs}) LECs are used.

From Fig.~\ref{fig_ph4}, we see that
the two-loop $2\pi$-exchange class (a), Fig.~\ref{fig_diagn5lo2pe}(a), generates a 
strong repulsive central force, while the spin-spin and tensor forces provided by this 
class are negligible. The fact that this class 
produces a relatively large contribution is not unexpected, since it is 
proportional to $c_i^2$. The $2\pi$-exchange contribution class (b), 
Fig.~\ref{fig_diagn5lo2pe}(b), creates a moderately repulsive central force as seen 
by its effect on $^1G_4$ and a noticeable tensor force as the impact on 
$^3G_5$ demonstrates. The $3\pi$-exchange class (a), Fig.~\ref{fig_diagn5lo3pe}(a), 
is negligible in $^1G_4$, but noticeable in $^3G_5$ and, therefore, it should 
not be neglected. This contribution is proportional to $c_i^2$, which 
suggests a non-negligible size but it is  typically  smaller than the 
corresponding $2\pi$-exchange contribution class (a). The $3\pi$-exchange 
class (b) contribution, Fig.~\ref{fig_diagn5lo3pe}(b), turns out to be negligible 
[see the difference between curve (4) and (5) in Fig.~\ref{fig_ph4}].
This may not be unexpected since it is a three-loop contribution with only 
leading-order vertices. Finally the relativistic $1/M_N^2$-corrections to the 
leading $2\pi$-exchange~\cite{Kai02} have a small but non-negligible 
impact, particularly in $^3G_5$.

\begin{figure*}
\vspace*{-1.2cm}
\hspace*{-0.7cm}
\scalebox{0.42}{\includegraphics{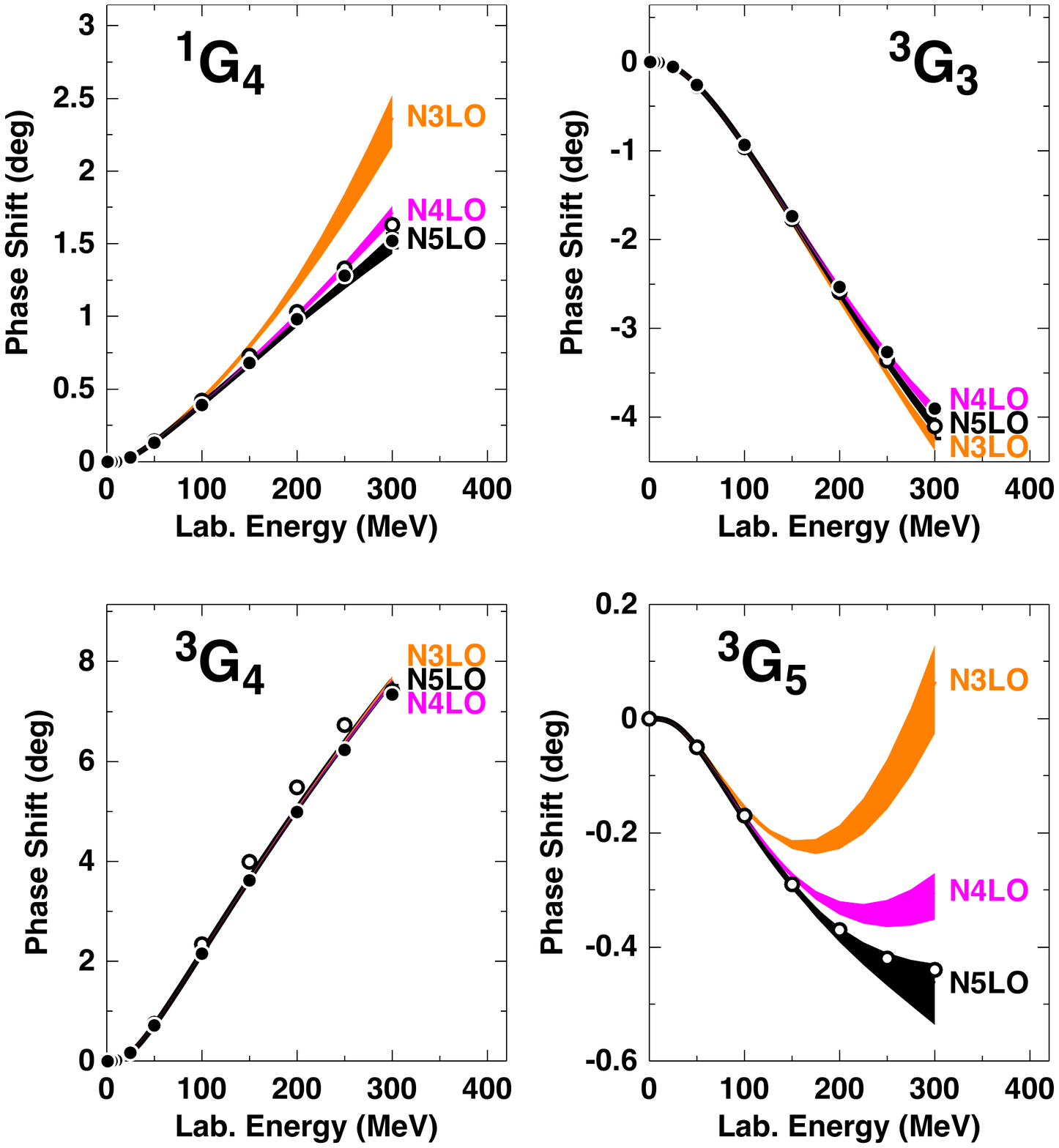}}
\hspace*{-1.2cm}
\scalebox{0.42}{\includegraphics{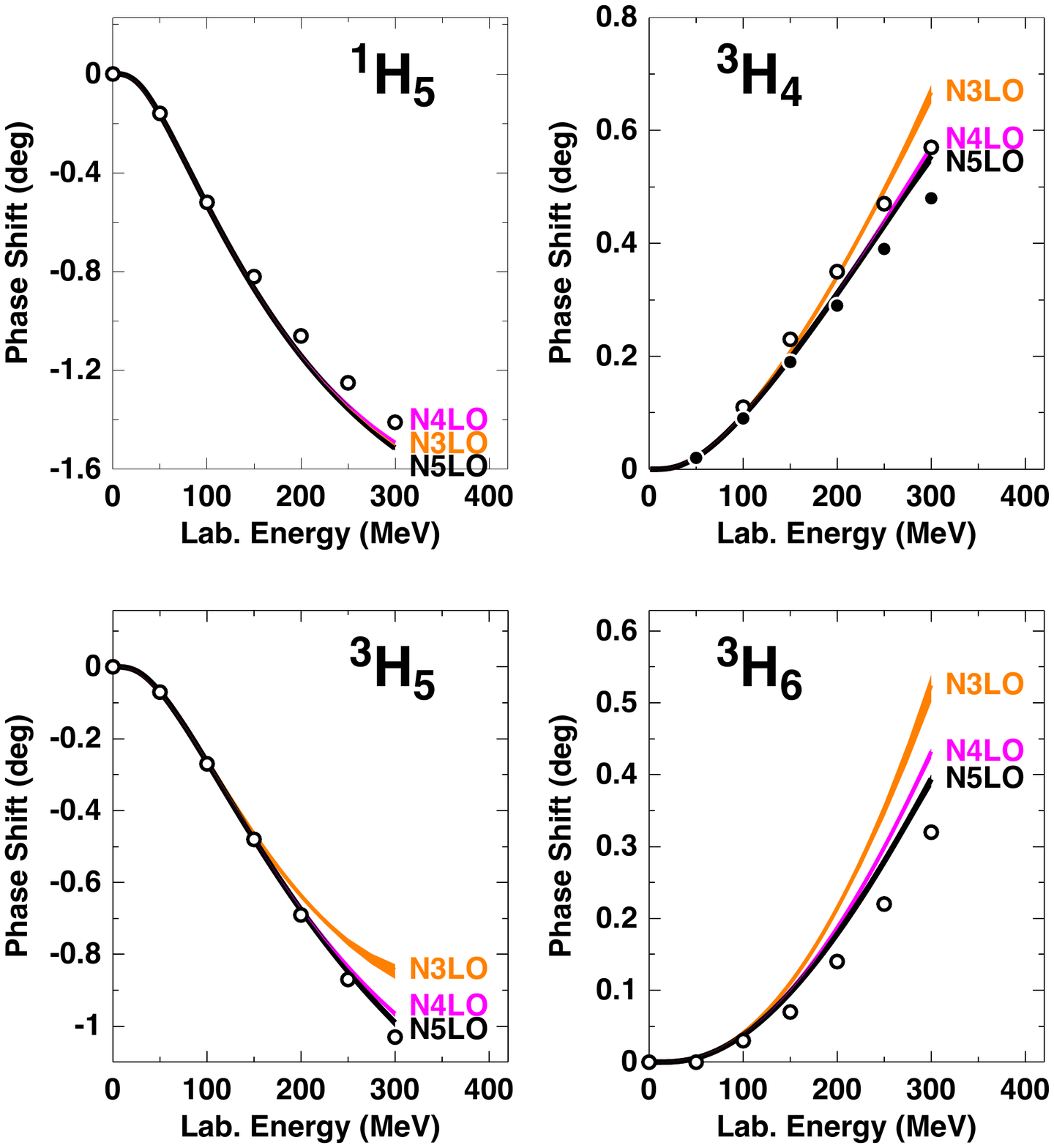}}
\vspace*{-2.3cm}
\caption{Phase-shifts of neutron-proton scattering in $G$ and $H$ waves at N$^3$LO, N$^4$LO, and  N$^5$LO. 
The colored bands show the variations of the 
predictions when the SFR cutoff $\tilde{\Lambda}$ is changed over the range 
700 to 900 MeV. The GW LECs are applied. Empirical phase shifts are as in Fig.~\ref{fig_ph4}.
(Figure reproduced from Ref.~\cite{Ent15b}.)
\label{fig_ph5}}
\end{figure*}

The N$^5$LO predictions for all $G$ and $H$ waves are displayed in Fig.~\ref{fig_ph5} 
in terms of colored bands that are generated by varying the SFR 
cutoff $\tilde{\Lambda}$ 
[cf.\ Eqs.~(\ref{eq_disp1}) and (\ref{eq_disp2})] 
between 700 and 900 MeV.
The figure clearly reveals again that, at N$^3$LO, the predictions are, in general, 
too attractive. As discussed, the N$^4$LO 
contribution, essentially, compensates this attractive surplus. 
N$^5$LO then adds additional repulsion 
bringing the final prediction right onto the data (i.e. empirical 
phase-shifts). Moreover, the N$^5$LO contribution is, in general, 
substantially smaller than the one at N$^4$LO, thus, showing a 
signature of convergence of the chiral expansion.

\begin{figure*}
\vspace*{-1.2cm}
\hspace*{-0.7cm}
\scalebox{0.42}{\includegraphics{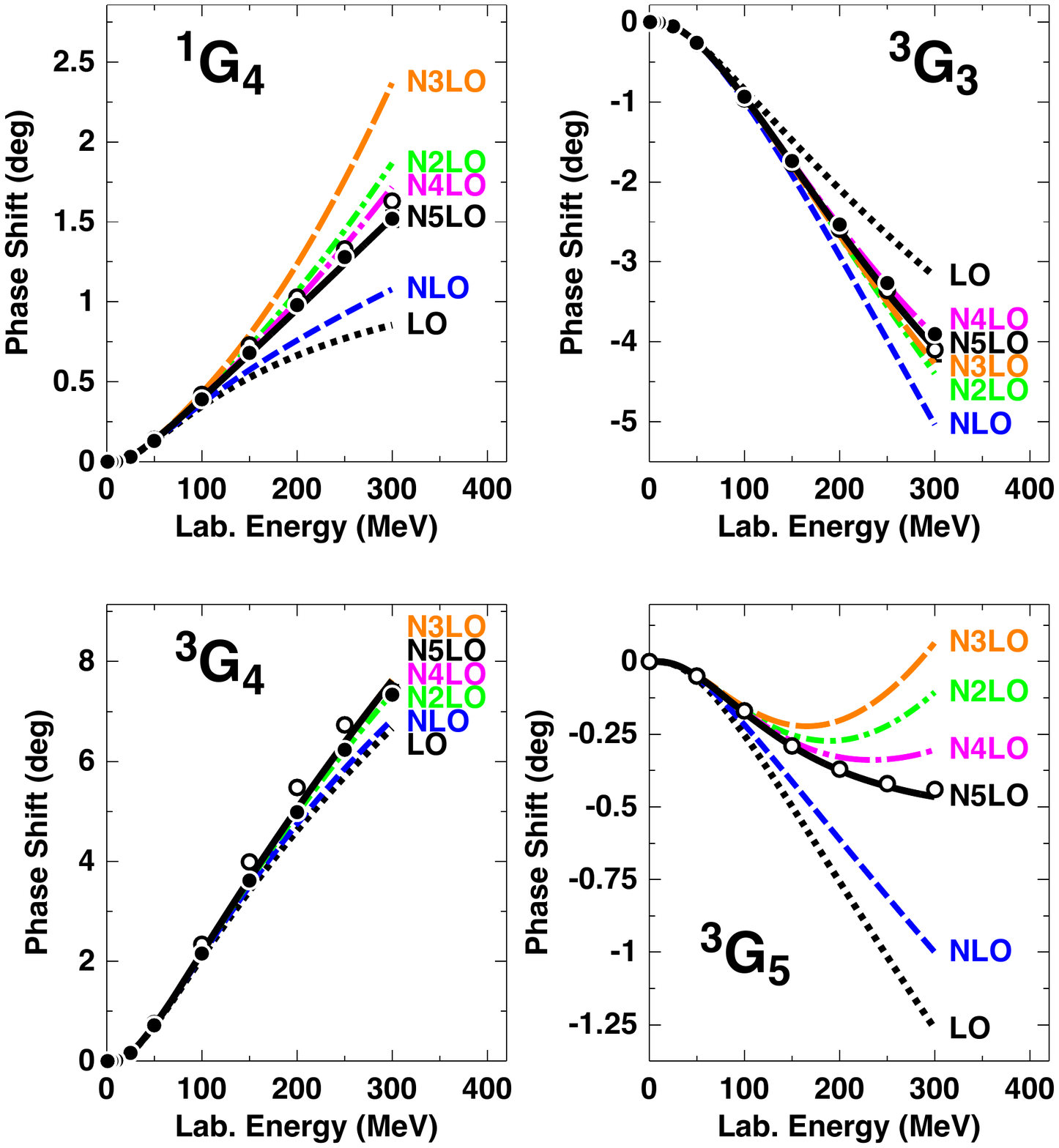}}
\hspace*{-1.2cm}
\scalebox{0.42}{\includegraphics{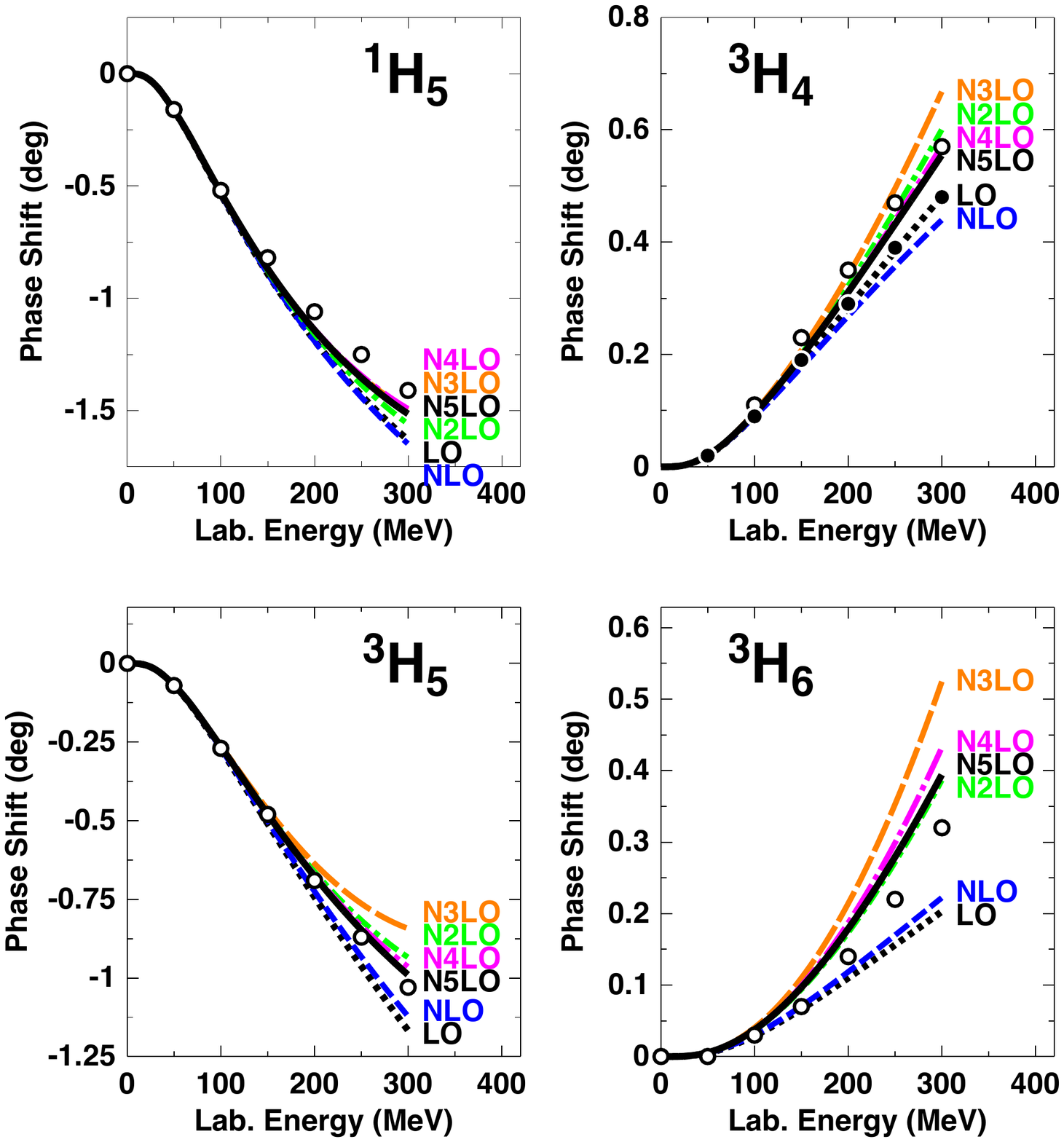}}
\vspace*{-2.3cm}
\caption{Phase-shifts of neutron-proton scattering in $G$ and $H$ waves at all orders 
from LO to N$^5$LO.  A SFR cutoff $\tilde{\Lambda}=800$ MeV is used
and the GW LECs are applied.
Empirical phase shifts are as in Fig.~\ref{fig_ph4}.
(Figure reproduced from Ref.~\cite{Ent15b}.)
\label{fig_ph6}}
\end{figure*}

To summarize, we present in Fig.~\ref{fig_ph6} a 
comparison between all orders from LO to N$^5$LO. Note that the difference between the LO prediction 
(one-pion-exchange, dotted line) and the data (filled and open circles) is to 
be provided by two- and three-pion exchanges, i.e. the intermediate-range part
of the nuclear force. How well that is accomplished is a crucial test for
any theory of nuclear forces.  NLO produces only a small contribution, 
but N$^2$LO creates substantial intermediate-range attraction (most clearly 
seen in $^1G_4$, $^3G_5$, and $^3H_6$). In fact, N$^2$LO is the largest 
contribution among all orders. This is due to the one-loop $2\pi$-exchange 
 triangle diagram which involves one $\pi\pi NN$-contact vertex 
proportional to $c_i$. This vertex represents correlated 2PE as well as 
intermediate $\Delta(1232)$-isobar excitation. It is well-known from the 
traditional meson theory of nuclear forces~\cite{Lac80,MHE87}
that these two features are crucial for a realistic and quantitative 2PE model.
Consequently, the one-loop $2\pi$-exchange at N$^2$LO is attractive and 
assumes a realistic size describing the intermediate-range attraction of the
nuclear force almost correctly. At N$^3$LO, more one-loop 2PE is added by the 
bubble diagram with two $c_i$-vertices, a contribution that seems to
overestimate the attraction. This attractive surplus is then compensated by
the prevailingly repulsive two-loop $2\pi$- and $3\pi$-exchanges that occur 
at N$^4$LO and N$^5$LO.

In this context, it is worth noting that also in conventional meson 
theory~\cite{MHE87} the one-loop models for the 2PE contribution always show 
some excess of attraction (cf. Figs.~7-9 of Ref.~\cite{EM02}). The same is 
true for the dispersion theoretic approach pursued by the Paris 
group~\cite{Lac80}. In conventional meson theory, the surplus 
attraction is reduced by heavy-meson exchange ($\rho$- and $\omega$-exchange) 
which, however, has no place in chiral effective field theory (as a 
finite-range contribution). Instead, in the latter approach, two-loop 
$2\pi$- and $3\pi$-exchanges provide the corrective action.

\section{Constructing complete chiral $NN$ potentials
\label{sec_pot2}}

%HERE1
Previously, we addressed the long- and medium-range 
parts of the nuclear interaction, which involve                                                                 
 pion-exchange contributions. Because of their long-range nature, these terms
control partial waves with high values of $L$ and are governed by                   
chiral symmetry.                                           
Of course, to obtain quantitative 
predictions of low-energy $NN$ scattering observables or nuclear properties,                           
all partial waves must be described realistically, in particular the most
central ones 
($L\leq 2$), which carry information on the dynamics at short range. 
The latter will be our next concern.

\subsection{$NN$ contact terms \label{sec_ct}}

It has been known for a long time that the bulk of the short-distance
behavior of the nuclear force can be explained with the introduction of
heavy bosons, e.g.~the 
$\omega(782)$. Applying Fourier transformation to the 
propagator of the meson, 
\begin{equation}
\int \frac{d^3q}{(2\pi)^3}  \;  \frac{e^{i{\vec q} \cdot {\vec r}}}{m^2_\omega
+ {\vec q}^2} \;
= \; \frac{1}{4\pi} \;
 \frac{e^{-m_\omega r}}{r} \; , 
\end{equation}
provides a qualitative description of the $NN$ force at short range. 

Since ChPT is an expansion valid for small values of the momentum, 
mesons such as 
the $\rho(770)$ or the $\omega(782)$ are outside its resolution power 
(notice that $\Lambda_\chi \approx m_{\rho,\omega}$). 
However, the propagator of the heavy boson under consideration can be handled 
with an expansion,                         
\begin{equation}
\frac{1}{m^2_\omega + Q^2} 
\approx 
\frac{1}{m^2_\omega} 
\left( 1 
- \frac{Q^2}{m^2_\omega}
+ \frac{Q^4}{m^4_\omega}
-+ \ldots
\right)
.
\end{equation}
One may then approach the short-range $NN$ interaction guided by
the expansion above, namely as a power series in 
$Q/m_\omega$. This is the origin of the contributions referred to as 
contact terms.

Contact terms play an important role in renormalization.
Contributions involving the exchange of more than one pion 
entail        
loop integrals, which produce            
polynomial terms whose coefficients can be divergent or 
scale dependent (cf.\ Appendix B of Ref.~\cite{ME11}). Contact terms are then crucial to 
remove those divergences or scale dependences and so they act as ``counter terms". 
        
Our procedure will involve partial-wave expansion of terms polynomial in $Q$, 
where $Q$ stands for        
the momentum transfer between the two               
nucleons, $q$, or their average momentum $k$
[see below Eq.~(\ref{eq_nnamp}) for their definitions].
In any case, for even $\nu$,
\begin{equation}
Q^\nu = 
f_{\frac{\nu}{2}}(\cos \theta) 
\, ,
\end{equation}
where $f_m$ stands for a polynomial of degree $m$
and $\theta$ is the scattering angle in the center-of-mass system. 
When expanding $Q^\nu$ in partial waves, we encounter 
the integral
\begin{equation}
I^{(\nu)}_L  
=\int_{-1}^{+1} Q^\nu P_L(\cos \theta) d\cos \theta 
=\int_{-1}^{+1}
f_{\frac{\nu}{2}}(\cos \theta) 
 P_L(\cos \theta) d\cos \theta 
\,,
\end{equation}
where $L$ is the orbital angular momentum and 
$P_L$ is a Legendre polynomial.
Since Legendre polynomials are orthogonal, 
\begin{equation}
I^{(\nu)}_L = 0  
\hspace*{.5cm}
\mbox{for}
\hspace*{.5cm}
L > \frac{\nu}{2} \, .
\end{equation}
Therefore, we can see that contact terms of a particular order
can only contribute up to some partial wave.   

Parity conservation forbids the presence of odd powers of $Q$. 
Therefore, the contact interaction can be formally written as 
\begin{equation}
V_{\rm ct} =
V_{\rm ct}^{(0)} + 
V_{\rm ct}^{(2)} + 
V_{\rm ct}^{(4)} + 
V_{\rm ct}^{(6)} 
+ \ldots \; ,
\label{eq_ct}
\end{equation}
with the superscript indicating a given order.      
       
Next, we display the contact $NN$ potentials at each order as 
they emerge from the             
Lagrangians shown at the end of Section~\ref{sec_Lpi}.

\subsubsection{Zeroth order (LO)}

From the Lagrangian
$\widehat{\cal L}^{(0)}_{NN}$,
Eq.~(\ref{eq_LNN0})
(which is part of 
$\widehat{\cal L}^{\Delta=0}$,
Eq.~(\ref{eq_LD0})), we can 
generate the contact potential 
\begin{equation}
V_{\rm ct}^{(0)}(\vec{p'},\vec{p}) =
C_S +
C_T \, \vec{\sigma}_1 \cdot \vec{\sigma}_2 \, ,
\label{eq_ct0}
\end{equation}
whose partial-wave contributions are 
\be
V_{\rm ct}^{(0)}(^1 S_0)          &=&  \widetilde{C}_{^1 S_0} =
4\pi\, ( C_S - 3 \, C_T )
\nonumber \\
V_{\rm ct}^{(0)}(^3 S_1)          &=&  \widetilde{C}_{^3 S_1} =
4\pi\, ( C_S + C_T ) \,.
\label{eq_ct0_pw}
\ee

\subsubsection{Second order (NLO)}

For this, we refer to 
$\widehat{\cal L}^{(2)}_{NN}$,
Eq.~(\ref{eq_LNN2})  
(part of 
$\widehat{\cal L}^{\Delta=2}$,
Eq.~(\ref{eq_LD2})). We now have 
\be
V_{\rm ct}^{(2)}(\vec{p'},\vec{p}) &=&
C_1 \, q^2 +
C_2 \, k^2 
\nonumber 
\\ &+& 
\left(
C_3 \, q^2 +
C_4 \, k^2 
\right) \vec{\sigma}_1 \cdot \vec{\sigma}_2 
\nonumber 
\\
&+& C_5 \left( -i \vec{S} \cdot (\vec{q} \times \vec{k}) \right)
\nonumber 
\\ &+& 
 C_6 \, ( \vec{\sigma}_1 \cdot \vec{q} )\,( \vec{\sigma}_2 \cdot 
\vec{q} )
\nonumber 
\\ &+& 
 C_7 \, ( \vec{\sigma}_1 \cdot \vec{k} )\,( \vec{\sigma}_2 \cdot 
\vec{k} ) \,.
\label{eq_ct2}
\ee
Notice that the constants $C_i$ which appear in these expressions 
are related to the coefficients $C_i'$
present in the Lagrangian
$\widehat{\cal L}^{(2)}_{NN}$,
Eq.~(\ref{eq_LNN2}), see 
Refs.~\cite{ORK96,EGM98} for details (not relevant for us at this point). 

One way to partial-wave decompose the potential above is the method proposed    
by Erkelenz, Alzetta, and Holinde~\cite{EAH71}.         
One obtains
\be
V_{\rm ct}^{(2)}(^1 S_0)          &=&  C_{^1 S_0} ( p^2 + {p'}^2 ) 
\nonumber \\
V_{\rm ct}^{(2)}(^3 P_0)          &=&  C_{^3 P_0} \, p p'
\nonumber \\
V_{\rm ct}^{(2)}(^1 P_1)          &=&  C_{^1 P_1} \, p p' 
\nonumber \\
V_{\rm ct}^{(2)}(^3 P_1)          &=&  C_{^3 P_1} \, p p' 
\nonumber \\
V_{\rm ct}^{(2)}(^3 S_1)          &=&  C_{^3 S_1} ( p^2 + {p'}^2 ) 
\nonumber \\
V_{\rm ct}^{(2)}(^3 S_1- ^3 D_1)  &=&  C_{^3 S_1- ^3 D_1}  p^2 
\nonumber \\
V_{\rm ct}^{(2)}(^3 D_1- ^3 S_1)  &=&  C_{^3 S_1- ^3 D_1}  {p'}^2 
\nonumber \\
V_{\rm ct}^{(2)}(^3 P_2)          &=&  C_{^3 P_2} \, p p' 
\label{eq_ct2_pw}
\ee
with
\be
C_{^1 S_0} 
&=&
4\pi\, \left( C_1 + \frac{1}{4} C_2 - 3 C_3 - \frac{3}{4} C_4 - 
C_6 - \frac{1}{4} C_7 \right)
\nonumber 
\\
C_{^3 P_0} 
&=&
4\pi\, \left( -\frac{2}{3} C_1 + \frac{1}{6} C_2 - \frac{2}{3} C_3 
+ \frac{1}{6} C_4 - \frac{2}{3} C_5
+ 2 C_6 - \frac{1}{2} C_7 \right)
\nonumber 
\\
C_{^1 P_1} 
&=&
4\pi\, \left( -\frac{2}{3} C_1 + \frac{1}{6} C_2 + 2 C_3 
- \frac{1}{2} C_4 
+ \frac{2}{3} C_6 - \frac{1}{6} C_7 \right)
\nonumber 
\\
C_{^3 P_1} 
&=&
4\pi\, \left( -\frac{2}{3} C_1 + \frac{1}{6} C_2 - \frac{2}{3} C_3 
+ \frac{1}{6} C_4 - \frac{1}{3} C_5
- \frac{4}{3} C_6 + \frac{1}{3} C_7 \right)
\nonumber 
\\
C_{^3 S_1} 
&=&
4\pi\, \left( C_1 + \frac{1}{4} C_2 + C_3 + \frac{1}{4} C_4 + 
\frac{1}{3} C_6 + \frac{1}{12} C_7 \right)
\nonumber 
\\
C_{^3 S_1- ^3 D_1} 
&=&
4\pi\, \left( -\frac{2\sqrt{2}}{3} C_6 - \frac{\sqrt{2}}{6} 
C_7 \right)
\nonumber 
\\
C_{^3 P_2} 
&=&
4\pi\, \left( -\frac{2}{3} C_1 + \frac{1}{6} C_2 - \frac{2}{3} C_3 
+ \frac{1}{6} C_4 + \frac{1}{3} C_5 \right) \,.
\label{eq_ct2_pwcoef}
\ee

\subsubsection{Fourth order (N$^3$LO)}

The contact potential of order four reads
\be
V_{\rm ct}^{(4)}(\vec{p'},\vec{p}) &=&
D_1 \, q^4 +
D_2 \, k^4 +
D_3 \, q^2 k^2 +
D_4 \, (\vec{q} \times \vec{k})^2 
\nonumber 
\\ &+& 
\left(
D_5 \, q^4 +
D_6 \, k^4 +
D_7 \, q^2 k^2 +
D_8 \, (\vec{q} \times \vec{k})^2 
\right) \vec{\sigma}_1 \cdot \vec{\sigma}_2 
\nonumber 
\\ &+& 
\left(
D_9 \, q^2 +
D_{10} \, k^2 
\right) \left( -i \vec{S} \cdot (\vec{q} \times \vec{k}) \right)
\nonumber 
\\ &+& 
\left(
D_{11} \, q^2 +
D_{12} \, k^2 
\right) ( \vec{\sigma}_1 \cdot \vec{q} )\,( \vec{\sigma}_2 
\cdot \vec{q})
\nonumber 
\\ &+& 
\left(
D_{13} \, q^2 +
D_{14} \, k^2 
\right) ( \vec{\sigma}_1 \cdot \vec{k} )\,( \vec{\sigma}_2 
\cdot \vec{k})
\nonumber 
\\ &+& 
D_{15} \left( 
\vec{\sigma}_1 \cdot (\vec{q} \times \vec{k}) \, \,
\vec{\sigma}_2 \cdot (\vec{q} \times \vec{k}) 
\right) .
\label{eq_ct4}
\ee
The corresponding partial-wave expressions at this order can be found 
in Appendix E of Ref.~\cite{ME11}.

\subsubsection{Sixth order (N$^5$LO)}

At sixth order, 26 new contact terms appear, bringing the total number to 50. These terms
as well as their partial-wave decomposition have been worked out in Ref.~\cite{EM03a}.
So far, these terms have not been used in the construction of $NN$ potentials.

\subsection{Definition of $NN$ potential \label{sec_pot}}

At this point, we have all the ``ingredients" required to describe the 
well-known phenomenology of the nuclear force at long, medium, and 
short distances. 
When approaching the most central waves, though, we are faced with
one more hurdle.      
As is known from the most elementary nuclear physics, 
the $NN$ system at
$L = 0$ admits                    
a bound state, the weakly-bound deuteron,
and large scattering lengths, which do not allow for
a perturbative treatment.   Moreover,
unlike what happens with $\pi$-$\pi$ and $\pi$-$N$ in the chiral limit,
the interaction of nucleons does not vanish      
 when $Q\rightarrow 0$. 
As argued by 
Weinberg~\cite{Wei91}, intermediate states with only nucleons are responsible
for the large increase of the 
scattering amplitude commonly referred to as                         
``infrared enhancement''. A way to circumvent this problem, as suggested by
Weinberg, is to calculate                    
the $NN$ potential perturbatively and then to apply it                
in a scattering equation 
to obtain the $NN$ amplitude.                       
This is the strategy we will adopt. 
          
The pion-exchange parts of the $NN$ potential were spelled out in 
Eqs.~(\ref{eq_VLO})-(\ref{eq_VN5LO}).
To obtain the complete potential, one just has to add to this the contact terms listed in Eq.~(\ref{eq_ct}). 
Thus, one has to do the following extensions
to some of the Eqs.~(\ref{eq_VLO})-(\ref{eq_VN5LO}):
\begin{eqnarray}
V_{\rm LO} & \longmapsto & V_{\rm LO} + V_{\rm ct}^{(0)} 
\label{eq_Vlo} \\
V_{\rm NLO} & \longmapsto & V_{\rm NLO} + V_{\rm ct}^{(2)} 
\label{eq_Vnlo} \\
V_{\rm N3LO} & \longmapsto & V_{\rm N3LO} + V_{\rm ct}^{(4)} 
\label{eq_Vn3lo} \\
V_{\rm N5LO} & \longmapsto & V_{\rm N5LO} + V_{\rm ct}^{(6)} \,,
\label{eq_Vn5lo}
\end{eqnarray}
and no changes to $V_{\rm NNLO}$ and $V_{\rm N4LO}$.

The potential $V$ as derived in previous sections is, in principal, an invariant amplitude and, thus, satisfies a relativistic scattering equation, for which we choose the
BbS equation~\cite{ME11}, which reads explicitly,
\begin{equation}
{T}({\vec p}~',{\vec p})= {V}({\vec p}~',{\vec p})+
\int \frac{d^3p''}{(2\pi)^3} \:
{V}({\vec p}~',{\vec p}~'') \:
\frac{M_N^2}{E_{p''}} \:  
\frac{1}
{{ p}^{2}-{p''}^{2}+i\epsilon} \:
{T}({\vec p}~'',{\vec p}) 
\label{eq_bbs2}
\end{equation}
with $E_{p''}\equiv \sqrt{M_N^2 + {p''}^2}$.
The use of a relativistic equation implies that               
relativistic corrections are already included to all orders (no additional corrections are needed
when increasing the EFT order). 

If we define  
\begin{equation}
\widehat{V}({\vec p}~',{\vec p})
\equiv 
\frac{1}{(2\pi)^3}
\sqrt{\frac{M_N}{E_{p'}}}\:  
{V}({\vec p}~',{\vec p})\:
 \sqrt{\frac{M_N}{E_{p}}}
\label{eq_minrel1}
\end{equation}
and
\begin{equation}
\widehat{T}({\vec p}~',{\vec p})
\equiv 
\frac{1}{(2\pi)^3}
\sqrt{\frac{M_N}{E_{p'}}}\:  
{T}({\vec p}~',{\vec p})\:
 \sqrt{\frac{M_N}{E_{p}}}
\,,
\label{eq_minrel2}
\end{equation}
where the factor $1/(2\pi)^3$ is simply a convenient choice,
the BbS equation assumes the form of the nonrelativistic
Lippmann-Schwinger (LS) equation,
\begin{equation}
 \widehat{T}({\vec p}~',{\vec p})= \widehat{V}({\vec p}~',{\vec p})+
\int d^3p''\:
\widehat{V}({\vec p}~',{\vec p}~'')\:
\frac{M_N}
{{ p}^{2}-{p''}^{2}+i\epsilon}\:
\widehat{T}({\vec p}~'',{\vec p}) \, .
\label{eq_LS}
\end{equation}
Since
$\widehat V$ 
satisfies Eq.~(\ref{eq_LS}),  
it may be regarded as a nonrelativistic potential. By the same arguments, 
$\widehat{T}$ 
may be regarded as the nonrelativistic 
T-matrix.
All technical aspects associated with the solution of the LS equation 
can be found in Appendix A of Ref.~\cite{Mac01}, including 
specific formulas for the $np$ and $pp$ phase shifts. 
Additional details 
concerning the relevant operators and their decompositions 
are given in section~4 of Ref.~\cite{EAH71}. Finally, computational methods
to solve the LS equation are found in Ref.~\cite{Mac93}.

\subsection{Regularization and non-perturbative renormalization}
\label{sec_reno}

Iteration of $\widehat V$ in the LS equation, Eq.~(\ref{eq_LS}),
requires cutting $\widehat V$ off for high momenta to avoid infinities.
This is consistent with the fact that ChPT
is a low-momentum expansion which
is valid only for momenta $Q \ll \Lambda_\chi \approx 1$ GeV.
Therefore, the potential $\widehat V$
is multiplied
with the regulator function $f(p',p)$,
\begin{equation}
{\widehat V}(\vec{ p}~',{\vec p})
\longmapsto
{\widehat V}(\vec{ p}~',{\vec p}) \, f(p',p) 
\end{equation}
with
\begin{equation}
f(p',p) = \exp[-(p'/\Lambda)^{2n}-(p/\Lambda)^{2n}] \,,
\label{eq_f}
\end{equation}
such that
\begin{equation}
{\widehat V}(\vec{ p}~',{\vec p}) \, f(p',p) 
\approx
{\widehat V}(\vec{ p}~',{\vec p})
\left\{1-\left[\left(\frac{p'}{\Lambda}\right)^{2n}
+\left(\frac{p}{\Lambda}\right)^{2n}\right]+ \ldots \right\} 
\,.
\label{eq_reg_exp}
\end{equation}
Typical choices for the cutoff parameter $\Lambda$ that
appears in the regulator are 
$\Lambda \approx 0.5 \mbox{ GeV} < \Lambda_\chi \approx 1$ GeV.
At N$^3$LO and N$^4$LO, an appropriate choice for $n$ is three.

We display Eq.~(\ref{eq_reg_exp}) to demonstrate that                        
the exponential cutoff may not impact the     
order at which we are working.         
If $n$ is sufficiently large, the regulating function generates terms beyond the given order.
Under the assumption of a reasonable 
convergence of the chiral expansion, these terms are sufficiently small 
not to impact                       
the accuracy at the present order. 
But note that
the form as given in Eq.~(\ref{eq_f}),
and not its expansion Eq.~(\ref{eq_reg_exp}), is used in actual calculations. We also mention in 
this context that 
the square-root factors
in Eqs.~(\ref{eq_minrel1}-\ref{eq_minrel2}) are not expanded, 
as their full structure ensures consistency with           
relativistic elastic unitarity.

It is pretty obvious that results for the $T$-matrix may
depend sensitively on the regulator and its cutoff parameter.
The removal of such regulator dependence is known as renormalization.
Proper renormalization of the chiral $NN$ interaction is a controversial issue, see
Section 4.5 of Ref.~\cite{ME11} for a more comprehensive discussion.

\begin{figure*}\centering
\vspace*{-2.5cm}
%\hspace{-1.0cm}
\includegraphics[scale=.5]{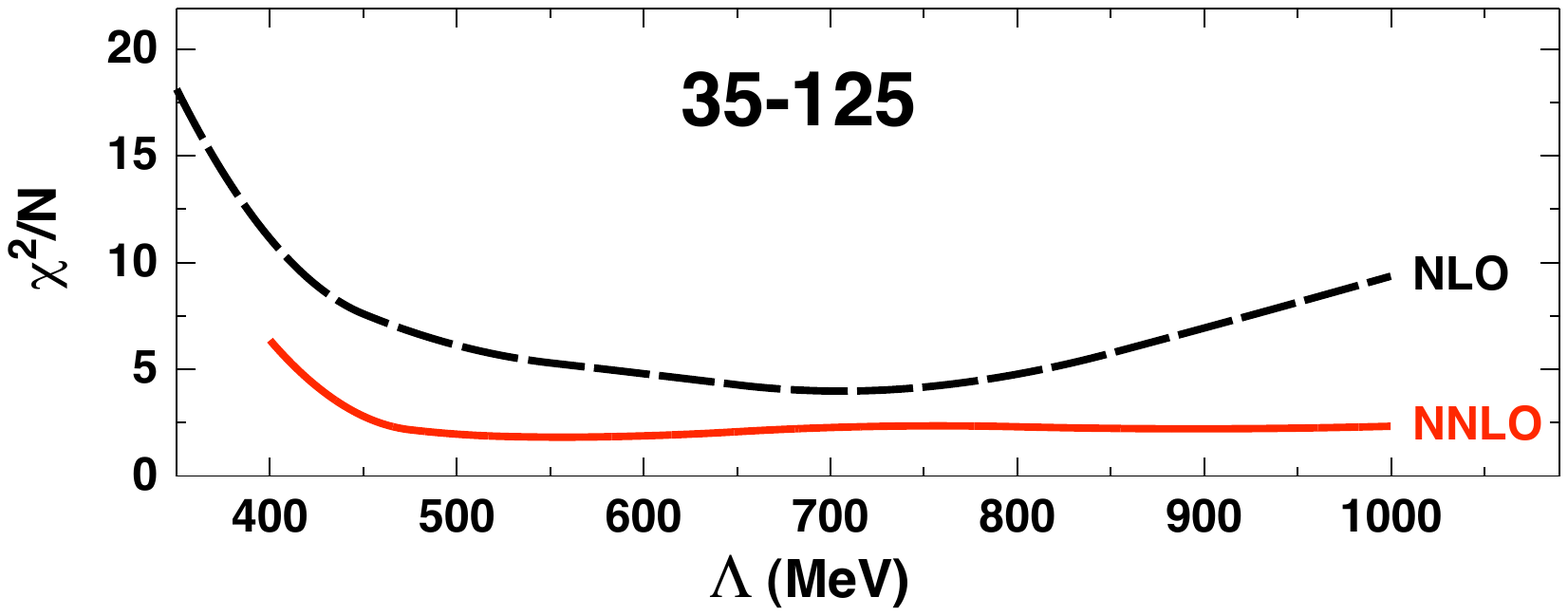}

\vspace*{-9.7cm}
\includegraphics[scale=.5]{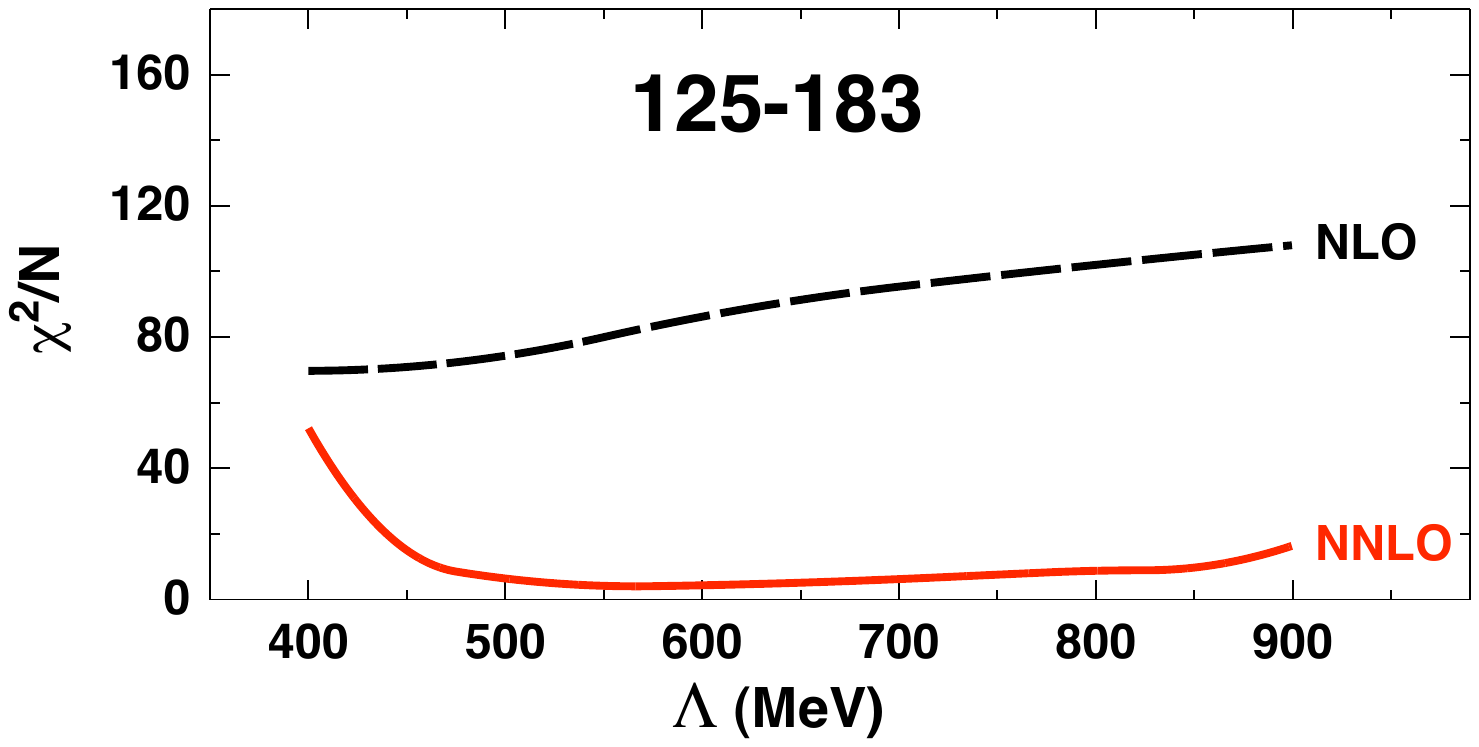}
\vspace*{-7.5cm}
\caption{$\chi^2$/datum for the reproduction of the $np$ data in the
energy range 35-125 MeV (upper frame) and 125-183 MeV (lower frame)
as a function of the cutoff parameter $\Lambda$ of the regulator function
Eq.~(\ref{eq_f}). The (black) dashed curves show the $\chi^2$/datum
achieved with $np$ potentials constructed at order NLO and the (red) solid curves are for NNLO.
(Figure reproduced from Ref.~\cite{Mar13}.)}
\label{fig_chi2}       % Give a unique label
\end{figure*}

For a successful EFT (in its domain of validity), one must be able to claim independence of the predictions on
the regulator. Also,                                         
truncation error must decrease as we go to higher and higher orders.
These are precisely the goals of renormalization.  

Lepage~\cite{Lep97} has stressed that the cutoff independence should be examined
for cutoffs below the hard scale and not beyond. Ranges of cutoff independence within the
theoretical error are to be identified using Lepage plots~\cite{Lep97}.
A systematic investigation of this kind has been conducted in Ref.~\cite{Mar13}.
In that work, the error of the predictions was quantified by calculating the $\chi^2$/datum 
for the reproduction of the neutron-proton ($np$) elastic scattering data
as a function of the cutoff parameter $\Lambda$ of the regulator function
Eq.~(\ref{eq_f}). Predictions by chiral $np$ potentials at 
order NLO and NNLO were investigated applying Weinberg counting for the counter terms ($NN$ contact terms).
The results from this study for the energy range 35-125 MeV are shown in the upper frame of Fig.~\ref{fig_chi2}
and for 125-183 MeV in the lower frame. 
It is seen that the reproduction of the $np$ data at these energies is generally poor
at NLO, while at NNLO the $\chi^2$/datum assumes acceptable values (a clear demonstration of
order-by-order improvement). Moreover, at NNLO one observes 
``plateaus'' of constant low $\chi^2$ for
cutoff parameters ranging from about 450 to 850 MeV. This may be perceived as cutoff independence
(and, thus, successful renormalization) for the relevant range of cutoff parameters.

\subsection{$NN$ potentials order by order}

\begin{figure*}[!t] 
\vspace*{-1cm}
\hspace*{-1.7cm}
\scalebox{0.49}{\includegraphics{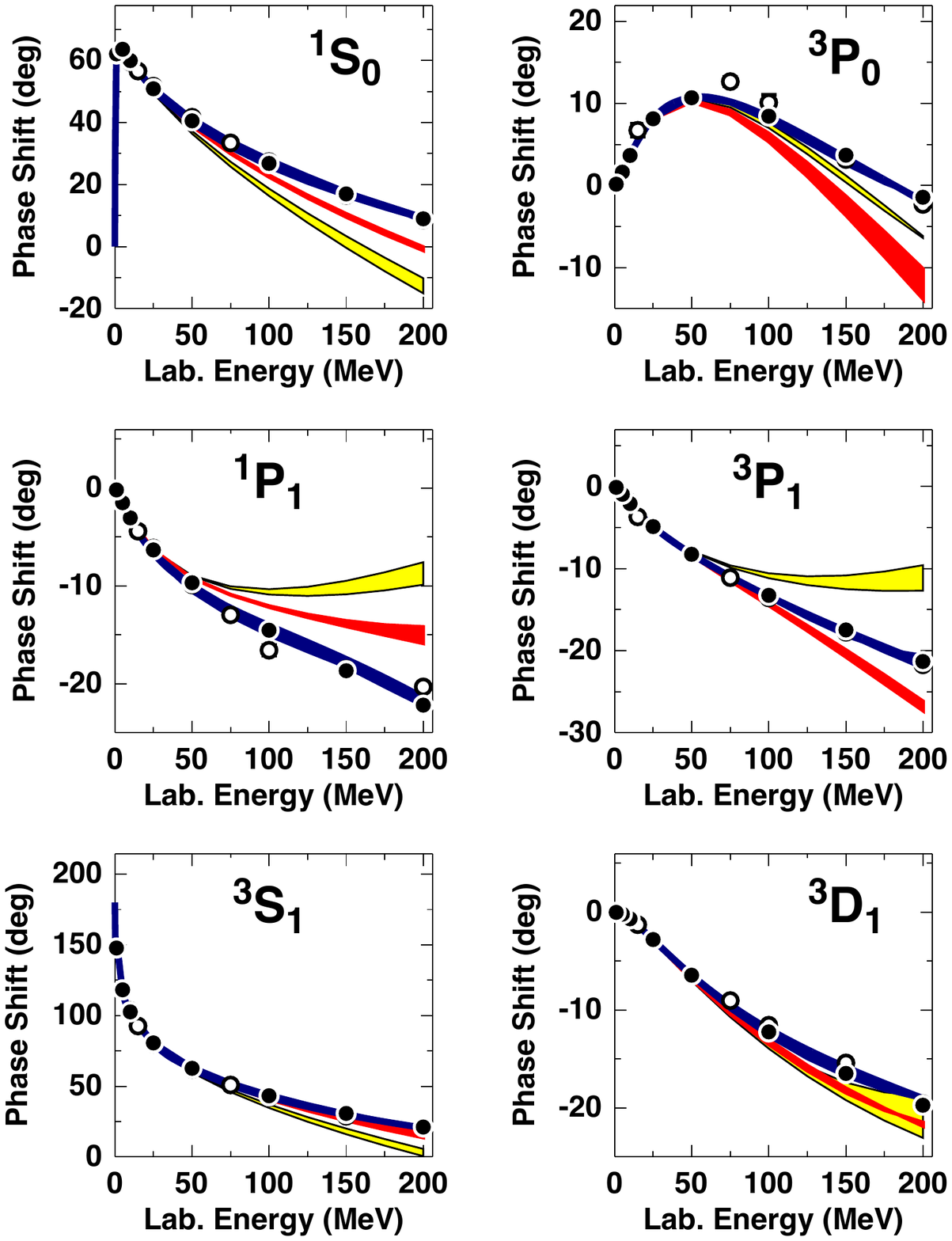}}
\hspace*{-2.8cm}
\scalebox{0.49}{\includegraphics{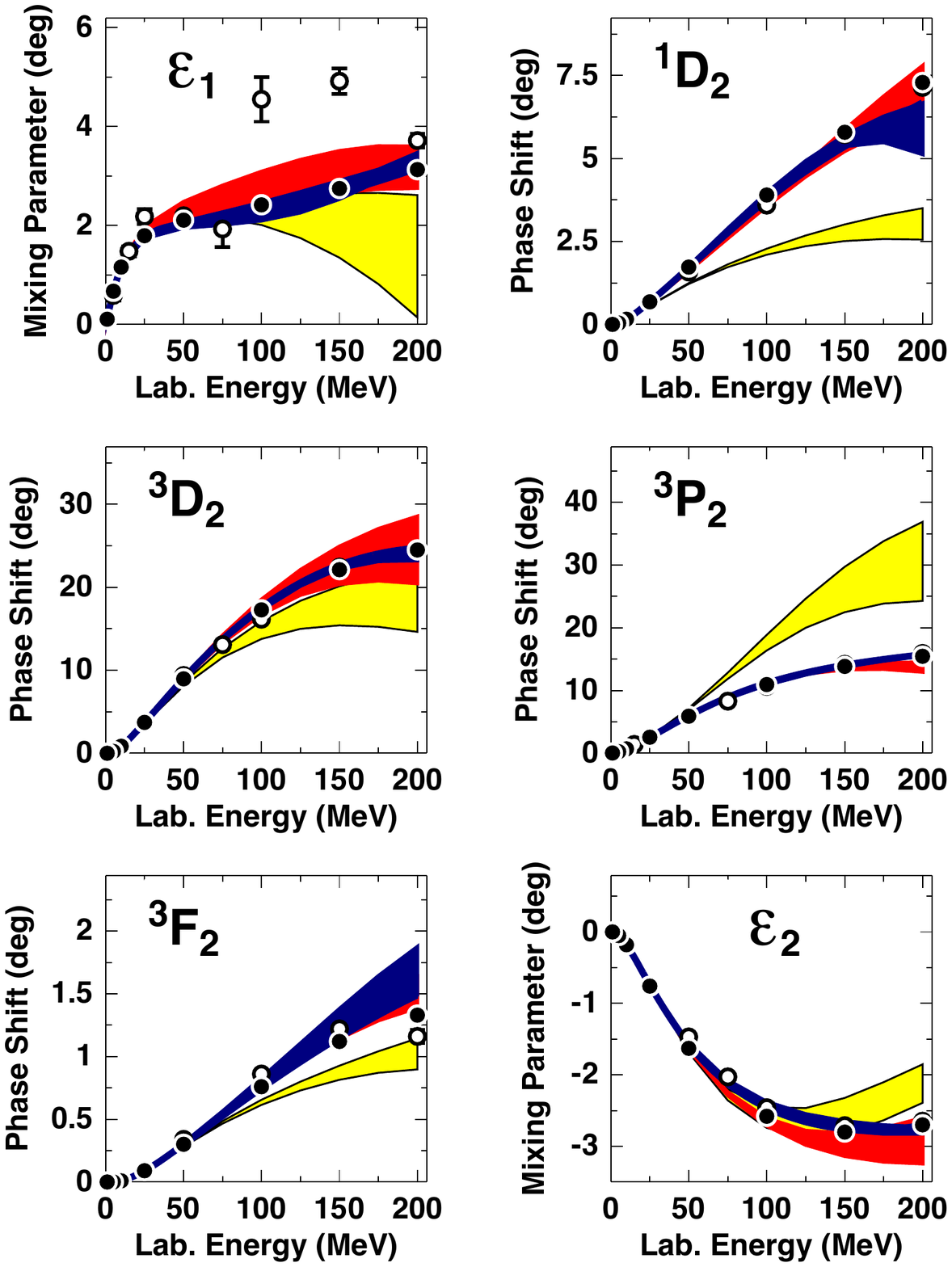}}
\vspace*{-2.5cm}
\caption{Phase shifts of neutron-proton scattering for the lower partial waves with $J\leq 2$. The 
yellow, red, and blue bands show the variations of the predictions with changing cutoffs 
between 450 and 600 at  NLO, N$^2$LO, and N$^3$LO, respectively.
The predictions by N$^4$LO potentials lie within the dark blue band and are, therefore,
not explicitly shown.
The solid dots and open circles are the results from the Nijmegen
multi-energy $np$ phase shift analysis~\protect\cite{Sto93} 
and the VPI/GW
single-energy $np$ analysis SM99~\protect\cite{SM99}, respectively.
(Figure reproduced from Ref.~\cite{Sam15}.)
\label{fig_ph7}}
\end{figure*}

As discussed, $NN$ potentials can be calculated at
various orders, cf.\ Eqs.~(\ref{eq_VLO})-(\ref{eq_VN5LO}) and Eqs.~(\ref{eq_Vlo})-(\ref{eq_Vn5lo}),
with better accuracy at higher orders. 
The convergence properties of the chiral expansion in the most central 
 partial waves
can be seen in Fig.~\ref{fig_ph7}. There, 
we display the $J\leq 2$ phase parameters for potentials constructed
at NLO, NNLO, and N$^3$LO with cutoffs ranging between 450 and 600 MeV. 
(The NLO and NNLO potentials are from Ref.~\cite{Mar13} and the N$^3$LO ones from Refs.~\cite{ME11,Cor14}.)
There is noticeable 
improvement in the agreement between the predictions and the empirical
phase shifts as the order increases.

For a more direct comparison of theory and experiment,
we can calculate observables, rather than phase shifts.
The quality of the agreement with experimental data is typically 
expressed in terms of                          
the $\chi^2$/datum, with a value close to unity indicating 
a nearly perfect agreement.

\begin{table*}[t]
\caption{Columns three and four show the
$\chi^2$/datum for the reproduction of the 1999 $np$ 
database (defined in Ref.~\cite{Mac01})
by families of $np$ potentials at NLO and NNLO constructed by the
Bochum group~\cite{EGM04}.
The $\chi^2$/datum is stated in terms of ranges
which result from a variation of the cutoff parameters
used in the regulator functions.
The values of these cutoff parameters in units of MeV are given in 
parentheses.
$T_{\rm lab}$ denotes the kinetic energy of the incident neutron
in the laboratory system.
\label{tab_chi2a}}
\smallskip
\begin{tabular*}{\textwidth}{@{\extracolsep{\fill}}cccc}
%\begin{tabular}{cccc}
\hline \hline 
\noalign{\smallskip}
 $T_{\rm lab}$ (MeV) &  \# of $np$ data & 
\multicolumn{2}{c}{\it --- Bochum $np$ potentials --- }\\
 Energy Bin & & NLO (550/700--400/500) & NNLO (600/700--450/500)
 \\
\hline \hline 
\noalign{\smallskip}
0--100&1058&4--5&1.4--1.9\\ 
100--190&501&77--121&12--32\\ 
190--290&843&140--220&25--69\\ 
\hline 
\noalign{\smallskip}
0--290&2402&67--105&12--27
\\ 
\hline \hline 
\end{tabular*}
\end{table*}

\begin{table*}[t]
\caption{Number of parameters needed for fitting the $np$ data
in the Nijmegen phase-shift analysis and by the high-precision CD-Bonn potential
{\it versus} the total number of $NN$ contact terms of EFT based potentials 
to different orders. 
\label{tab_par}}
\smallskip
\begin{tabular*}{\textwidth}{@{\extracolsep{\fill}}ccc|cccc}
%\begin{tabular}{ccc|ccc}
\hline \hline
\noalign{\smallskip}
     & Nijmegen PWA93     & CD-Bonn pot.\       & 
               \multicolumn{4}{c}{ --- EFT contact potentials~\cite{EM03a} --- }\\
     & Ref~\cite{Sto93} & Ref.~\cite{Mac01} 
     & $Q^0$ & $Q^2$ & $Q^4$ & $Q^6$ \\
\hline \hline
\noalign{\smallskip}
$^1S_0$         & 3 & 4 & 1&2 & 4 & 6 \\
$^3S_1$         & 3 & 4 & 1&2 & 4 & 6 \\
\hline
\noalign{\smallskip}
$^3S_1$-$^3D_1$ & 2 & 2 & 0&1 & 3 & 6 \\
\hline
\noalign{\smallskip}
$^1P_1$         & 3 & 3 & 0&1 & 2 & 4 \\
$^3P_0$         & 3 & 2 & 0&1 & 2 & 4 \\
$^3P_1$         & 2 & 2 & 0&1 & 2 & 4 \\
$^3P_2$         & 3 & 3 & 0&1 & 2 & 4 \\
\hline
\noalign{\smallskip}
$^3P_2$-$^3F_2$ & 2 & 1 & 0&0 & 1 & 3 \\
\hline
\noalign{\smallskip}
$^1D_2$         & 2 & 3 & 0&0 & 1 & 2 \\
$^3D_1$         & 2 & 1 & 0&0 & 1 & 2 \\
$^3D_2$         & 2 & 2 & 0&0 & 1 & 2 \\
$^3D_3$         & 1 & 2 & 0&0 & 1 & 2 \\
\hline
\noalign{\smallskip}
$^3D_3$-$^3G_3$ & 1 & 0 & 0&0 & 0 & 1 \\
\hline
\noalign{\smallskip}
$^1F_3$         & 1 & 1 & 0&0 & 0 & 1 \\
$^3F_2$         & 1 & 2 & 0&0 & 0 & 1 \\
$^3F_3$         & 1 & 2 & 0&0 & 0 & 1 \\
$^3F_4$         & 2 & 1 & 0&0 & 0 & 1 \\
\hline
\noalign{\smallskip}
$^3F_4$-$^3H_4$ & 0 & 0 & 0&0 & 0 & 0 \\
\hline
\noalign{\smallskip}
$^1G_4$         & 1 & 0 & 0&0 & 0 & 0 \\
$^3G_3$         & 0 & 1 & 0&0 & 0 & 0 \\
$^3G_4$         & 0 & 1 & 0&0 & 0 & 0 \\
$^3G_5$         & 0 & 1 & 0&0 & 0 & 0 \\
\hline\hline
\noalign{\smallskip}
Total         & 35  & 38 & 2 & 9 & 24 & 50 \\
\hline\hline
\noalign{\smallskip}
\end{tabular*}
\end{table*}

In Table~\ref{tab_chi2a}, we report 
the $\chi^2$/datum for the comparison beween the world $np$ data
below 290 MeV and the predictions of $np$ potentials at 
NLO and NNLO by the Bochum group~\cite{EGM04}.
The NLO potentials generate a very large $\chi^2$/datum (between 67 and 105),
while the NNLO potentials give values between 12 and 27, consistent with the findings of Ref.~\cite{Mar13}
shown in Fig.~\ref{fig_chi2}.
%HERE2
It is promising to see that there is order-by-order improvement, but                           
the $np$ data at NLO and NNLO are not reproduced with sufficient            
quality.\footnote{For an optimized NNLO potential see Ref.~\cite{Eks13} and for local NLO and NNLO potentials see Ref.~\cite{Gez14}.}

The most natural strategy is then to proceed to the next order, as suggested already in       
2002~\cite{EM02a,EM02}.                 
The first N$^3$LO  potential followed shortly 
after~\cite{EM03}.

At N$^3$LO ($Q^4$), 24 contact terms bring in a total of 24 parameters
which impact partial waves with $L\leq 2$, while at NLO and NNLO there are only
9 contacts with $L\leq 1$
(cf.~Section~\ref{sec_ct} and Table~\ref{tab_par}).
These LECs are free constants
employed to parametrize 
the short-range phenomenology.                                     
%HERE3
Table~\ref{tab_par} shows how many terms with a certain power of $Q$ participate 
in a given $NN$ state.                  
One can see from the Table that 
contacts appear for the first time in $D$-waves  
at N$^3$LO. This is one important mechanism behind the considerable improvement in the 
reproduction of the $NN$ data at this order. 
Because the 
$D$-states are somewhat in between central and peripheral waves,                  
contact terms, in addition to the one- and two-pion exchanges, are important to describe 
the $D$-phases correctly. Moreover, 
 at N$^3$LO, every 
$P$-wave also benefits from an additional contact term, leading to further improvement, 
especially in $^3P_0$ and $^3P_1$ at incident laboratory energies greater than 100 MeV
(cf.\ Fig.~\ref{fig_ph7}).

Table~\ref{tab_par} also displays
the number of free parameters
used in the Nijmegen partial wave analysis (PWA93)~\cite{Sto93}
and in the high-precision CD-Bonn potential~\cite{Mac01}.
For $S$ and $P$ waves, that number 
is approximately equal to the one required by EFT at N$^3$LO ($Q^4$). 
{\bf Interestingly, we find in EFT a retroactive motivation         
for the phenomenology which became popular in the 1990's to construct high-precision potentials.}

\begin{table*}[t]
\caption{Columns three to five display the
$\chi^2$/datum for the reproduction of the 1999 
$np$ database
(defined in Ref.~\cite{Mac01})
by various $np$ potentials.
For the chiral potentials, 
the $\chi^2$/datum is stated in terms of ranges
which result from a variation of the cutoff parameters
used in the regulator functions.
The values of these cutoff parameters 
in units of MeV
are given in parentheses.
$T_{\rm lab}$ denotes the kinetic energy of the incident nucleon
in the laboratory system.
\label{tab_chi2b}}
\smallskip
%\begin{tabular}{ccccc}
\begin{tabular*}{\textwidth}{@{\extracolsep{\fill}}ccccc}
\hline \hline 
\noalign{\smallskip}
 $T_{\rm lab}$ (MeV)
 & \# of {\boldmath $np$}  data
 & {\it Idaho} N$^3$LO
 & {\it Bochum} N$^3$LO
 & Argonne  $V_{18}$ 
\\
    Energy Bin
 & 
 & (500--600)~\cite{EM03}
 & (600/700--450/500)~\cite{EGM05} 
 & Ref.~\cite{WSS95}  
\\
\hline \hline 
\noalign{\smallskip}
0--100&1058&1.0--1.1&1.0--1.1&0.95\\ 
100--190&501&1.1--1.2&1.3--1.8&1.10\\ 
190--290&843&1.2--1.4&2.8--20.0&1.11\\ 
\hline 
\noalign{\smallskip}
0--290&2402&1.1--1.3&1.7--7.9&1.04
\\ 
\hline \hline 
\end{tabular*}
\end{table*}

\begin{table*}[b]
\caption{Same as Table~\ref{tab_chi2b} but for 
{\boldmath\bf $pp$}.
\label{tab_chi2c}}
\smallskip
%\begin{tabular}{ccccc}
\begin{tabular*}{\textwidth}{@{\extracolsep{\fill}}ccccc}
\hline \hline 
\noalign{\smallskip}
 $T_{\rm lab}$ (MeV)
 & \# of {\boldmath $pp$}  data
 & {\it Idaho} N$^3$LO
 & {\it Bochum} N$^3$LO
 & Argonne  $V_{18}$ 
\\
    Energy Bin
 & 
 & (500--600)~\cite{EM03}
 & (600/700--450/500)~\cite{EGM05} 
 & Ref.~\cite{WSS95}  
\\
\hline \hline  
\noalign{\smallskip}
0--100&795&1.0--1.7&1.0--3.8&1.0 \\ 
100--190&411&1.5--1.9&3.5--11.6&1.3 \\ 
190--290&851&1.9--2.7&4.3--44.4&1.8 \\ 
\hline 
\noalign{\smallskip}
0--290&2057&1.5--2.1&2.9--22.3&1.4 
\\ 
\hline \hline
\end{tabular*}
\end{table*}

Thanks to the larger number of parameters,  N$^3$LO 
potentials can be constructed which are of about the same quality as the high-precision $NN$ 
potentials of the 1990's~\cite{Mac01,Sto94,WSS95}.
This fact is clearly revealed in
the $\chi^2/$datum for the fit of the $np$ and $pp$ data below
290 MeV shown in Table~\ref{tab_chi2b} and \ref{tab_chi2c}, respectively.
%HERE4
Table~\ref{tab_chi2b}, which is pretty self-explanatory, displays the                         
 $\chi^2$/datum for various chiral potentials as well as 
 the Argonne potential, 
compared with the world $np$ data below 290 MeV. 

As we turn now to $pp$, note first that
the $\chi^2$ for $pp$ data are typically
larger than for $np$
because of the higher precision of $pp$ data (Table~\ref{tab_chi2c}).
Thus, the Argonne $V_{18}$ produces
a $\chi^2$/datum = 1.4 for the world $pp$ data
below 290 MeV and the best Idaho N$^3$LO $pp$ potential obtains
1.5. The fit by the best Bochum 
N$^3$LO $pp$ potential results in
a $\chi^2$/datum = 2.9 and the worst  produces 22.3.
In view of these poor $\chi^2$,
the Bochum group has recently launched an attempt towards improving their chiral potentials~\cite{EKM15a,EKM15b}. However, as in their previous work~\cite{EGM05}, they have fitted their new potentials only to $NN$ phase shifts and not to the $NN$ data. The $\chi^2$ for the reproduction of the $NN$ data by the new Bochum potentials are not available and, thus, 
no reliable statement about the quality of the new potentials can be made.
In the 1990's, the Nijmegen group has pointed out repeatedly that for high quality potentials
it is insufficient to fit phase shifts only. A seemingly ``good'' fit of phase shifts can be misleading and can result in a poor $\chi^2$ for the reproduction of the data.

Concerning alternative N$^3$LO potentials, we note that a minimally non-local $NN$ potential of this kind has been constructed in Ref.~\cite{Pia15} which produces a $\chi^2$/datum of about 1.3 for the $pp$ plus $np$ data.

Now turning to N$^4$LO: Based upon the derivation of the 2PE and 3PE contributions to the $NN$ interaction at N$^4$LO by Entem {\it et al.}~\cite{Ent15a} presented in Sec.~\ref{sec_n4lo} and applied in peripheral scattering in Sec.~\ref{sec_pertNN}, $NN$ potential at N$^4$LO have recently been developed~\cite{Ent15a,EKM15b}. Note that the lower partial waves, which are crucial for a quantitative reproduction of the $NN$ data, are ruled by the contact terms. The number of contacts at N$^4$LO ($Q^5$) is the same as at N$^3$LO ($Q^4$). Thus, the N$^4$LO potentials are not very different
from the N$^3$LO ones. Note also that the high quality of some of the N$^3$LO 
potentials~\cite{ME11,EM03,Pia15} 
leaves little room for improvements. 

A further increase in accuracy (if needed)
could be achieved at N$^5$LO ($Q^6$), where the number of contact terms advances to 
50 (Table~\ref{tab_par})~\cite{EM03a}. 
As discussed in Sec.~\ref{sec_n5lo}, the dominant 2PE and 3PE contributions at N$^5$LO
have been derived~\cite{Ent15b}.
Thus, all the mathematical material for the construction of N$^5$LO potentials is available. 
However, it is
debatable if there is a need for them.

%HERE5
\section{Nuclear many-body forces \label{sec_manyNF}}

Two-nucleon forces derived from chiral EFT as described above 
have been applied, often successfully, in the many-body system.                                  
On the other hand, over the past several years we have learnt that, for some few-nucleon
reactions and nuclear structure issues, 3NFs cannot be neglected.              
The most well-known cases are the so-called $A_y$ puzzle of $N$-$d$ scattering~\cite{EMW02},
the ground state of $^{10}$B~\cite{Cau02}, and the saturation of nuclear matter~\cite{Cor14,Sam15}.
As we observed previously, 
the EFT approach generates          
consistent two- and many-nucleon forces in a natural way 
(cf.\ the overview given in Fig.~\ref{fig_hi}).
We now shift our focus to chiral three- and four-nucleon forces.

\subsection{Three-nucleon forces}

Weinberg~\cite{Wei92} was the first to discuss     
nuclear three-body forces. Not long after that, 
the first 3NF at NNLO was derived by van Kolck~\cite{Kol94}.

For a 3NF, we have $A=3$ and $C=1$ and, thus, Eq.~(\ref{eq_nu})
implies
\begin{equation}
\nu = 2 + 2L + 
\sum_i \Delta_i \,.
\label{eq_nu3nf}
\end{equation}
We will use this equation to analyze 3NF contributions
order by order.

\subsubsection{Next-to-leading order}

The lowest possible power is obviously $\nu=2$ (NLO), which
is obtained for no loops ($L=0$) and 
only leading vertices
($\sum_i \Delta_i = 0$). 
As discussed by Weinberg~\cite{Wei92} and van Kolck~\cite{Kol94}, 
the contributions from these diagrams
vanish at NLO. So, the bottom line is that there is no genuine 3NF contribution at NLO.
The first non-vanishing 3NF appears at NNLO.

\subsubsection{Next-to-next-to-leading order}

\begin{figure}[t]\centering
%\vspace{-10.0cm}
%\hspace*{-2.0cm}
\scalebox{0.7}{\includegraphics{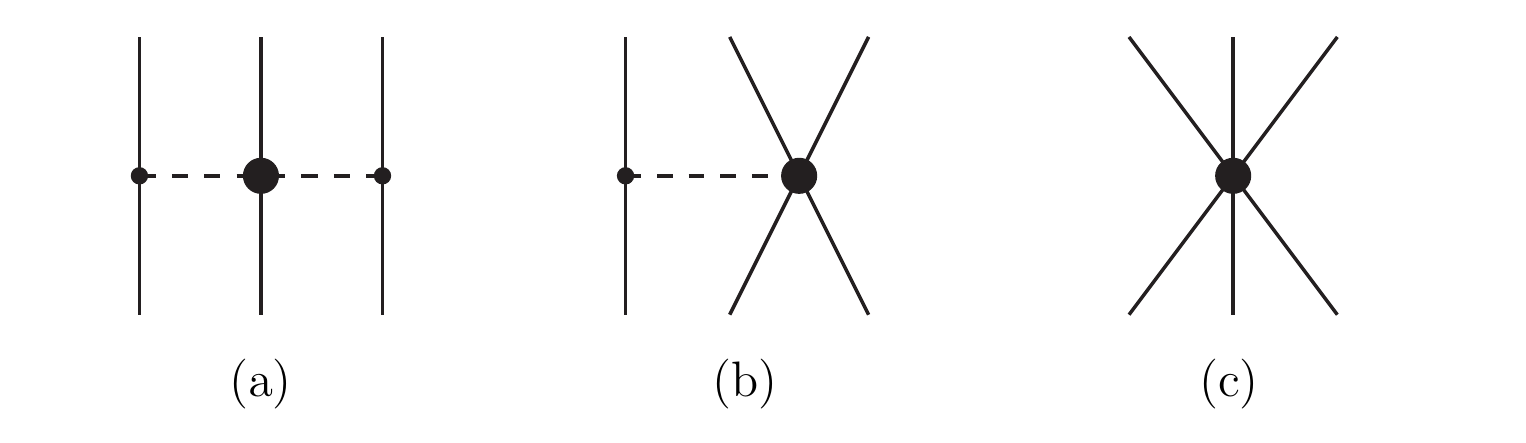}}
\vspace{-0.5cm}
\caption{The three-nucleon force at NNLO
with (a) 2PE, (b) 1PE, and (c) contact diagrams.
Notation as in Fig.~\ref{fig_hi}.
(Figure reproduced from Ref.~\cite{Sam15}.)}
\label{fig_3nf_nnlo}
\end{figure}

The power $\nu=3$ (NNLO) is obtained when
there are no loops ($L=0$) and 
$\sum_i \Delta_i = 1$, i.e., 
$\Delta_i=1$ for one vertex 
while $\Delta_i=0$ for all other vertices.
There are three topologies which fulfill this condition,
known as the 2PE, 1PE,
and contact graphs~\cite{Kol94,Epe02b}
(Fig.~\ref{fig_3nf_nnlo}).

The 2PE 3N-potential is derived to be
\begin{equation}
V^{\rm 3NF}_{\rm 2PE} = 
\left( \frac{g_A}{2f_\pi} \right)^2
\frac12 
\sum_{i \neq j \neq k}
\frac{
( \vec \sigma_i \cdot \vec q_i ) 
( \vec \sigma_j \cdot \vec q_j ) }{
( q^2_i + m^2_\pi )
( q^2_j + m^2_\pi ) } \;
F^{ab}_{ijk} \;
\tau^a_i \tau^b_j
\label{eq_3nf_nnloa}
\end{equation}
with $\vec q_i \equiv \vec{p_i}' - \vec p_i$, 
where 
$\vec p_i$ and $\vec{p_i}'$ are the initial
and final momenta of nucleon $i$, respectively, and
\begin{equation}
F^{ab}_{ijk} = \delta^{ab}
\left[ - \frac{4c_1 m^2_\pi}{f^2_\pi}
+ \frac{2c_3}{f^2_\pi} \; \vec q_i \cdot \vec q_j \right]
+ 
\frac{c_4}{f^2_\pi}  
\sum_{c} 
\epsilon^{abc} \;
\tau^c_k \; \vec \sigma_k \cdot [ \vec q_i \times \vec q_j] \; .
\label{eq_3nf_nnlob}
\end{equation}  
%HERE6
It is interesting to observe that there are clear analogies between this force and earlier          
2PE 3NFs already proposed decades ago, particularly the Fujita-Miyazawa~\cite{FM57} and
the Tucson-Melbourne (TM)~\cite{Coo79} forces. 

The 2PE 3NF does not introduce additional fitting constants, 
since the LECs $c_1$, $c_3$, and $c_4$ are already present in the 2PE 2NF.
These LECs are constrained by $NN$ and $\pi N$ data.

The other two 3NF contributions shown in Fig.~\ref{fig_3nf_nnlo}
are easily derived by taking the last two terms of the $\Delta=1$ Langrangian, Eq.~(\ref{eq_LD1}),
into account. The 1PE contribution is
\begin{equation}
V^{\rm 3NF}_{\rm 1PE} = 
-D \; \frac{g_A}{8f^2_\pi} 
\sum_{i \neq j \neq k}
\frac{\vec \sigma_j \cdot \vec q_j}{
 q^2_j + m^2_\pi }
( \mbox{\boldmath $\tau$}_i \cdot \mbox{\boldmath $\tau$}_j ) 
( \vec \sigma_i \cdot \vec q_j ) 
\label{eq_3nf_nnloc}
\end{equation}
%HERE7 
and the 3N contact potential is given by 
\begin{equation}
V^{\rm 3NF}_{\rm ct} = E \; \frac12
\sum_{i \neq j \neq k}
 \mbox{\boldmath $\tau$}_i \cdot \mbox{\boldmath $\tau$}_j  \; .
\label{eq_3nf_nnlod}
\end{equation}
These 3NF potentials introduce 
two additional constants, $D$ and $E$, which can be constrained in             
 more than one way.                                    
One may use 
the triton binding energy and the $nd$ doublet scattering
length $^2a_{nd}$ as done in     
 Ref.~\cite{Epe02b}. Alternative choices include 
the binding
energies of $^3$H and $^4$He~\cite{Nog06} or
an optimal global fit of the properties of light nuclei~\cite{Nav07}.
Another method makes use of
the triton binding energy and the Gamow-Teller matrix element of tritium $\beta$-decay~\cite{Mar12}.
When the values of $D$ and $E$ are determined, the results for other
observables involving three or more nucleons are true theoretical predictions.

Applications of the leading 3NF include                          
few-nucleon 
reactions, spectra of light- and medium-mass 
nuclei~\cite{Hag12a,Hag12b},
and nuclear and neutron matter~\cite{Cor14,Sam15}, often with 
satisfactory results. Some problems, though, remain unresolved, such as 
the well-known `$A_y$ puzzle' in nucleon-deuteron                
scattering~\cite{EMW02,Epe02b}.
Predictions which employ only 2NFs underestimate 
the analyzing power in $p$-$^3$He scattering
to a larger degree than in $p$-$d$. 
Although the $p$-$^3$He $A_y$ improves considerably (more than in the $p$-$d$ case) when the leading 3NF is          
included~\cite{Viv10}, the disagreement with the data is not fully removed. 
Also, predictions for light nuclei are not quite satisfactory~\cite{Nav07}.

In summary, the leading 3NF of ChPT is an outstanding contribution. It gives validation to, 
and provides a better framework for, 
3NFs which were proposed already 5 decades ago; it alleviates existing problems            
in few-nucleon reactions and the spectra of
light nuclei.
Nevertheless, we still face several challenges.                   
With regard to the 2NF, we have discussed earlier that it is necessary 
to go to order 4 for high-quality predictions. 
Thus, the 3NF at N$^3$LO must be considered simply as a matter of consistency with the 2NF sector. 
At the same time, one hopes that its 
inclusion may result in further improvements with the aforementioned unresolved problems.

\begin{figure}[t]\centering
\vspace*{-0.5cm}
%\hspace*{-1.5cm}
\scalebox{1.0}{\includegraphics{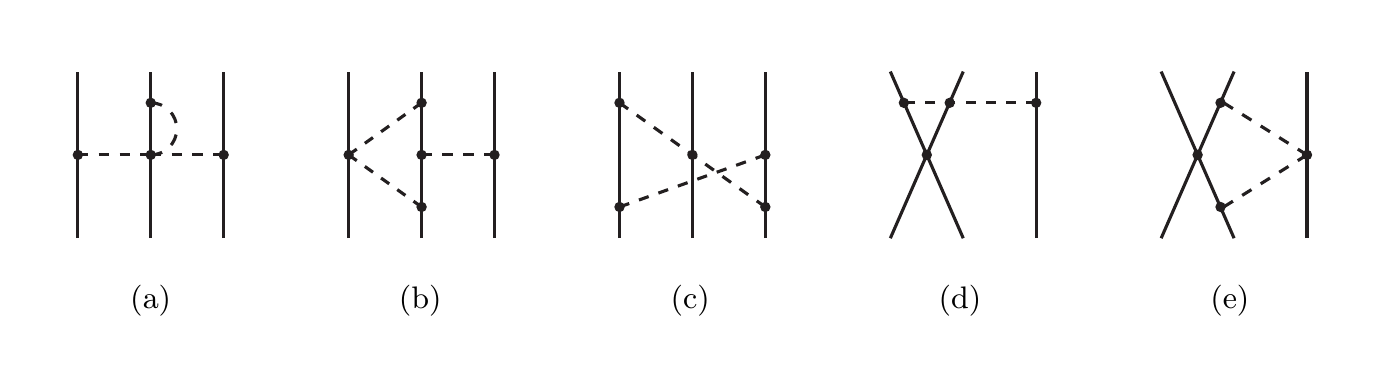}}
\vspace*{-0.75cm}
\caption{Leading one-loop 3NF diagrams at N$^3$LO.
We show one representative example for each of five topologies,
which are: (a) 2PE, (b) 1PE-2PE, (c) ring, (d) contact-1PE, (e) contact-2PE.
Notation as in Fig.~\ref{fig_hi}.}
\label{fig_3nf_n3lo}
\end{figure}

\subsubsection{Next-to-next-to-next-to-leading order}

At N$^3$LO, there are loop and tree diagrams.
For the loops (Fig.~\ref{fig_3nf_n3lo}), we have
$L=1$ and, therefore, all $\Delta_i$ have to be zero
to ensure $\nu=4$. 
Thus, these one-loop 3NF diagrams can include
only leading order vertices, the parameters of which
are fixed from $\pi N$ and $NN$ analysis.
One sub-group of these diagrams (the 2PE graphs, cf.\ Fig.~\ref{fig_3nf_n3lo})
has been calculated by Ishikawa and Robilotta~\cite{IR07},
and the other topologies
have been evaluated by the Bochum-Bonn group~\cite{Ber08}.
The N$^3$LO 2PE 3NF has been applied in the calculation
of nucleon-deuteron observables in Ref.~\cite{IR07} 
causing little impact.
Very recently, the long-range part of the chiral N$^3$LO 3NF has been
tested in the triton~\cite{Ski11} and in three-nucleon scattering~\cite{Wit12}
yielding only moderate effects. The long- and short-range parts of this
force have been used in neutron matter calculations
(together with the N$^3$LO 4NF) producing relatively large contributions
from the 3NF~\cite{Tew12}. Thus, the ultimate assessment of the N$^3$LO 3NF is still
outstanding and will require more few- and many-body applications.

\subsubsection{The 3NF at N$^4$LO}

\begin{figure}[t]\centering
\vspace*{-0.5cm}
\scalebox{1.0}{\includegraphics{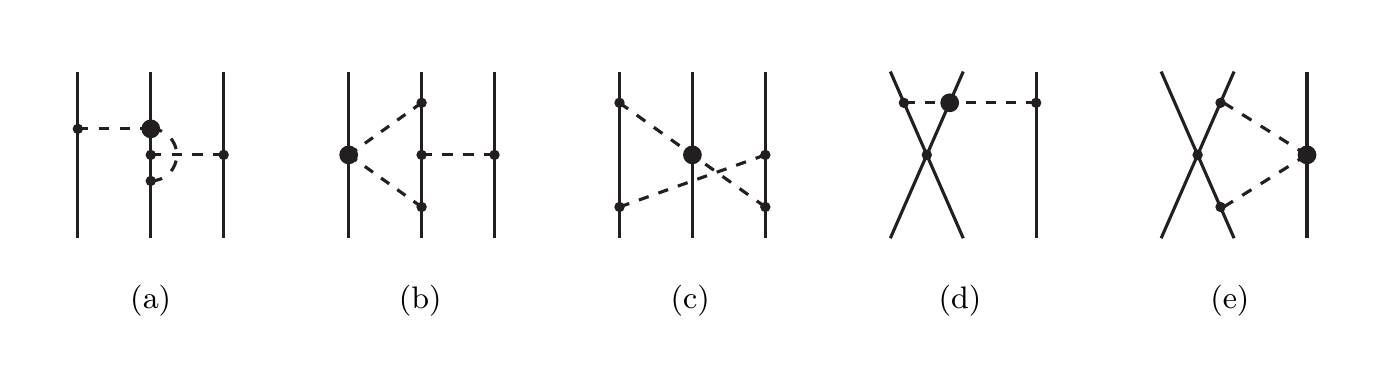}}
\vspace*{-0.75cm}
\caption{Sub-leading one-loop 3NF diagrams which appear at N$^4$LO
with topologies similar to Fig.~\ref{fig_3nf_n3lo}.
Notation as in Fig.~\ref{fig_hi}.}
\label{fig_3nf_n4loloops}
\end{figure}

\begin{figure}[t]\centering
\vspace*{-0.5cm}
\scalebox{1.0}{\includegraphics{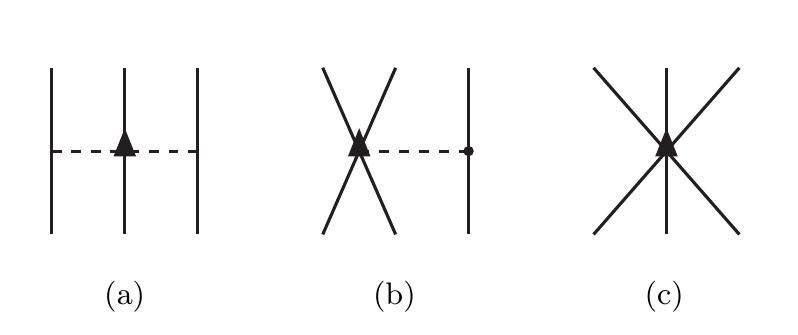}}
\vspace*{-0.3cm}
\caption{3NF tree graphs at N$^4$LO ($\nu=5$) denoted by: (a) 2PE, (b) 1PE-contact, and (c) contact. Notation as in Fig.~\ref{fig_hi}.}
\label{fig_3nf_n4lotrees}
\end{figure}

In the meantime, one may go ahead and look
at the next order of 3NFs, which is N$^4$LO or $\nu=5$.
The loop contributions that occur at this order
are obtained by replacing in the N$^3$LO loops
one vertex by a $\Delta_i=1$ vertex (with LEC $c_i$), Fig.~\ref{fig_3nf_n4loloops},
which is why these loops may be more sizable than the N$^3$LO loops.
The 2PE, 1PE-2PE, and ring topologies have been evaluated~\cite{KGE12,KGE13} so far.
In addition, we have three `tree' topologies (Fig.~\ref{fig_3nf_n4lotrees}), which include
a new set of 3N contact interactions that has recently been derived
by the Pisa group~\cite{GKV11}. Contact terms are typically simple (as compared
to loop diagrams) and their coefficients are essentially free. 
Therefore, it would be an
attractive project to test some terms (in particular, the spin-orbit terms) 
of the N$^4$LO contact 3NF~\cite{GKV11} in calculations of few-body reactions (specifically,
the $p$-$d$ and $p$-$^3$He $A_y$) and spectra of light nuclei.

\subsection{Four-nucleon forces}

\begin{figure}[t]\centering
%\vspace*{-1.5cm}
\scalebox{0.9}{\includegraphics{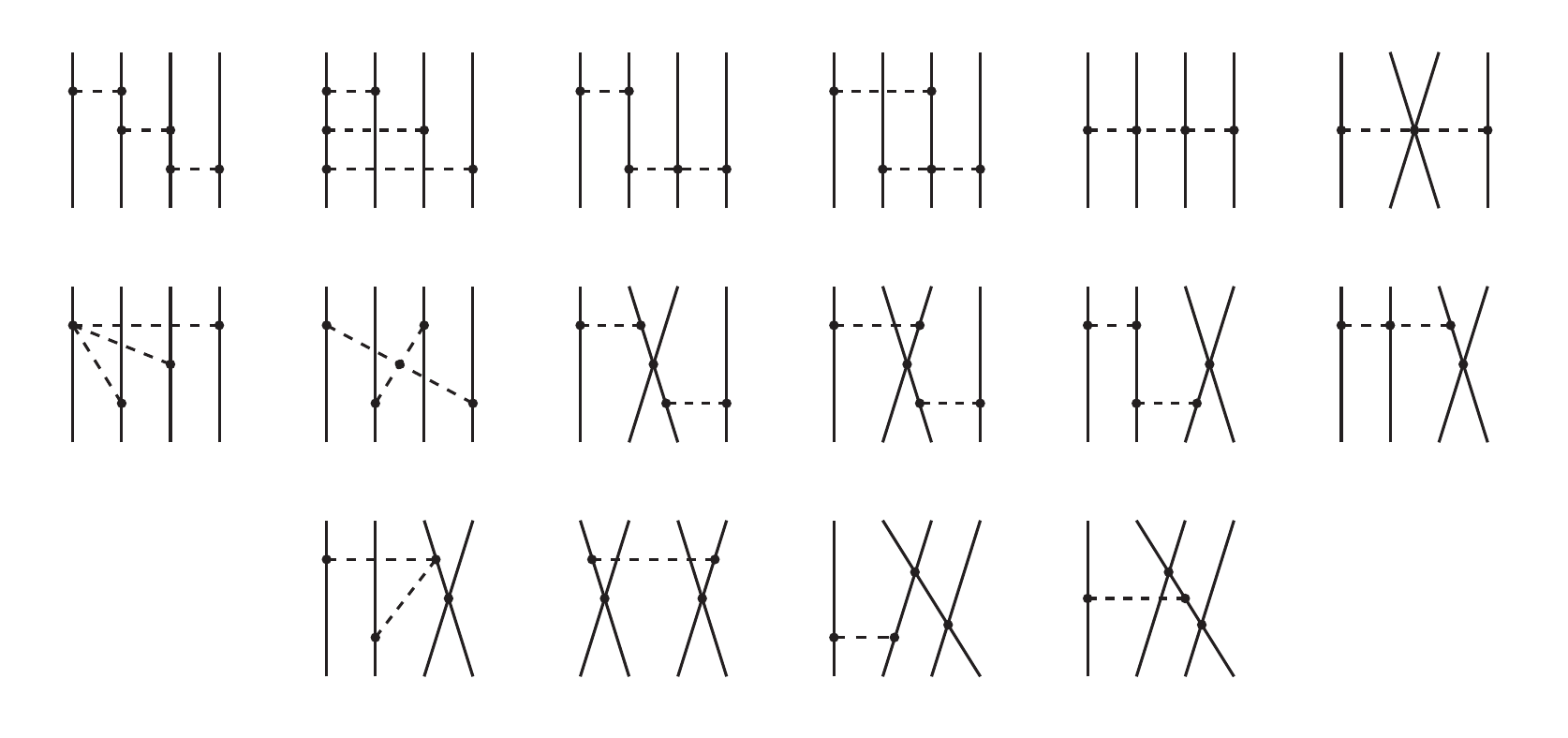}}
\vspace*{-0.5cm}
\caption{Leading four-nucleon force at N$^3$LO.
(Figure reproduced from Ref.~\cite{ME11}.)}
\label{fig_4nf_n3lo}
\end{figure}

For connected ($C=1$) $A=4$ diagrams, Eq.~(\ref{eq_nu}) yields
\begin{equation}
\nu = 4 + 2L + 
\sum_i \Delta_i \,.
\label{eq_nu4nf}
\end{equation}
%HERE8
We then see that the first (connected) non-vanishing 4NF is generated at $\nu = 4$ (N$^3$LO), with                   
all vertices of leading type, Fig.~\ref{fig_4nf_n3lo}. 
This 4NF contribution has no loops and introduces no novel parameters~\cite{Epe07}.
(See Ref.~\cite{ME11} for a more detailed discussion on these diagrams.) 

For a reasonably convergent series, terms                      
of order $(Q/\Lambda_\chi)^4$ must be small, and therefore              
chiral 4NF are predicted to be very weak.                    
This expectation was confirmed in a recent calculation 
of the $^4$He binding energy including 
the leading 4NF (Fig.~\ref{fig_4nf_n3lo}). Its effect was found to be 
 a few 100 keV~\cite{Roz06}, to be compared with the actual size of the 
binding energy, 28.3 MeV. Although                 
obtained with the help of several approximations, this                
preliminary predictions supports the notion that 4NF may indeed be negligeable. 

The effects of the leading chiral 4NF in symmetric nuclear matter and pure neutron matter have been
worked out by Kaiser {\it et al.}~\cite{Kai12,KM16}.

\section{Applications in the nuclear many-body problem \label{sec_manybody}}

In this Section, we will address some recent applications of the few-nucleon
forces which were derived and discussed in previous sections. In particular, we will 
concentrate on applications where the analysis is conducted in the spirit of exploring 
order-by-order convergence of the predictions. 

It should be clear from the historical perspectives presented at the opening of this article that our present            
knowledge of nuclear forces in free space and in the few-nucleon system is 
the result of decades of struggle.                                                          
The nature of the nuclear force in a dense medium is an even more complex problem, 
as it involves aspects of the forces that cannot be constrained
through free-space $NN$ scattering or the properties of the (relatively ``simple") few-nucleon system. 

Although predictions for finite nuclei are the ultimate test for 
many-body predictions, 
infinite nuclear matter is an alternative and convenient testing ground for many-body theories. By ``nuclear matter" we mean an infinite system 
of nucleons acted on by their mutual strong forces and no electromagnetic interactions. Nuclear matter 
is characterized by its energy per particle as a function of density and other quantities as appropriate 
(e.g.\ temperature, isospin asymmetry, spin asymmetry). Such relation is known as the nuclear matter equation of state (EoS). 
The translational invariance of the system facilitates theoretical calculations. At the same time, adopting     
what is known as 
``local density approximation", one may use the EoS directly in calculations of finite systems, as we will discuss 
 below.

When proton and neutron densities are different (that is, in the presence of isospin asymmetry),
the energy per particle becomes a function of both the total density and the relative concentrations
of neutrons and protons. The EoS of isospin-asymmetric matter naturally introduces
the symmetry energy, similarly to the appearance of the symmetry term in the well-known           
Bethe-Weiz{\" a}cker formula. As will be discussed later in more details, the symmetry energy is typically
approximated as the difference between the energy per particle of symmetric nuclear matter and
pure neutron matter as a function of density.
Of particular contemporary interest is the EoS of highly neutron-rich matter, all the way to
nearly pure neutron matter. This quantity is important for understanding wide ranging questions
in modern nuclear physics, from the properties of rare isotopes to those of neutron stars.
On the one hand, the density dependence of the symmetry energy is known to correlate strongly
with the neutron skin thickness of a heavy nucleus. On the other hand, the pressure in neutron-rich
matter is the main input for the structure equations of compact stars.
Therefore, microscopic predictions together with empirical constraints from observables that
are sensitive to the equation of state are an ideal combination to learn about the in-medium behavior
of nuclear forces, particularly in isospin-asymmetric medium. We also recall that the EoS is an
important part of the input of transport models describing heavy-ion collisions and thus can be
constrained through analyses of carefully selected observables in ion-ion scattering.
Concerning non-terrestrial observations, partnership between nuclear physics and astrophysics
is increasingly important as better constraints on the high-density part of the equation of state
become available through more accurate measurements of neutron star masses. In 
summary, studies of nucleonic matter are especially timely and important, as they support
rich on-going and future experimental effort, both in terrestrial laboratories and the cosmos. 

\subsection{Order-by-order predictions of the energy per nucleon in nuclear and neutron matter} 
\label{obo}

The problem shared by all non-EFT based approaches is that it is essentially 
impossible to estimate reliably the uncertainty associated with a particular 
prediction. 
On the other hand, EFT provides a well-defined framework to calculate observables where the 
truncation error decreases systematically as higher orders are included. Earlier in this article, we have seen
that such task can be accomplished quite successfully at the level of $NN$ phase shifts. 

In this section, we will review and discuss recent calculations of the energy per particle in infinite matter at different orders
of chiral EFT~\cite{Sam15}.                                                                   
The discussion will also emphasize the importance of error quantification and how it 
should be addressed in chiral EFT. 

Estimates of theoretical uncertainties~\cite{furnstahl15} for calculations of the equation of state have mostly
focused on varying the low-energy constants and resolution scale at which nuclear dynamics are
probed~\cite{coraggio13,krueger,Cor14,bogner05,hebeler11,gezerlis13}. In a recent work~\cite{Sam15} we layed the foundation for order-by-order calculations 
of nuclear many-body systems by presenting consistent NLO and N$^2$LO chiral nuclear forces
whose relevant short-range three-body forces are fit to A = 3 binding energies and the lifetime of
the triton. We then assessed the accuracy with which infinite nuclear matter properties and the isospin
asymmetry energy can be predicted from order-by-order calculations in chiral effective field theory.

Uncertainty originates from:
\begin{itemize}
\item 
The choice of the many-body method (a source of error not inherent to EFT).
\item 
Error in the determination of the low-energy constants (LECs). Short-range LECs ($NN$) and
long-range LECs ($\pi N$) must be considered separately.
\item 
Regulator dependence.
\item 
Truncation error.
\end{itemize} 
In the following, we will address those items  briefly  but systematically.

A variety of many-body methods are available and have been used extensively in nuclear
matter predictions. They include: the coupled-cluster method, many-body perturbation theory, variational Monte
Carlo or Green’s function Monte Carlo methods.
In computing the EoS, we employ the nonperturbative particle-particle ladder approximation.
In the traditional hole-line expansion, it represents the leading-order contribution. To quantify the
uncertainty carried by this choice, it is insightful to compare with Refs.~\cite{cc1,cc2}. In Ref.~\cite{cc1},
the authors report on coupled-cluster calculations in symmetric nuclear matter including particle-particle (pp) and
hole-hole (hh) diagrams (as well as an exact treatment of the Pauli operator). The overall effect, as seen from
comparing the first and last entries in Table II of Ref.~\cite{cc1}, is very small around saturation density,
consistent with Table II in Ref.~\cite{Cor14}, and grows to 1.5 MeV at the highest Fermi momentum included
in the study. Note that these calculations adopt the N$^3$LO potential~\cite{EM03} (with $\Lambda$=500 MeV) and no
three-nucleon forces. On the other hand, in Ref.~\cite{cc2} coupled-cluster calculations in nucleonic matter
were performed at N$^2$LO with two- and three-body forces and with the inclusion of selected triples
clusters, namely correlations beyond pp and hh ladders. The effect of these contributions is found
to be negligible in neutron matter and about 1 MeV per nucleon in symmetric matter in the density
range under consideration~\cite{cc2}. 
In the light of the above considerations, we conclude that a realistic estimate of the impact of
using a nonperturbative approach beyond pp correlations is about 1 MeV in nuclear matter around
saturation density and much smaller in neutron matter. As we show below, such uncertainties are
significantly smaller than those associated with variations in the cutoff scale.

In order to quantify the error associated with possible variations of the (short-range) $NN$ LECs,
we refer to recent findings from the Granada group~\cite{Gran}. They applied 205 samples of smooth local
potentials, all with $\chi ^2$/datum of approximately 1, and found a variation of 15 keV in the triton
binding energy.
From our part, we performed Brueckner-Hartree-Fock calculations in nuclear matter using
local high-precision potentials from the Nijmegen group~\cite{Sto94} and observed an uncertainty of 0.6
MeV in the energy per particle at normal density.
In summary, we conclude that the uncertainty arising from the error in the $NN$data has negligible
impact on the many-body system.
Concerning the (long-range) $\pi N$ LECs, they are likely to impact mostly peripheral partial
waves (namely, those high partial waves where no contact terms are present). At NLO and N$^2$LO, that means 
$D$-waves and higher, whereas at N$^3$LO no contacts exist in $F$-waves and higher. Therefore,
we expect variations of the $\pi N$ LECs (within the range allowed by $\pi N$ scattering data) to have
only minor impact in nuclear matter, since its sensitivity is limited to peripheral partial waves.
Nevertheless, we stress that a systematic investigation with consideration of $\pi N$ LECs uncertainty
consistently in the 2NF and the 3NF, has not yet been done and is part of our future plans.

\begin{figure}[!t]
\vspace*{-3.5cm}
\hspace*{-1cm}
\includegraphics[width=10cm]{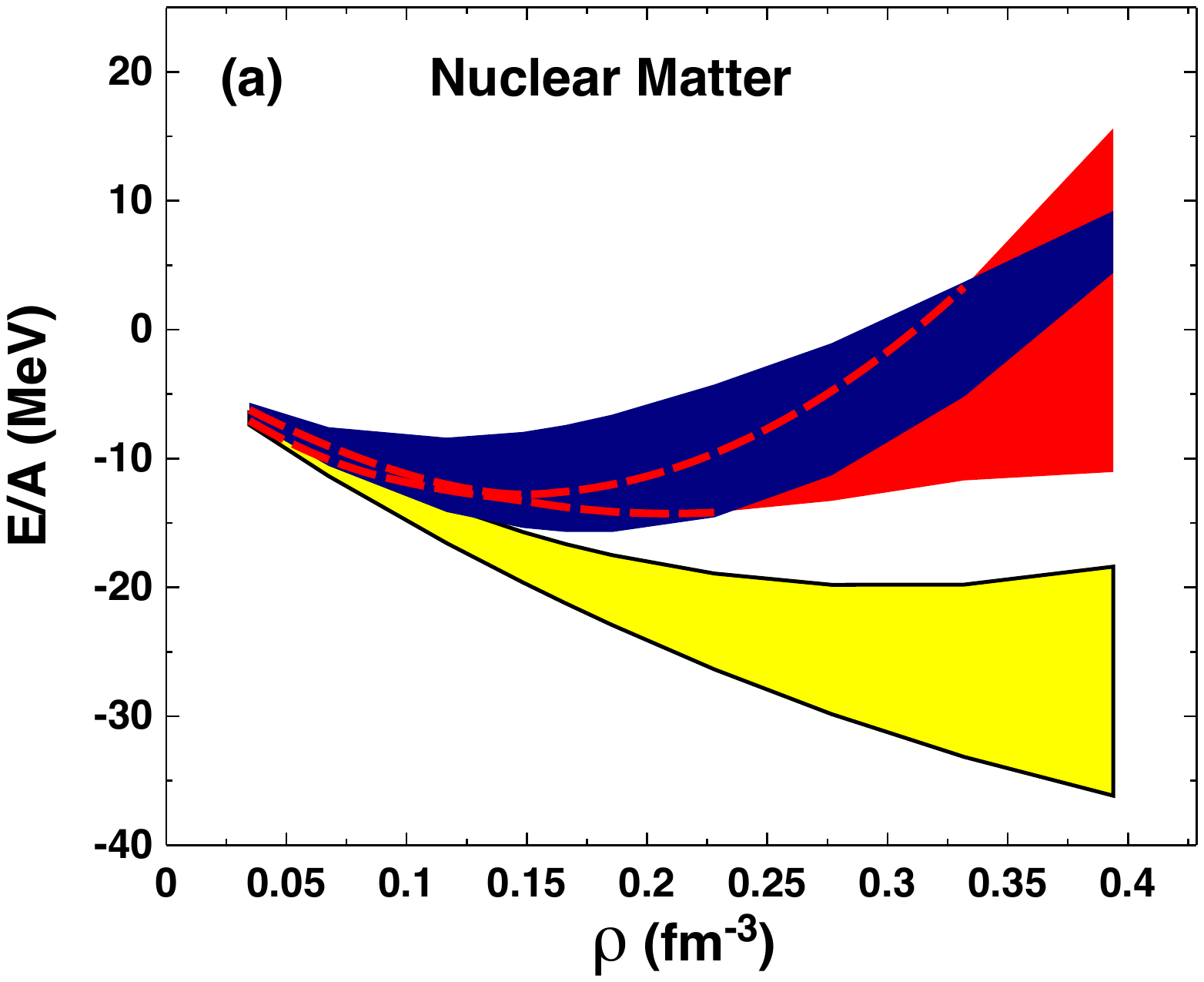}\hspace{-.5in}
\hspace*{-1cm}
\includegraphics[width=10.34cm]{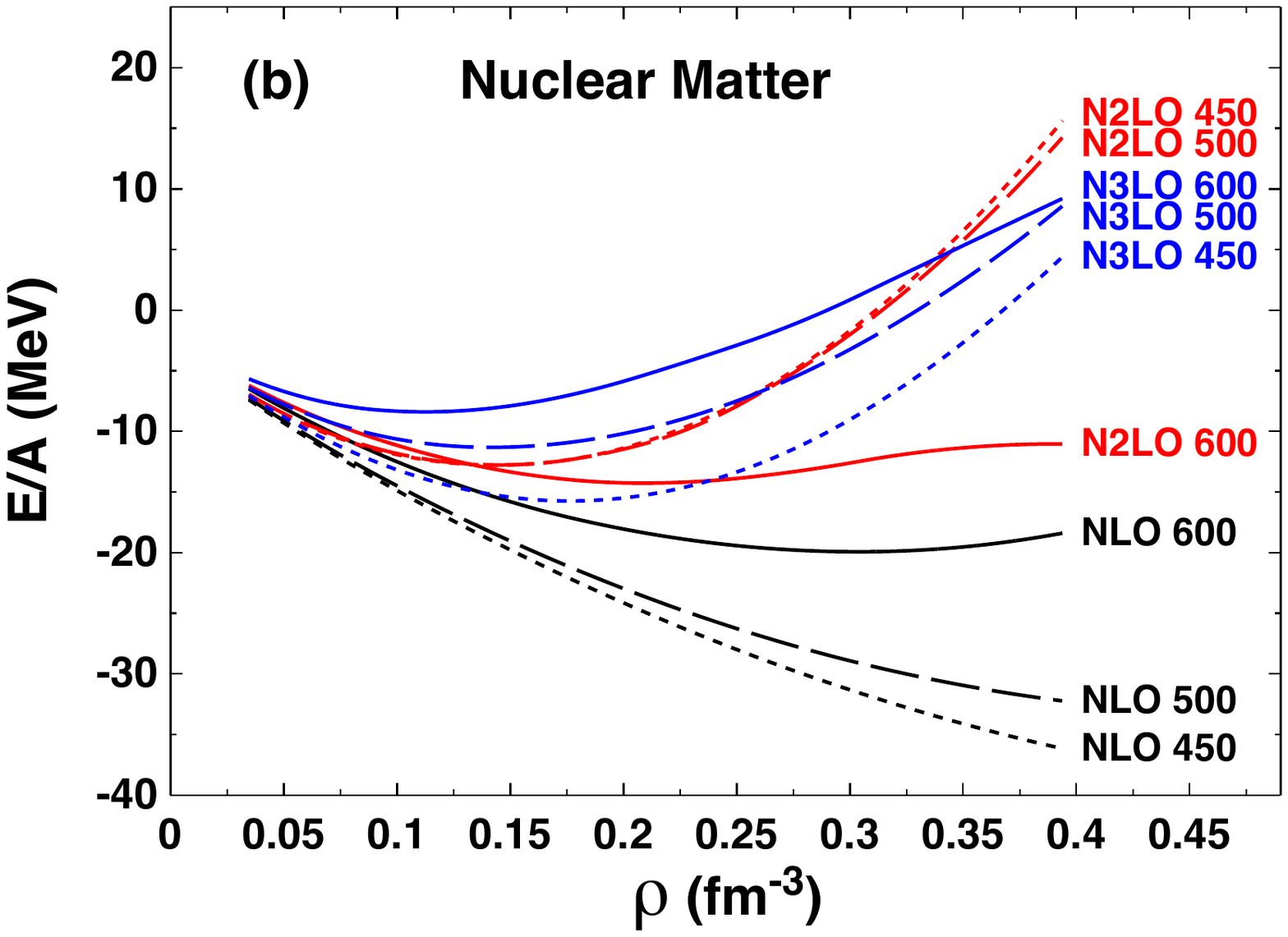}\hspace{.1in}
\vspace*{-3.5cm}
\caption{Energy/nucleon (E/A) in symmetric nuclear matter
as a function of density, $\rho$. Left frame: The yellow and
red bands represent the uncertainties in the predictions due to cutoff variations
as obtained in complete calculations at NLO and N$^2$LO, respectively.
The blue band is the result of a calculation employing N$^3$LO $NN$ potentials together with
N$^2$LO 3NFs. 
The dashed lines show the upper or lower limits of hidden bands. Right frame: predictions at 
the specified order and cutoff value.
(Figure reproduced from Ref.~\cite{Sam15}.)} 
\label{snm}
\end{figure}

Keeping in mind the uncertainty considerations
made above, we now move to nuclear and neutron matter predictions. 
Our results for the energy per particle as a function of the nuclear density are shown in Fig.~\ref{snm} 
for symmetric nuclear matter. We note that the particle-particle ladder approximation employed
in the present work is in good agreement with the perturbative results available at N$^3$LO from
Ref.~\cite{Cor14} including up to third-order pp diagrams. In Fig.~\ref{snm}, the shaded bands
in yellow and red represent the spread of our complete calculations conducted at NLO and N$^2$LO,
respectively. The blue band is the result of a calculation that employs N$^3$LO $NN$ potentials together
with N$^2$LO 3NFs. In all cases shown, the cutoff is varied over the range 450-600 MeV. As noted
before, the N$^3$LO 3NFs and 4NFs are at present omitted, and the resulting convergence pattern
gives an estimate on the theoretical uncertainty of the calculation (and not necessarily of the chiral 
effective field theory expansion per se).
We note that at NLO the potentials
constructed at lower cutoff scales do not exhibit saturation until very high densities. On the other
hand, for the 600 MeV cutoff potential the $^1S_0$ partial wave (together with the $^3S_1$ partial wave)
is sufficiently repulsive to enable saturation at a relatively smaller density. We observe that the
convergence pattern for the low-cutoff ($\Lambda$=450-500 MeV) potentials is significantly better than
for the 600 MeV potential. Overall there is a large spread from cutoff variations both at NLO and
N$^2$LO beyond nuclear matter saturation density. Moreover, the bands at these two orders do not
overlap, suggesting that their width is not a suitable representation of the uncertainty. Although the
(not yet complete) N$^3$LO calculation reveals a strong reduction of the cutoff dependence, it is important
to notice that an uncertainty of about 8 MeV remains at saturation density. While we do not expect
much of a change in nuclear matter predictions from 4NFs~\cite{Epe07,Roz06,krueger}, it is quite possible that the
inclusion of N$^3$LO 3NFs might reduce either the cutoff dependence or improve the convergence
pattern. This will be an interesting subject for future investigations.

\begin{figure}[!t]
\centering         
\vspace*{-4.5cm}
\includegraphics[width=11cm]{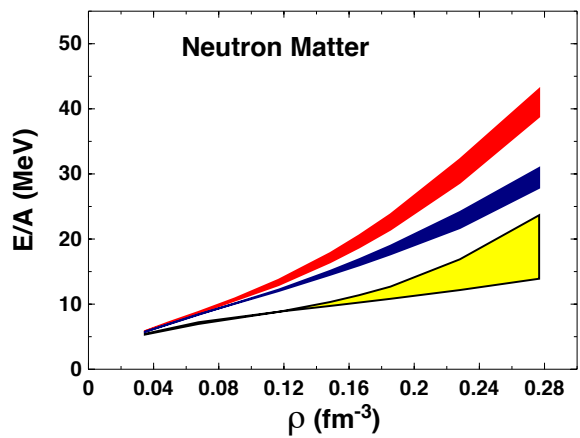}\hspace{.1in}
\vspace*{-3.5cm}
\caption{ As in Fig.~\ref{snm} for pure neutron matter. 
(Figure reproduced from Ref.~\cite{Sam15}.)} 
\label{nm}
\end{figure}

\begin{figure}[!t] 
\centering         
\vspace*{-3.5cm}
\scalebox{0.5}{\includegraphics{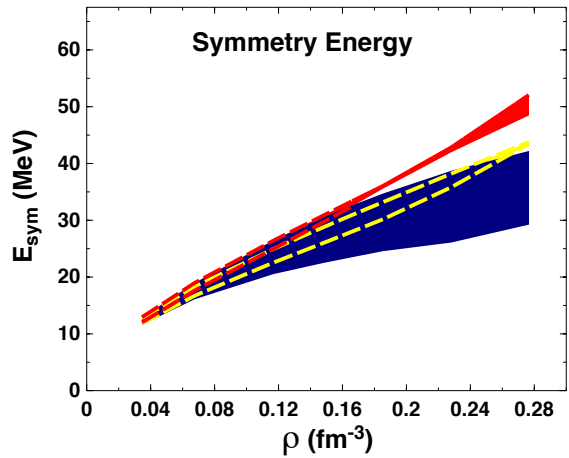}}
\vspace*{-3.5cm}
\caption{The symmetry energy, $E_{\rm sym}$, as a function of density, $\rho$.
Meaning of bands and dashed lines as in Fig.~\ref{snm}.
(Figure reproduced from Ref.~\cite{Sam15}.)} 
\label{esym}
\end{figure}

The results for neutron matter are presented in Fig.~\ref{nm}, where the bands have
the same meaning as in Fig.~\ref{snm}. Note that the range of densities under consideration is smaller
for neutron matter in order to keep the Fermi momentum below the cutoff in all cases. We see
a large spread at NLO for the largest densities considered, whereas the band has only moderate
size at the next order and remains small for our N$^3$LO calculation. Similar to what was observed
in symmetric nuclear matter, the bands at NLO and N$^2$LO do not overlap in neutron matter.                         
In addition, the N$^3$LO band does not generally overlap with the N$^2$LO band. Therefore, the variation
obtained by changing the cutoff does not seem to provide a reliable representation of the uncertainty
at the given order. A better way to estimate such uncertainty is to consider the difference between
the predictions at two consecutive orders.

In Fig.~\ref{esym} we present the results for the symmetry energy, which is defined as the strength
of the quadratic term in an expansion of the energy per particle in asymmetric matter with respect
to the asymmetry parameter $\alpha$: 
\begin{equation}
\bar E (\rho,\alpha) \approx \bar E (\rho,\alpha = 0) + E_{\rm sym}\alpha^2 + {\cal{O}}(\alpha ^4) \; , 
\label{sym}
\end{equation}
where $\bar E = E/A$ is the energy per particle and $\alpha= (\rho_n - \rho_p)/(\rho_n+\rho_p)$. 
The nearly linear behavior of $\bar E(\rho,\alpha)$ with $\alpha ^2$ has been confirmed by many 
microscopic calculations (see for instance 
Refs.~\cite{AS,Drischler}, but see also Ref.~\cite{Kai15}). 
It is a common approximation to neglect powers beyond
$\alpha ^2$ in the expansion above and thus defining 
the symmetry energy as the difference between the energy per particle in neutron
matter and symmetric nuclear matter.

As mentioned at the beginning of this section, systematic efforts are ongoing to set better empirical constraints on the
symmetry energy, through both laboratory and astrophysical measurements. It is therefore important
to have an understanding of the theoretical uncertainty affecting calculations of this quantity.
The spread due to the change of the cutoff values in our NLO, N$^2$LO, and N$^3$LO calculations is represented
by the three bands as before. As observed previously for symmetric matter, the spread due
to cutoff variations remains large at N$^2$LO, with some minimal overlap with the NLO band. The
N$^3$LO band reflects the large cutoff sensitivity previously observed in symmetric matter. Again,
we conclude that the spread generated by changing the cutoff does not in general provide a reliable
estimate of the theoretical uncertainty.

\subsection{Spin-polarized neutron matter} 
\label{spin}

Polarized neutron matter (NM) is an interesting system for various reasons. Among them is                  
the impact that spin instabilities in the interior of stellar matter would have on                     
neutrino interactions and thus the star cooling mechanism.           

Spin polarization is also of interest in symmetric or nearly  
symmetric nuclear matter (SNM). For instance, for the purpose of scattering from
polarized nuclei, one may define a 
spin dependent optical potential which, for the spin degree of freedom, plays
the same role as the Lane potential \cite{Lane}  for the isospin degree of freedom.
Such {\it spin symmetry potential} can be obtained from the difference between 
the single-particle potentials for spin-up and spin-down nucleons in polarized SNM. 

To address the most general case, one must include                               
both spin and isospin polarizations. From the astrophysics point of view, stellar matter contains
a small, but not insignificant 
proton fraction. With regard to experiments in terrestrial laboratories, 
the spin dependence of the nuclear interaction in nuclear matter can be explored through 
collective exitations, such as giant resonances.  
Most typically, a nucleus with non-zero spin is 
also isospin asymmetric, making it necessary to include both spin and isospin polarizations.  
For those reasons, in previous work~\cite{Sam10,Sam11,SK07} 
we explored matter with different densities 
of neutrons and protons where each type of nucleon can have arbitrary degree of spin polarization.
We obtained predictions employing the Dirac-Brueckner-Hartree-Fock approach to nuclear matter and a relativistic
one-boson-exchange $NN$ potential and did not see any indications                             
of a phase transition to a spin-polarized state. We note that all models which start from the bare $NN$ force 
and apply it in the          
medium (see, for instance, 
Ref.~\cite{pol18}) end up with similar conclusions. In contrast, approaches based on parametrizations of Skyrme forces, or other phenomenological forces, report different findings.  
For instance, 
 with the {\it SLy4} and {\it SLy5} forces and the Fermi liquid 
formalism                                                                                          
  a phase transition  to the antiferromagnetic state is predicted in asymmetric matter 
at a critical density equal to about 2-3 times normal density \cite{IY}.
Qualitative disagreement is also encountered with other approaches such as                     
relativistic Hartree-Fock models based on effective meson-nucleon Lagrangians. For instance, in Ref.~\cite{pol12} 
it was reported that the onset of 
a ferromagnetic transition in neutron matter, and its critical density, are crucially determined by the inclusion of isovector mesons and the 
nature of their couplings. 

The brief review given above summarizes the findings of many useful and valid calculations. However, the 
problem common to all of them is that it is
essentially impossible to estimate, in a statistically meaningful way, the uncertainties
associated with a particular prediction, or to quantify the error related to the approximations 
applied in a particular model. 
Therefore, in this section, we apply the same philosophy as in Sec.~\ref{obo} to study the equation of state of 
polarized neutron matter at different orders of ChPT. 

Based on the literature mentioned above,                              
a phase transition to a polarized phase (at least up to normal densities) seems unlikely, 
although the validity of such conclusion must be assessed in the context of EFT errors.
Furthermore, polarized neutron matter is a very interesting system for several reasons.
Because of the large neutron-neutron scattering length, NM displays behaviors similar to those 
of a unitary Fermi gas. In fact, 
up to nearly normal density, (unpolarized) neutron matter is found to display the behavior of an                  
$S$-wave superfluid~\cite{Carls03,Carls12}. 
The possibility of simulating low-density NM with 
ultracold atoms near a Feshbach resonance~\cite{Bloch08} has also been discussed.
When the system is totally polarized, 
it has been observed to behave like a weakly interacting Fermi gas~\cite{krueg14}. 
Here, we wish to explore to which extent and up to which densities we are in agreement with such conclusions, and 
how this and other observations depend on the chiral order and the resolution scale. 

In contrast with previous calculations, 
our recent work summarized here 
contains the following novelties:                                                            
\begin{itemize}
\item We consider both cutoff dependence and truncation error for the purpose of uncertainty 
quantification of chiral EFT.
Although incomplete in the 3NF at N$^3$LO, our calculations are a substantial step in that direction.
We note, further, that the contribution from the 3NF at N$^3$LO was found to be very small in neutron 
matter for the potentials in our perview~\cite{krueger}, about -0.5 MeV at normal density. Here, we consider
neutron matter or highly neutron-rich matter.
\item For the first time, we present results for both spin and isospin asymmetries within the framework of chiral forces.
These tools are necessary to assess, for instance, the sensitivity 
of the results (particularly, the potential onset of a phase transition) to the presence of a non-zero proton fraction. 
\end{itemize}

For a detailed description of the formalism, the reader is referred to Ref.~\cite{SMK15}.     
Here, we will just summarize some definitions which are necessary for the discussion which follows.

In a spin-polarized and isospin asymmetric system with fixed total density, $\rho$,               
the partial densities of each species are               
\begin{equation}
\rho_n=\rho_{nu}+\rho_{nd}\; , \; \; \; 
\rho_p=\rho_{pu}+\rho_{pd}\;, \; \; \; 
\rho=\rho_{n}+\rho_{p} \; ,           
\label{rho} 
\end{equation}
where $u$ and $d$ refer to up and down spin-polarizations, respectively, of protons ($p$) or neutrons ($n$). 
The isospin and spin asymmetries, $\alpha$, $\beta_n$, and $\beta_p$,  are defined in a natural way: 
\begin{equation}
\alpha=\frac{\rho_{n}-\rho_{p}}{\rho} \;, \; \; \;
\beta_n=\frac{\rho_{nu}-\rho_{nd}}{\rho_n} \;, \; \; \; 
\beta_p=\frac{\rho_{pu}-\rho_{pd}}{\rho_p} \;. 
\label{alpbet} 
\end{equation}
The density of each individual component can be related to the total density by 
%\begin{equation}
\begin{displaymath}
\rho_{nu}=(1 + \beta_n)(1 + \alpha){\rho \over 4} \;, \; \; 
%\label{rhonu}
%\end{equation}
%\begin{equation}
\rho_{nd}=(1 - \beta_n)(1 + \alpha){\rho \over 4}\; ,\;  \; 
\end{displaymath}
\begin{equation}
%\label{rhond}
%\end{equation}
%\begin{equation}
\rho_{pu}=(1 + \beta_p)(1 - \alpha){\rho \over 4}\; ,\; \; 
%\label{rhopu}
%\end{equation}
%\begin{equation}
\rho_{pd}=(1 - \beta_p)(1 - \alpha){\rho \over 4}\; , 
\label{rhopnud}
\end{equation}
where each partial density is related to the corresponding Fermi momentum 
through $\rho_{\tau \sigma}$ =$ (k_F^{\tau \sigma})^3/(6\pi^2)$. 
The {\it average} Fermi momentum  and the total density are related in the usual way as 
$\rho= (2 k_F^3)/(3 \pi ^2)$.

\begin{figure}[!t] 
\centering 
\vspace*{-3cm}
\hspace*{-1.0cm}
\scalebox{0.55}{\includegraphics{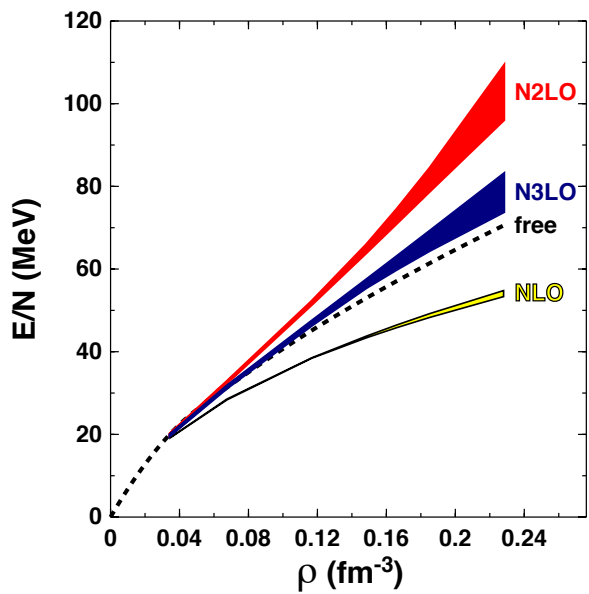}} 
\vspace*{-4cm}
\caption{                                        
Energy per neutron in fully polarized neutron matter as a function 
of density. The yellow and red bands represent the uncertainities due to cutoff variations obtained
in the complete calculations at NLO and N$^2$LO, respectively. The blue band is the result of the 
same cutoff variations applied to our exploratory N$^3$LO calculation, see text for details. 
The dotted curve shows the energy of the free Fermi gas.
(Figure reproduced from Ref.~\cite{SMK15}.)
} 
\label{pnm}
\end{figure}

We show in Fig.~\ref{pnm} the energy per particle in fully polarized neutron matter
as a function of density. The yellow and red bands represent the predictions of 
complete calculations at second and third 
order, respectively, of chiral effective field theory, while the blue band shows the predictions obtained with the exploratory N$^3$LO calculation as described
above. 
For each band, the width is obtained by changing the cutoff between 450 MeV and 600 MeV.

At N$^2$LO and N$^3$LO, 
cutoff dependence is generally moderate up to saturation density. At NLO, the cutoff dependence is 
practically negligible throughout.
In unpolarized
neutron matter, on the other hand, the largest cutoff dependence was seen at NLO~\cite{Sam15}. This suggests that, in unpolarized NM,
the larger cutoff sensitivity at NLO is mostly due to singlet states, particularly $^1S_0$, which are absent from the polarized system.
At the same time, 3NFs do not appear at NLO, implying that most of the cutoff dependence in polarized NM
at N$^2$LO and N$^3$LO is caused by the 3NF contributions. 

Clearly, the variations associated with changing the cutoff are not a good 
indicator of the uncertainty at a given order of chiral effective field theory, 
as the results from one order to the other do not overlap.
Furthermore, the predictions do not show a good convergence pattern, although 
some indication of slow convergence can be seen when moving from N$^2$LO to our N$^3$LO calculation.

As can be concluded from Table~\ref{tab2}, the predictions from 
the N$^3$LO calculation are close to the free Fermi gas energy, at least up
to saturation densities. Our results with the N$^3$LO~\cite{EM03} ($\Lambda$=500 MeV) potential are in good agreement 
 with those from Ref.~\cite{krueg14} using the same potential as well as three- and four-nucleon forces at 
 N$^3$LO.                                                                   

\begin{figure}[!t] 
\centering 
\vspace*{-4.5cm}
%\hspace*{-1.0cm}
\scalebox{0.6}{\includegraphics{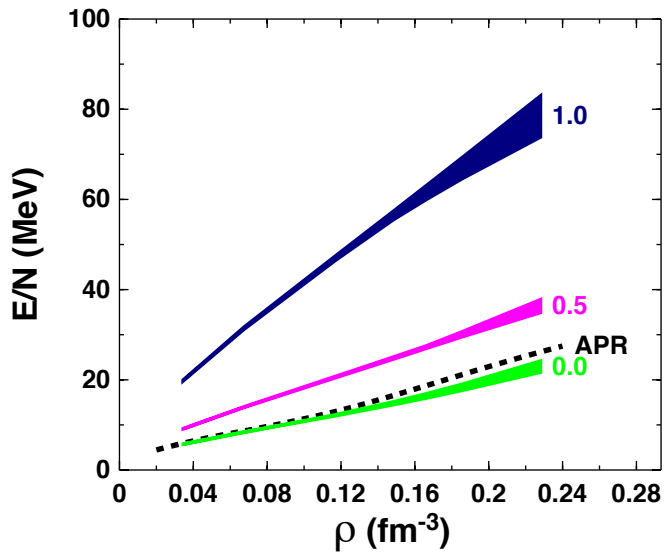}} 
\vspace*{-4.5cm}
\caption{                                        
Energy per neutron in pure neutron matter as a function of density at N$^3$LO. From lowest to highest curve:
unpolarized NM; partially polarized NM, with $\beta_n$=0.5; fully polarized NM ($\beta_n$=1). 
The width of each band shows the uncertainty from varying the cutoff between 
450 and 600 MeV. The black dotted line shows the predictions for the equation of state of unpolarized
neutron matter from Ref.~\cite{APR}. 
(Figure reproduced from Ref.~\cite{SMK15}.)
} 
\label{beta}
\end{figure}

In Fig.~\ref{beta}, for our N$^3$LO calculation, we compare predictions (along with their cutoff
variations) of the energy per neutron in: unpolarized NM (green band), partially polarized NM (pink band), and
fully polarized NM (blue band). 
For the partially polarized case, the value of $\beta_n$ is equal to 0.5, corresponding to 75\% 
of the neutrons being polarized in one direction and 25\% in the opposite direction, see Eqs.~(\ref{alpbet}).
Clearly, a lesser degree of spin asymmetry (as compared to the ferromagnetic case) yields considerably
less repulsion. 
There is definitely no sign of a phase transition, particularly to a ferromagnetic state, nor an indication that 
such transition may occurr at higher densities. This is consistent with what we observed earlier~\cite{Sam11} 
with meson-theoretic interactions. 

As a baseline comparison, we also include, for the unpolarized case, predictions based on a different
approach, shown by the black dotted line in Fig.~\ref{beta}.
These are taken from Ref.~\cite{APR} and are based on the Argonne $v_{18}$ 
two-nucleon interaction 
plus the Urbana IX three body-force, using variational methods. The predictions are overall in reasonable agreement with our green band, although those from 
Ref.~\cite{APR} show more repulsion as compared to the softer chiral interactions.

\begin{figure}[!t] 
\centering 
\vspace*{-4.5cm}
%\hspace*{-1.0cm}
\scalebox{0.6}{\includegraphics{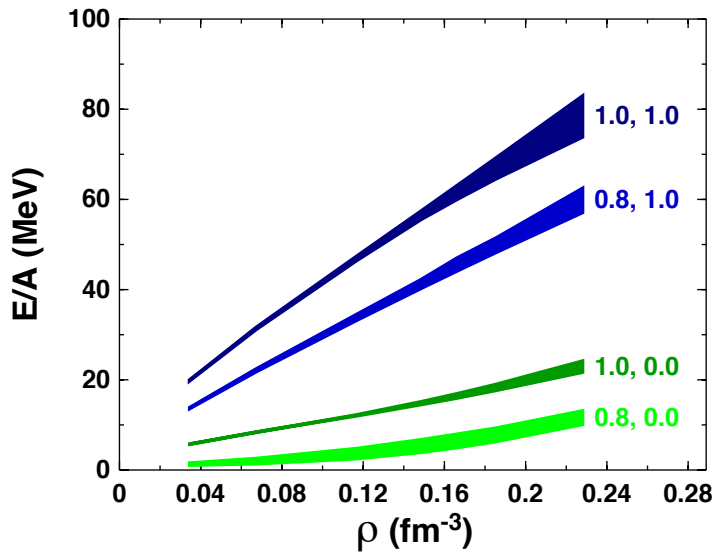}} 
\vspace*{-4.5cm}
\caption{                                       
Energy per nucleon in neutron-rich matter 
as a function of density at N$^3$LO and different conditions of isospin and spin polarization.
The (brighter blue) band labeled as ``0.8, 1.0" displays the results for neutron-rich matter with a 
proton fraction equal to 10\% ($\alpha$=0.8) and fully polarized neutron ($\beta_n$=1.0).                      
The (brighter green) band labeled as ``0.8, 0.0" refers to neutron-rich matter with the same proton fraction 
and no polarization ($\beta_n$=0.0). The protons are unpolarized. 
For comparison, we also include the bands (darker blue and darker green) already shown in the previous figure, 
which refer to pure neutron matter ($\alpha$=1) with fully polarized ($\beta_n$=1) or 
unpolarized ($\beta_n$=0) neutrons.   
The bands are obtained varying the cutoff between 450 and 600 MeV.
(Figure reproduced from Ref.~\cite{SMK15}.)
} 
\label{alpha}
\end{figure}

Most typically, models which do predict spin instability of neutron matter find the phase transition to 
occurr at densities a few times normal density. Such high densities are outside the domain of chiral perturbation
theory. 
With some effective forces, though, it was found~\cite{pol19} that a small fraction of protons can significantly reduce the 
onset of the threshold density for a phase transition to a spin-polarized state of neutron-rich matter.
We explored this scenario by adding a small fraction of protons to fully polarized or unpolarized neutrons.
From Eqs.~(\ref{rho})-(\ref{rhopnud}), a proton fraction of 10\% is obtained with $\alpha$=0.8.
The results are displayed in 
Fig.~\ref{alpha}, where a crossing of the bands labeled with ``0.8, 1.0" and ``0.8, 0.0", respectively,
would indicate a phase transition. Thus we conclude that such transition is not predicted with chiral 
forces. By extrapolation, a transition to a polarized state would also appear very unlikely at higher densities. 

\begin{table}                
\centering
\begin{tabular}{|c||c|c|}
\hline
Density (fm$^{-3}$) & $\Lambda$ (MeV) & $E_{FFG}/E$ \\
\hline     
  0.15 & 450 & 0.95 \\
       & 500 & 0.92  \\
       & 600 & 0.95  \\
  0.17 & 450 & 0.95  \\
       & 500 & 0.91  \\
       & 600 & 0.93  \\
\hline
\end{tabular}
\caption                                                    
{Ratio of the energy per particle of a free Fermi gas 
to the energy per particle of polarized neutron matter 
around saturation density at N$^3$LO (as described in the text)
and for different values of the cutoff.} 
\label{tab2}
\end{table}

To summarize this section, 
we have calculated the equation of state of (fully and partially) polarized neutron-rich matter.
We performed complete calculations at second and third order of chiral effective field
theory and calculations employing the N$^3$LO 2NF plus the leading 3NF.
Results with both spin and isospin asymmetries have been presented for the first time with chiral forces 
in Refs.~\cite{SMK15}. 

In all calculations, the 
cutoff dependence is moderate and definitely underestimates the uncertainty of each order.  
Concerning the latter, we do not see a satisfactory convergence pattern. The missing
3NFs are most likely not the main cause of uncertainty at N$^3$LO, since               
Ref.~\cite{krueg14} has demonstrated that large cancelations take place between 
the 2$\pi$-exchange 3NF and the $\pi$-ring 3NF at N$^3$LO, while other 3NF contributions are very small (about
0.1-0.2 MeV). 
Clearly a calculation at 
N$^4$LO is necessary to get a realistic indication of the EFT error at 
N$^3$LO. Such effort is in progress.                         
If such calculation displays a reasonable convergence pattern, it will be strong evidence that 
polarized neutron matter, indeed, behaves nearly like a free Fermi gas, at least up to normal densities.

In our N$^3$LO calculation,                                         
the energies of the unpolarized system at normal density are close to 16 MeV for all cutoffs, 
whereas those in the polarized case are approximately 60 MeV. Thus, even in the presence
of the large uncertainties discussed above, 
a phase transition to a ferromagnetic state can be excluded.
This conclusion remains valid in the presence of a small proton fraction.

\subsection{Uncertainty analysis for predictions of the neutron skin in $^{208}$Pb at different orders
of chiral effective field theory} 
\label{skin}  

As mentioned at the beginning of Sec.~\ref{sec_manybody}, intense effort is going on to obtain reliable empirical information for the less
known aspects of the EoS. Heavy-ion (HI) reactions are a popular way to seek constraints on the symmetry 
energy, through analyses of observables that are sensitive to the difference between the pressure in        
nuclear and neutron matter.           
Isospin diffusion data in HI collisions together with analyses based on isospin-dependent transport
models, provide information on the slope of the symmetry energy.                          
For a recent review on available constraints from a broad spectrum of experiments, see Ref.~\cite{Tsang+}. 

Concerning the lower densities,     
isospin-sensitive observables can also be identified among the properties of normal nuclei. 
The neutron skin of neutron-rich nuclei is a powerful isovector observable, being sensitive to the   
slope of the symmetry energy, which determines to which extent neutrons are
pushed outwards to form the skin~\cite{FSskin15}.
Parity-violating electron scattering experiments are now a realistic option        
to determine neutron distributions with unprecedented accuracy. 
These experiments at low momentum transfer are especially suitable to probe neutron densities,        
because the $Z^0$ boson couples primarily to neutrons~\cite{Hor}. 
From the first electroweak observation of the neutron skin in a neutron-rich heavy nucleus, a values of           
0.33$^{+0.16}_{-0.18}$ for the neutron skin of 
$^{208}$Pb was determined~\cite{Jlab}, but                                                      
the next PREX experiment aims to measure the skin                 
 within an uncertainty smaller by a factor of 3 (see Ref.~\cite{Jlab} and references therein). 

From the theoretical point of view, we stress once again that microscopic calculations with statistically meaningful 
uncertainties are essential to guide experiments. Therefore, 
following the spirit of Ref.~\cite{Sam15}, 
it is the purpose of this section to systematically examine and discuss predictions of the neutron skin in       
$^{208}$Pb at different orders of chiral EFT and changing resolution scale.

It is well established that the neutron skin thickness correlates with the derivative of the symmetry energy. 
The latter is often represented through the $L$ parameter, 
\begin{equation}
L = 3 \rho_0 \Big (\frac{\partial E_{sym}(\rho)}{\partial \rho}\Big )_{\rho_0} \approx 
 3 \rho_0 \Big (\frac{\partial e_{n.m.}(\rho)}{\partial \rho}\Big )_{\rho_0} \; , 
\label{L} 
\end{equation} 
which originates from an expansion of the symmetry energy around the saturation point, $\rho_0$.        
The second (approximate) equality is due to the vanishing of the first derivative               
of the energy per particle in SNM at $\rho_0$, leaving a term proportional to the pressure in neutron matter.
Nevertheless, $L$ depends sensitively on the saturation density, which can be quite different from model
to model, particularly when considering different chiral orders and regulators.                       
In other words, theoretical predictions of $L$ carry larger EFT uncertainties than the ones of just neutron matter
pressure at some {\it fixed} density. 
To explore this point further, 
we will also compare predictions and uncertainties with those obtained using a phenomenological EoS 
for SNM consistent with the empirical saturation point.

\subsubsection{Predictions with microscopic EoS for NM and SNM}
 
We calculate proton and neutron density distributions with a method described in an earlier work \cite{AS03}. 
The method is based on an energy functional derived from the semi-empirical mass formula, where the volume and  
symmetry terms are contained in the isospin-asymmetric equation of state. Thus, we write the 
energy of a (spherical) nucleus as 
\begin{equation}
E(Z,A) = \int d^3 r~ e(\rho,\alpha)\rho(r) + 
\int d^3 r f_0(|\nabla \rho|^2 + \beta 
|\nabla \rho_I|^2) + I_C \; ,         
\label{drop} 
\end{equation} 
where $I_C$ stands for the Coulomb term. 
In the above equation, 
$\rho$ and $\rho_I$ are the usual isoscalar and isovector densities, given by $\rho_n +\rho_p$ and 
$(\rho_n -\rho_p)$, respectively, $\alpha$ is the neutron asymmetry
parameter, $\alpha=\rho_I/\rho$, and $e(\rho,\alpha)$ is the energy per particle in 
isospin-asymmetric nuclear matter. 
The constant $f_0$ in Eq.~(\ref{drop}) is approximately 70 MeV fm$^5$, whereas  the magnitude of  $\beta$ is about    
1/4~\cite{Furn}. 
(Even with variations of $\beta$ between -1 and +1, we found that
the contribution from that term was negligibly small, so we disregarded its 
contribution.) 

The symmetry energy, $E_{\rm sym}$, has been defined in Eq.~(\ref{sym}). 
As discussed earlier, it is customary to retain only the term quadratic in $\alpha$ in 
Eq.~(\ref{sym}). 

The proton and neutron density functions are obtained by minimizing the value
of the energy, Eq.~(\ref{drop}), with respect to the paramaters of Thomas-Fermi distributions
for proton and neutron densities.
 Although simple, this method has the advantage of allowing a         
very direct connection between the EoS and the properties of finite nuclei. Furthermore, microscopic structure  
calculations for $A$ = 208 are presently not possible. In Ref.~\cite{AS03}, our method was shown to yield 
realistic predictions for $^{40}$Ca, $^{90}$Zr, and $^{208}$Pb with some of the Bonn meson-exchange 
potentials~\cite{Mac89}.

\begin{figure}[!t]
\centering 
\vspace*{-3.5cm}
\includegraphics[width=11cm]{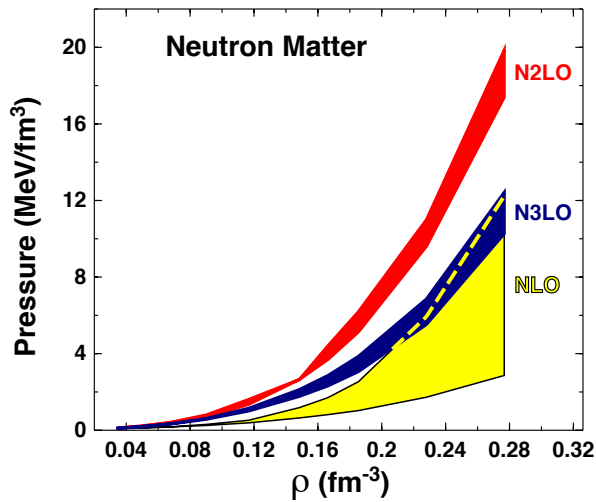}\hspace{.1in}
\vspace*{-3.5cm}
\caption{ Pressure in pure neutron matter
as a function of density, $\rho$. The yellow and
red bands represent the uncertainties in the predictions due to cutoff variations
as obtained in complete calculations at NLO and N$^2$LO, respectively.
The blue band is the result of a calculation employing N$^3$LO $NN$ potentials together with
N$^2$LO 3NFs. 
The dashed line shows the upper limit of the yellow band.     
(Figure reproduced from Ref.~\cite{FSskin15}.)                                        
} 
\label{pr}
\end{figure}

\begin{figure}[!t]
\centering 
\vspace*{-3cm}
\includegraphics[width=11cm]{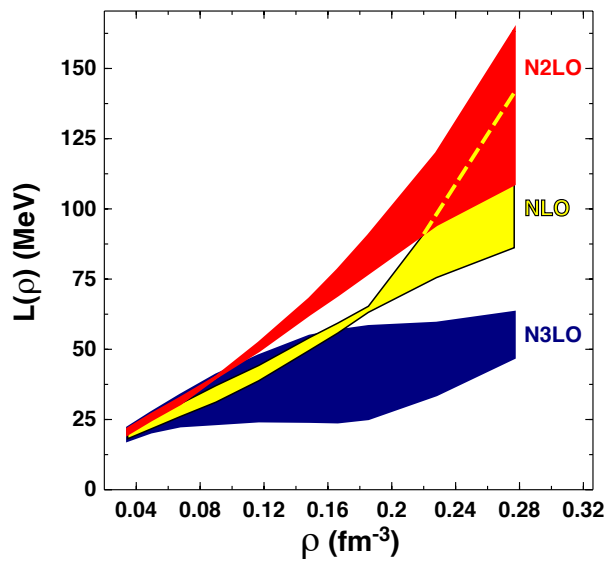}\hspace{.1in}
\vspace*{-3.5cm}
\caption{ The $L$ parameter as a function of density, as defined in Eq.~(\ref{L2}). 
(Figure reproduced from Ref.~\cite{FSskin15}.)
} 
\label{Lrho}
\end{figure}

In the figures which follow, the size of each band is obtained from variations of the cutoff
between 450 and 600 MeV in the regulator applied to the 2NF and the 3NF.                 
In Fig.~\ref{pr}, the pressure in neutron matter is shown. The yellow and                          
red bands represent the uncertainties in the predictions due to cutoff variations
as obtained in complete calculations at NLO and N$^2$LO, respectively.
The blue band is the result of a calculation employing N$^3$LO $NN$ potentials together with
3NFs at N$^2$LO.                
The pressure is proportional to the slope 
of the various curves which make up the corresponding bands shown in Fig.~\ref{pnm}.      
We observe moderate cutoff dependence except at NLO and a slow convergence tendency with 
increasing order. 

As already pointed out,                         
the $L$ parameter, defined as in Eq.~(\ref{L}), is sensitive to the                                         
characteristics of the equation of state of symmetric matter through 
$\rho_0$. The latter                                                             
changes dramatically from 
order to order as well as with changing cutoff, which can be clearly seen from Fig.~\ref{snm}.            
In Fig.~\ref{Lrho}, we show the $L$ parameter as a function of density, {\it i.~e.} 
\begin{equation} 
L(\rho) = 3 \rho \Big (\frac{\partial E_{sym}(\rho')}{\partial \rho'}\Big )_{\rho'=\rho}\; ,       
\label{L2} 
\end{equation} 
which reflects the difference between the pressures 
in NM and in SNM at each density.                                
The derivative of the EoS of SNM comes in through the symmetry energy 
and determines larger uncertainties than those seen in Fig.~\ref{pr}.     

The predictions for the skin thickness of $^{208}$Pb are summarized in Table~\ref{tab1}, along
with the corresponding values of the $L$ parameter at the appropriate saturation density, 
different in each case and also reported in Table~\ref{tab1}. 
Note that we do not show predictions at NLO because, at this low order, only the EoS with 
the largest cutoff (of 600 MeV) displays some (late) saturating behavior, cf.~Fig.~\ref{snm}. 
The upper and lower errors are the distances of the largest and smallest values (when changing
the cutoff) from the average.

\begin{table}                
\centering
\begin{tabular}{|c||c|c|c|}
\hline
Order & $S$(fm) & $L(\rho_0)$(MeV) & $\rho_0$(fm$^{-3}$)  \\
\hline     
\hline
N$^2$LO & $0.21^{+0.04}_{-0.02}$ & $77.4^{+31.2}_{-16.2}$ & $0.167^{+0.043}_{-0.022}$  \\
   &       &   &  \\ 
N$^3$LO & $0.17^{+0.02}_{-0.01}$ & $39.9^{+17.2}_{-15.7}$ & $ 0.144^{+0.032}_{-0.032} $ \\
\hline
\end{tabular}
\caption                                                    
{Neutron skin thickness, $S$, in $^{208}$Pb at the specified order of chiral EFT
as explained in the text. The corresponding values of the $L$ parameter and the              
saturation density are given in the last two columns. 
} 
\label{tab1}
\end{table}

The truncation error at order $\nu$ of chiral EFT is the difference between the predictions at orders
$\nu$+1 and $\nu$. Thus, from the Table, we can estimate this error at N$^2$LO to be about 0.04 fm.
A similar estimate at N$^3$LO would require knowledge of the prediction at 
N$^4$LO, which is not available. Assuming a (pessimistic) truncation error at 
N$^3$LO of similar size as the one at                               
N$^2$LO, we then summarize our predictions for the skin as 0.17$\pm$ 0.04 fm, where the error is likely to
be smaller assuming a reasonable convergence rate. 
[In fact, if one takes the cutoff variation as a realistic estimate of the error (as it is    
approximately the case at N$^2$LO, cf. Table~\ref{tab1}), then our N$^3$LO prediction carries an error of 0.02 fm.]

\subsubsection{Using a  phenomenological EoS for symmetric nuclear matter}
 
The nearly linear correlation between skins and neutron matter pressure typically observed in phenomenological investigations of skins~\cite{Typel,Brown} refers to a family of models with the same, or very similar, SNM properties
which differ mostly in the slope of neutron matter. This scenario  
can be simulated, for instance, by combining an empirical SNM equation of state together with different (microscopic) 
NM EoS, thus separating out the role of neutron matter pressure and removing any model dependence originating from the details of the saturation point.    

We repeated the calculations adopting, this time, the empirical EoS from 
Ref.~\cite{eos} for SNM. The latter is obtained from a Skyrme-type energy density functional and has a realistic saturation point at $\rho_0$=0.16 fm$^{-3}$ with 
energy per particle equal to -16.0 MeV.
The corresponding findings are displayed in Table~\ref{skintab}. For this test, we also show the results
at NLO, since the saturation point can be defined for all cases.
Although the midvalues are reasonably consistent with those in Table~\ref{tab1}, the uncertainties are 
much smaller, particularly for the $L$ parameter, as to be expected based on the previous observations.     
The much smaller uncertainty at N$^3$LO reflects the negligible cutoff dependence of neutron matter 
pressure at that order, see Fig.~\ref{pr}. 

\begin{table}                
\centering
\begin{tabular}{|c||c|c|}
\hline
Order & $S$(fm) & $L(\rho_0)$(MeV) \\
\hline     
\hline
NLO & $0.126^{+0.004}_{-0.003}$ & $20.4^{+8.8}_{-6.3}$ \\
   &       &     \\ 
N$^2$LO & $0.20^{+0.01}_{-0.01}$ & $70.6^{+4.1}_{-8.0}$ \\
   &       &     \\ 
N$^3$LO & $0.172^{+0.002}_{-0.005}$ & $44.9^{+3.8}_{-5.4}$ \\
\hline
\end{tabular}
\caption                                                    
{As Table \ref{tab1}, but employing a phenomenological model for the EoS of SNM. See text for details. 
} 
\label{skintab}
\end{table}

With similar considerations as above with respect to the truncation error, we define the uncertainty 
at N$^2$LO as the difference between the prediction at this order and the one at the next order, which 
gives approximately 0.03. Assuming a similar uncertainty at N$^3$LO, we estimate the skin thickness 
at N$^3$LO, when adopting an empirical parametrization for the EoS of SNM, to be 
0.17$\pm$0.03. We note, again, that this reflects the uncertainty in pure neutron matter at the low densities 
probed by the skin. 
Such uncertainty is small, consistent with the low-density behavior seen in Fig.~\ref{pr}.

We observe that our final estimate is consistent with the value reported in Ref.~\cite{EPJ14},
where the skin is obtained through correlations from Ref.~\cite{Brown}, and including a study based on 
the liquid drop model. 
This strengthens our confidence in the method we adopt to obtain the skin. 

To summarize, the neutron skin is an important isospin-sensitive ``observable'',           
essentially determined by the difference in pressure between symmetric and 
neutron matter.                                                                 
We calculated the neutron skin of $^{208}$Pb with two- and three-body chiral interactions.
The neutron and proton density functions are obtained in a simple approach based on the 
semi-empirical mass formula. 
We observed that, 
in fully microscopic calculations, model dependence from the details of SNM at the saturation point 
does impact predictions of the symmetry pressure and, to a lesser extent, the neutron skin. 

At the low densities typically probed by studies of the skin, EFT theoretical uncertainties for the skin
are small on a scale set by a realistic experimental uncertainty, 
particularly at the higher orders of chiral effective field theory. 

Calculations at N$^4$LO are needed for a better quantification of the truncation error at 
 N$^3$LO, and thus a reliable comparison of the EFT error 
with the target uncertainty set by future PREX experiments. Concerning the latter, from Ref.~\cite{Jlab}
we learn that the target uncertainty of PREX II is a factor of 3 smaller than the one
from the first PREX experiment, thus approximately $\pm$0.05. If accomplished, this will allow
to discriminate between theoretical predictions, along with the measured central value. For instance,
the present EFT predictions would not be consistent with 
a measurement such as 0.33 (the current central value) $\pm$0.05.

\section{Conclusions
\label{sec_concl}}

The past 20 years have seen great progress in our understanding of nuclear forces
in terms of low-energy QCD. Key to this development was the realization that
low-energy QCD is equivalent to an effective field theory (EFT) which              
has become known as chiral perturbation theory (ChPT).
In this framework, two- and many-body forces
 emerge on an equal footing and
the empirical fact that nuclear many-body forces are
substantially weaker than the two-nucleon force
is explained naturally.

We presented the current status of the development of chiral nuclear forces and discussed open 
questions and future challenges.
We also reviewed some representative examples for typical applications of chiral forces in many-body systems. For this we chose, specifically, nuclear and neutron-rich matter,
including isospin and spin asymmetries, as well as an analysis of neutron skin
thickness predictions in a neutron-rich nucleus.       

Chiral forces have also been applied in {\it ab initio} calculations of finite nuclei
(structure and reactions).
Because of lack of space, we could not discuss this topic in this review and, therefore,
we like to refer the interested reader to the comprehensive 
literature~\cite{Gez14,Hag12a,Hag12b,Bin14,BNV13,Hag14,Car15,Eks15,Car16,Nav16,Hag16}.

The importance of error quantification has finally been recognized in theoretical nuclear physics.   
We explored various sources of uncertainty systematically and noticed that
the largest uncertainty comes from the truncation error of the chiral expansion
(as given by the difference between the predictions at two consecutive orders).
We also found that the predictions up to N$^3$LO (fourth order) for many-body observables
carry a truncation error that is, in general, substantially larger than the error of the empirical information, rendering the predictions inconclusive. Thus, in many applications of chiral EFT it may be necessary to proceed beyond fourth order.
In any case, the convergence of the chiral expansion is one of the most important
issues to which more work needs to be devoted in the near future.

%%%%%%%%%%%%%%%%%%%%%%%%%%%%%%%%%%%%%%%%%%

\ack
The authors acknowledge support by the US Department of Energy under Grant No.\ DE-FG02-03ER41270.
We thank our collaborators L. Coraggio, D. R. Entem, J. W. Holt, 
N. Itaco, N. Kaiser, L. E. Marcucci, 
and Y. Nosyk for their contributions to the work reviewed in this article.

%%%%%%%%%%%%%%%%%%%%%%%%%%%%%%%%%%%%%%%%%%


\section*{References}
\begin{thebibliography}{999} 
\bibitem{Rei68}
Reid R V 1968 {\it Ann. Phys. (N.Y.)} {\bf 50} 411
\bibitem{BS64} 
Bryan R A and Scott B L 1964 {\it Phys. Rev.} {\bf 135} B434
\nonum
Bryan R A and Scott B L 1969 {\it Phys. Rev.} {\bf 177} 1435 
\bibitem{Erk74} 
Erkelenz K 1974 {\it Phys. Rep.} {\bf 13C} 191
\bibitem{HM75} 
Holinde K and Machleidt R 1975 {\it Nucl. Phys.} A {\bf 247} 495
\nonum
Holinde K and Machleidt R 1976 {\it Nucl. Phys.} A {\bf 256} 479
\bibitem{Mac89} 
Machleidt R 1989 {\it  Adv. Nucl. Phys.} {\bf 19} 189
\bibitem{Lac80} 
Lacombe M, Loiseau B, Richard J M, Vinh Mau R,
C\^{o}t\'{e} J, Pires P and de Tourreil R 1980 
{\it Phys. Rev.} C {\bf 21} 861
\bibitem{MHE87} 
Machleidt R, Holinde K and Elster Ch 1987
{\it Phys.\ Rep.} {\bf 149} 1
\bibitem{Wei79} Weinberg S 1979 {\it Physica} {\bf 96A} 327
\bibitem{Wei91} Weinberg S 1991 {\it Nucl.\ Phys.} B {\bf 363} 3
\bibitem{Wei92} Weinberg S 1992 {\it Phys.\ Lett.} B {\bf 295} 114
\bibitem{ME11}
Machleidt R and Entem D R 2011 {\it Phys. Rep.} {\bf 503} 1
\bibitem{EHM09} 
Epelbaum E, Hammer H-W and Mei\ss ner U-G 2009
{\it Rev. Mod. Phys.} {\bf 81} 1773
\bibitem{Mei16} 
Mei\ss ner U-G 2016
{\it Phys. Scr.} {\bf 91} 033005
\bibitem{Mac14}
Machleidt R 2014 {\it Scholarpedia} {\bf 9}(1) 30710
\bibitem{Mac16}
Machleidt R 2016 {\it Symmetry} {\bf 8} 26
\bibitem{Org15}
Orginos K, Parre\~{n}o A, Savage M J, Beane S R, Chang E and Detmold W 2015
{\it Phys. Rev.} D {\bf 92} 114512
\bibitem{Hat12}
Hatsuda T 2012
{\it J. Phys. Conf. Ser.} {\bf 381} 012020
\bibitem{Sch03} 
Scherer S 2003 {\it Adv. Nucl. Phys.} {\bf 27} 277 
\bibitem{PDG} 
Olive K A {\it et al} (Particle Data Group) 2014 
{\it Chin. Phys.} C {\bf 38} 0900001
\bibitem{CCWZ} 
Coleman S, Wess J and Zumino B 1969 {\it Phys.\ Rev.} {\bf 177} 2239
\nonum
Callan C G, Coleman S, Wess J and Zumino B 1969 
{\it Phys. Rev.} {\bf 177} 2247
\bibitem{Fet00} 
Fettes N, Mei\ss ner U-G, Moj\v{z}i\v{s} M and Steininger S 2000 
{\it Ann.\ Phys.\ (N.Y.)} {\bf 283} 273
\nonum
Fettes N, Mei\ss ner U-G, Moj\v{z}i\v{s} M and Steininger S 2001
{\it Ann.\ Phys.\ (N.Y.)} {\bf 288} 249
\bibitem{KGE12}
Krebs H, Gasparyan A and Epelbaum E 2012
{\it Phys. Rev.} C {\bf 85} 054006
\bibitem{ORK96}
Ord\'o\~nez C, Ray L and van Kolck U 1996
{\it Phys.\ Rev.} C {\bf 53} 2086
\bibitem{Kol94} 
van Kolck U 1994 {\it Phys. Rev.} C {\bf 49} 2932
\bibitem{Epe02b} 
Epelbaum E, Nogga A, Gl\"ockle W, Kamada H, Mei\ss ner U-G and Witala H 2002
{\it Phys. Rev.} C {\bf 66} 064001
\bibitem{Kai00a} Kaiser N 2000 {\it Phys.\ Rev.} C {\bf 61} 014003
\bibitem{Kai00b} Kaiser N 2000 {\it Phys.\ Rev.} C {\bf 62} 024001
\bibitem{EM03} 
Entem D R and Machleidt R 2003
{\it Phys.\ Rev.} C {\bf 68} 041001
\bibitem{Ent15a}
Entem D R, Kaiser N, Machleidt R and Nosyk Y 2015 
{\it Phys. Rev.} C {\bf 91} 014002
\bibitem{KGE13}
Krebs H, Gasparyan A and Epelbaum E 2013
{\it Phys. Rev.} C  {\bf 87} 054007
\bibitem{GKV11}
Girlanda L, Kievsky A and Viviani M 2011
{\it Phys. Rev.} C {\bf 84} 014001
\bibitem{Ent15b}
Entem D R, Kaiser N, Machleidt R and Nosyk Y 2015 
{\it Phys. Rev.} C {\bf 92} 064001
\bibitem{EM03a}
Entem D R and Machleidt R unpublished
\bibitem{Liu10}
Liu J {\it et al} 2010 {\it Phys. Rev. Lett.} {\bf 105} 181803
\bibitem{Pav00}
Pavan M M, Arndt R A, Strakovsky I I and Workman R L 2000
{\it Physica Scripta} {\bf T87} 65
\bibitem{Arn00}
Arndt R A, Strakovsky I I, Workman R L and Pavan M M 2000
{\it Physica Scripta} {\bf T87} 62
\bibitem{KBW97} 
Kaiser N, Brockmann R and Weise W 1997
{\it Nucl.\ Phys.} A {\bf 625} 758
\bibitem{EGM04} 
Epelbaum E, Gl\"ockle W and Mei\ss ner U-G 2004
{\it Eur.\ Phys.\ J.} A {\bf 19} 401
\bibitem{Kai01a} 
Kaiser N 2001
{\it Phys.\ Rev.} C {\bf 64} 057001
\bibitem{Kai01} 
Kaiser N 2001
{\it Phys.\ Rev.} C {\bf 63} 044010
\bibitem{Kai02} 
Kaiser N 2002
{\it Phys. Rev.} C {\bf 65} 017001
\bibitem{EAH71} 
Erkelenz K, Alzetta R and Holinde K 1971
{\it Nucl. Phys.} A {\bf 176} 413
\bibitem{Mac93}
Machleidt R 1993 {\it Computational Nuclear Physics 2 -- Nuclear Reactions}
ed K Langanke {\it et al} (New York: Springer) 
pp~1-29
\bibitem{Mac01} 
Machleidt R 2001 {\it Phys. Rev.} C {\bf 63} 024001
\bibitem{Arn06} 
Arndt R A, Briscoe W J, Strakovsky I I and Workman R L 2006
{\it Phys. Rev.} C {\bf 74} 045205
\bibitem{Koc86} 
Koch R 1986 {\it Nucl. Phys.} A {\bf 448} 707
\bibitem{Sto93} 
Stoks V G J, Klomp R A M, Rentmeester M C M and de Swart J J 1993 
{\it Phys.\ Rev.} C {\bf 48} 792
\bibitem{SM99} 
Arndt R A, Strakovsky I I and Workman R L 1999
{\it SAID Partial-Wave Analysis Facility} Data Analysis Center,
The George Washington University, Solution SM99 (Summer 1999)
\bibitem{EM02} 
Entem D R and Machleidt R 2002
{\it Phys. Rev.} C {\bf 66} 014002
\bibitem{SP07}  
Briscoe W J, Strakovsky I I and Workman R L 2007
{\it SAID Partial-Wave Analysis Facility} Data Analysis Center,
The George Washington University,  Solution SP07 (Spring 2007)
\bibitem{EGM98}
Epelbaum E, Gl\"ockle W and Mei\ss ner U-G 1998 
{\it Nucl.\ Phys.} A {\bf 637} 107
\bibitem{Lep97}
Lepage G P 1997
How to Renormalize the Schr\"odinger Equation
arXiv:nucl-th/9706029
\bibitem{Mar13}
Marji E {\it et al} 2013 {\it Phys. Rev.} C {\bf 88} 054002
\bibitem{EM02a} 
Entem D R and Machleidt R 2002
{\it Phys.\ Lett.} B {\bf 524} 93
\bibitem{Sto94} 
Stoks V G J, Klomp R A M, Terheggen C P F and de Swart J J 1994 
{\it Phys.\ Rev.} C {\bf 49} 2950
\bibitem{WSS95} 
Wiringa R B, Stoks V G J and Schiavilla R 1995
{\it Phys.\ Rev.} C {\bf 51} 38
\bibitem{EGM05} 
Epelbaum E, Gl\"ockle W and Mei\ss ner U-G 2005
{\it Nucl.\ Phys.} A {\bf 747} 362
\bibitem{EKM15a}
Epelbaum E, Krebs H and Mei\ss ner U-G 2015 
{\it Eur. Phys. J.} A {\bf 51} 53
\bibitem{EKM15b}
Epelbaum E, Krebs H and Mei\ss ner U-G 
2015 {\it Phys. Rev. Lett.} {\bf 115} 122301
\bibitem{EMW02} 
Entem D R, Machleidt R and Witala H 2002
{\it Phys. Rev.} C {\bf 65} 064005
\bibitem{Cau02}
Caurier E,  Navratil P, Ormand W E and Vary J P 2002
{\it Phys. Rev.} C {\bf 66} 024314
\bibitem{Cor14}
Coraggio L, Holt J W, Itaco N, Machleidt R, Marcucci L E and Sammarruca F 2014
{\it Phys. Rev.} C {\bf 89} 044321
\bibitem{Eks13}
Ekstr\H{o}m {\it et al} 2013 {\it Phys. Rev. Lett.} {\bf 110} 192502
\bibitem{Gez14}
Gezerlis A {\it et al} 2014
{\it Phys. Rev.} C {\bf 90} 054323
\bibitem{Pia15}
Piarulli M, Girlanda L, Schiavilla R, Navarro Perez R,  Amaro J E and Ruiz Arriola E 2015
{\it Phys. Rev.} C {\bf 91} 024003
\bibitem{Sam15}
Sammarruca F, 
Coraggio L, Holt J W, Itaco N, Machleidt R and Marcucci L E 2015
{\it Phys. Rev.} C {\bf 91} 054311
\bibitem{FM57}
Fujita J-I and Miyazawa H 1957 
{\it Prog. Theor. Phys.} {\bf 17} 360 
\bibitem{Coo79}
Coon S A, Scadron M D, McNamee P C, Barrett B R, Blatt D W E and McKeller B H J 1979
{\it Nucl. Phys.} A {\bf 317} 242
\bibitem{Nog06}
Nogga A, Navratil P, Barrett B R and Vary J P 2006
{\it Phys.\ Rev.} C {\bf 73} 064002
\bibitem{Nav07}
Navratil P, Gueorguiev V G, Vary J P, Ormand W E and Nogga A 2007
{\it Phys. Rev. Lett.} {\bf 99} 042501
\bibitem{Mar12}
 Marcucci L E, Kievsky A, Rosati S, Schiavilla R and Viviani M 2012
{\it Phys. Rev. Lett.} {\bf 108} 052502
\bibitem{Hag12a}
Hagen G, Hjorth-Jensen M, Jansen G R, Machleidt R and Papenbrock T 2012
{\it Phys. Rev. Lett.} {\bf 108} 242501
\bibitem{Hag12b}
Hagen G, Hjorth-Jensen M, Jansen G R, Machleidt R and Papenbrock T 2012
{\it Phys. Rev. Lett.} {\bf 109} 032502
\bibitem{Viv10}
Viviani M, Giarlanda L, Kievsky A and Marcucci L E 2013
{\it Phys. Rev. Lett.} {\bf 111} 172302
\bibitem{IR07} 
Ishikawa S and Robilotta M R 2007 
{\it Phys. Rev.} C {\bf 76} 014006
\bibitem{Ber08}
Bernard V,  Epelbaum E, Krebs H and Mei\ss ner U-G 2008
{\it Phys. Rev.} C {\bf 77} 064004
\nonum
Bernard V,  Epelbaum E, Krebs H and Mei\ss ner U-G 2011
{\it Phys. Rev.} C {\bf 84} 054001
\bibitem{Ski11}
Skibinski R {\it et al} 2011
{\it Phys. Rev.} C {\bf 84} 054005
\bibitem{Wit12}
Witala H {\it et al} 2013
{\it Few-Body Syst.} {\bf 54} 897
\bibitem{Tew12}
Tews I, Kr\"uger T, Hebeler K and Schwenk A 2013 
{\it Phys. Rev. Lett.} {\bf 110} 032504
\bibitem{Epe07}
Epelbaum E 2007
{\it Eur. Phys. J.} A {\bf34} 197
\bibitem{Roz06}
Rozpedzik D, Golak J, Skibinski R, Witala H, Gl\"ockle W, Epelbaum  E, Nogga A and Kamada H 2006
{\it Acta Phys. Polon.} B {\bf 37} 2889
(arXiv:nucl-th/0606017)
\bibitem{Kai12}
Kaiser N 2012 {\it Eur. Phys. J.} A {\bf 48} 135
\bibitem{KM16}
Kaiser N and Milkus R 2016 {\it Eur. Phys. J.} A {\bf 52} 4
\bibitem{furnstahl15} 
Furnstahl R J, Phillips D R and Wesolowski S 2015
{\it J. Phys.} G {\bf 42} 034028
\bibitem{coraggio13} 
Coraggio L, Holt J W, Itaco N, Machleidt R and Sammarruca F 2013 
{ \it Phys. Rev.} C {\bf 87} 014322
\bibitem{krueger} 
Kr{\"u}ger T, Tews I, Hebeler K and Schwenk A 2013
{ \it Phys. Rev.} C {\bf 88} 025802
\bibitem{bogner05} 
Bogner S K, Schwenk A, Furnstahl R J and Nogga A 2005 
{\it Nucl. Phys.} A {\bf 763} 59
\bibitem{hebeler11} 
Hebeler K, Bogner S K, Furnstahl R J, Nogga A and Schwenk A 2011
 { \it Phys. Rev.} C {\bf 83} 031301
\bibitem{gezerlis13} 
Gezerlis A, Tews I, Epelbaum E, Gandolfi S, Hebeler K, Nogga A and Schwenk A 2013 
{ \it Phys. Rev. Lett.} {\bf 111} 032501
\bibitem{cc1} 
Baardsen G {\it et al} 2013 {\it Phys. Rev.} C {\bf 88} 054312
\bibitem{cc2}  
Hagen G {\it et al} 2014 {\it Phys. Rev.} C {\bf 89} 014319
\bibitem{Gran}  
Navarro Perez R, Garrido E,  Amaro J E and Ruiz Arriola E 2014 
{\it Phys. Rev.} C {\bf 90} 047001
\bibitem{AS} 
Alonso D and Sammarruca F 2003
 {\it Phys. Rev.} C {\bf 67} 054301                    
\bibitem{Drischler} 
Drischler C, Soma V and Schwenk A 2014
 {\it Phys. Rev.} C {\bf 89} 025806
 \bibitem{Kai15}
 Kaiser N 2015 {\it Phys. Rev.} C {\bf 91} 065201
\bibitem{Sam10} 
Sammarruca F 2010 {\it Phys. Rev.} C {\bf 82} 027307          
\bibitem{Lane} 
Lane A M 1962 {\it Nucl. Phys.} {\bf 35} 676              
\bibitem{Sam11} 
Sammarruca F 2011 {\it Phys. Rev.} C {\bf 83} 064304            
\bibitem{SK07} 
Sammarruca F and Krastev P 2007
 {\it Phys. Rev.} C {\bf 75} 034315               
\bibitem{pol18} 
Vida{\~n}a I and Bombaci I 2002
 {\it Phys. Rev.} C {\bf 66} 045801             
\bibitem{IY} 
Isayev A A and Yang J 2004 {\it Phys. Rev.} C {\bf 70} 064310           
\bibitem{pol12} 
Marcos S, Niembro R, Quelle M L and Navarro J 1991
 {\it Phys. Lett.} B {\bf 271} 277       
\bibitem{Carls03} 
Carlson J, Chang S-Y, Pandharipande V R and Schmidt K E 2003
 {\it Phys. Rev. Lett.} {\bf 91} 050401           
\bibitem{Carls12} 
Carlson J, Gandolfi S and Gezerlis A 2012
 {\it PTEP} 01A209         
\bibitem{Bloch08} 
Bloch I, Dalibard J and Zwerger W 2008
 {\it Rev. Mod. Phys.} {\bf 80} 885        
\bibitem{krueg14}  
Kr{\" u}ger T, Hebeler K and Schwenk A 2014
 {\it Phys. Lett.} B {\bf 744} 18 and references therein 
\bibitem{APR} 
Akmal A, Pandharipande V R and Ravenhall D J 1998
 {\it Phys. Rev.} C {\bf 59} 1804        
\bibitem{pol19}  
Isayev A A and Yang J 2004
 {\it Phys. Rev.} C {\bf 69} 025801            
\bibitem{SMK15} 
Sammarruca F, Machleidt R and Kaiser N 2015
 {\it Phys. Rev.} C {\bf 92} 054327
\bibitem{Tsang+} 
Tsang M B {\it et al} 2012 {\it Phys. Rev.} C {\bf 86} 015803 and references therein 
\bibitem{FSskin15}
Sammarruca F 2015 {\it Symmetry} {\bf 7} 1646 and references therein    
\bibitem{Hor} 
Horowitz C J, Pollock S J, Souder P A and Michaels R 2001
 {\it Phys. Rev.} C {\bf 63} 025501             
\bibitem{Jlab} 
Abrahamyan S {\it et al} (PREX Collaboration) 2012 {\it Phys. Rev. Lett.}  {\bf 108} 112502          
\bibitem{AS03} 
Alonso D and Sammarruca F 2003
 {\it Phys. Rev.} C {\bf 68} 054305             
\bibitem{Furn} 
Furnstahl R J 2002 {\it Nucl. Phys.} A {\bf 706} 85                                       
\bibitem{Typel} 
Typel S and Brown B A 2001 {\it Phys. Rev.} C {\bf 64} 027302
\bibitem{Brown} 
Brown B A 2000 {\it Phys. Rev. Lett.} {\bf 85} 5296 
\bibitem{eos} 
Alam N, Agrawal B K, De J N, Samaddar S K and Col\`{o} G 2014 
{\it Phys. Rev.} C {\bf 90} 054317
\bibitem{EPJ14} 
Hebeler K and Schwenk A 2014 
{\it Eur. Phys. J.} A {\bf 50} 14011

\bibitem{Bin14}
Binder S, Langhammer J, Calci A and Roth R 2014
{\it Phys. Lett.} B {\bf 736} 119
\bibitem{BNV13}
Barrett B R, Navratil P and Vary J P 2013
{\it Prog. Part. Nucl. Phys.} {\bf 69} 131
\bibitem{Hag14}
Hagen G, Papenbrock T, Hjorth-Jensen M and Dean D J 2014
{\it Rept. Prog. Phys.} {\bf 77} 096302
\bibitem{Car15}
Carlson J, Gandolfi S, Pederiva F, Pieper S C, Schiavilla R, Schmidt K E and Wiringa R B 2015
{\it Rev Mod. Phys.} {\bf 87} 1067
\bibitem{Eks15}
Ekstr\H{o}m A {\it et al} 2015
{\it Phys. Rev.} C {\bf 91} 051301
\bibitem{Car16}
Carlsson B D {\it et al} 2016
{\it Phys. Rev.} X {\bf 6} 011019
\bibitem{Nav16}
Navrátil P, Quaglioni S, Hupin G, Romero-Redondo C and Calci A
2016 {\it Phys. Scr.} {\bf 91} 053002
\bibitem{Hag16}
Hagen G, Hjorth-Jensen M, Jansen G R and Papenbrock T
2016 {\it Phys. Scr.} {\bf 91} 063006
\end{thebibliography}
\end{document}